\documentclass[12pt,fullpage]{article}
\usepackage{amsmath,graphicx,amsfonts,amssymb,sansseriftitles,natbib,bm}
\usepackage[title]{appendix}
\usepackage{algorithm,algorithmic}
\usepackage{caption}
\usepackage[section]{placeins}
\usepackage{lscape}
\usepackage{xcolor}
\usepackage[colorlinks=true,citecolor=blue,linkcolor=blue,urlcolor=blue]{hyperref}
\usepackage{subfigure}
\usepackage{setspace}
\usepackage{ragged2e}
\usepackage{multirow,booktabs}

\input epsf

\newcommand{\bi}{\begin{itemize}}
\newcommand{\ei}{\end{itemize}}
\newcommand{\be}{\begin{enumerate}}
\newcommand{\ee}{\end{enumerate}}

\newcommand {\bea}{\begin{eqnarray}}
\newcommand {\eea}{\end{eqnarray}}
\newcommand {\Pb}{\text{P}}
\newcommand {\Ex}{\text{E}}
\newcommand {\Var}{\text{Var}}

\newcommand {\bs}{{\bf s}}

\newcommand {\bp}{{\bf p}}

\newcommand {\bz}{{\bf z}}
\newcommand {\bI}{{\bf I}}

\newcommand {\bmu}{{\boldsymbol \mu}}
\newcommand {\bnu}{{\boldsymbol \nu}}

\newcommand {\btheta}{{\boldsymbol \theta}}
\newcommand {\bvartheta}{{\boldsymbol \vartheta}}
\newcommand {\bSigma}{{\bsym \Sigma}}
\newcommand {\bSigmaOld}{{\bsym \Sigma}_{\mbox{\tiny old}}}
\newcommand {\bmuold}{{\bsym \mu}_{\mbox{\tiny old}}}

\newcommand {\bSigmaNew}{{\bsym \Sigma}_{\mbox{\tiny new}}}
\newcommand {\bmuNew}{{\bsym \mu}_{\mbox{\tiny new}}}

\newcommand{\balpha}{{\bsym{\alpha}}}

\newcommand {\sA}{s_{\mbox{\tiny A}}}
\newcommand {\sB}{s_{\mbox{\tiny B}}}

\newcommand {\VA}{J^{\mbox{\tiny A-BR}}}

\newcommand {\piA}{\pi_{\mbox{\tiny A}}}
\newcommand {\piB}{\pi_{\mbox{\tiny B}}}
\newcommand {\pA}{p_{\mbox{\tiny A}}}
\newcommand {\pB}{p_{\mbox{\tiny B}}}
\newcommand {\PA}{P_{\mbox{\tiny A}}}

\newcommand {\VVA}{V_{\mbox{\tiny A}}}
\newcommand {\VVB}{V_{\mbox{\tiny B}}}

\newcommand {\JpiApib}{J^{{\tiny \pi_{\mbox A}, \pi_{\mbox B}}}}
\newcommand {\JpisApibs}{J^{{\tiny \pi_{\mbox A}^*, \pi_{\mbox B}^*}}}

\newcommand {\JpiApibs}{J^{{\tiny \pi_{\mbox A}, \pi_{\mbox B}^*}}}
\newcommand {\JpisApib}{J^{{\tiny \pi_{\mbox A}^*, \pi_{\mbox B}}}}

\def\bsym#1{{\boldsymbol{#1}}}

\newtheorem{rem}{Remark}
\newtheorem{assum}{Assumption}

\begin{document}

\title{\vspace{20mm}\sffamily{\bfseries An Empirical Bayes Approach for Estimating Skill Models for Professional Darts Players } } 


\author{\sffamily Martin B. Haugh \\
\sffamily Department of Analytics, Marketing \& Operations \\
\sffamily Imperial College Business School, Imperial College \\
\texttt{m.haugh@imperial.ac.uk}
\vspace{.75cm}
\and \sffamily Chun Wang \\
\sffamily  Department of Management Science and Engineering\\
\sffamily School of Economics and Management, Tsinghua University \\
\texttt{wangchun@sem.tsinghua.edu.cn}}

\date{ \today}

\maketitle

\centerline {\large \bf Abstract} \baselineskip 14pt
\noindent
We perform an exploratory data analysis on a data-set for the top 16 professional darts players from the 2019 season. We use this data-set to fit player skill models which can then be used in dynamic zero-sum games (ZSGs) that model real-world matches between players. We propose an empirical Bayesian approach based on the Dirichlet-Multinomial (DM) model that overcomes limitations in the data. Specifically we introduce two DM-based skill models where the first model borrows strength from other darts players and the second model borrows strength from other regions of the dartboard. We find these DM-based models outperform simpler benchmark models with respect to Brier and Spherical scores, both of which are proper scoring rules. We also show in ZSGs settings that the difference between DM-based skill models and the simpler benchmark models is practically significant. Finally, we use our DM-based model to analyze specific situations that arose in real-world darts matches during the 2019 season.


\noindent
\newpage


\section{Introduction}
\label{sec:intro}

The game of darts has experienced a surge in popularity in recent years and now has a substantial presence in many countries including the U.K., Germany, the Netherlands, and Australia, among others. Indeed, a recent headline in \cite{Economist2020} explains how darts has moved from pastime to prime time and how it is quite common today to have tens of thousands of fans attend darts tournaments. The article also notes how in recent years darts has become the second-most-watched sport over the Christmas period on Sky Sports in the U.K., coming second only to football. Further growth is expected and professional tournaments now take place in locations such as Shanghai and Madison Square Garden in New York - locations that are far removed from the U.K. and the Netherlands, the traditional hot-beds of darts. Even more recently, in January 2024 the 16-year old Luke Littler became the youngest-ever player to reach the final of the PDC World Championships. Beyond the sports pages, his exploits were covered in the New York Times, Bloomberg News and even on the front page of the Financial Times among others. The game is also becoming increasingly attractive to women as evidenced by the exploits of Fallon Sherrock who in 2019 became the first woman to win a match at the PDC World Darts Championship and in fact made it to the third round of that tournament.

In this paper we conduct an exploratory data analysis of a darts data-set based on 16 professional players' dart throws in the 2019 season. Our ultimate goal is to build skill models that accurately reflect the players' skill levels and that can be used as inputs for zero-sum games (ZSGs) that model real-world darts matches. There are two major issues with the data-set, however. First, the data is too coarse and we do not have enough data for some player / target-region combinations. Second, while we know the target area for each dart in the data-set, we do not know the precise location in the target area at which it was aimed. These issues result in several problems. For example, estimating a player's skill model in certain regions of the dartboard becomes challenging and there is also difficulty in inferring whether the mean outcome of a dart differs from the intended target, i.e. whether or not there is a {\em bias}.

We propose to partially resolve these issues via an empirical Bayes approach that utilizes the Dirichlet-Multinomial (DM) distribution to borrow strength from (a) all players or (b) all regions (of the dartboard) when fitting each player's skill model. Our overall approach to building skill models consists of two stages. The first stage utilizes the DM model to convert the raw data-counts into pseudo-counts for the various target regions of the dartboard. This ensures there is sufficient data-coverage for all players in the important target areas of the board.
The second stage is only required if we want to consider ZSG settings where (say) every square millimeter of the dartboard is a possible target. It proceeds by feeding the pseudo-counts into a bivariate-normal skill model. Our bivariate-normal skill model extends \cite{Tibshirani} (hereafter TPT) by (i) partitioning the dartboard by target area to reflect the fact that a professional player's skill level varies considerably by target area and (ii) allowing for the precise target of the dart to also be inferred. As with TPT, we fit the bivariate-normal skill models via the EM algorithm. We show via the use of proper scoring rules that the DM-based models outperform benchmark models from both a statistical and practical viewpoint.

Our DM approach resolves several issues but bias and correlation identification issues remain with the bivariate-normal skill model from the second stage. We finesse these issues in the context of ZSGs by considering so-called ``single-action'' and ``multi-action'' players. The single-action player is only allowed to aim at a single point in each possible target region, (e.g. treble 20), whereas the multi-action player can aim at any point in the target area. The single-action player only needs the skill model given directly by the DM-based pseudo-counts in the first stage above whereas the multi-action player needs the bivariate-normal skill model from the second stage. We show that the single-action player is only at a very small disadvantage to the multi-action player in real-world darts matches and in fact even this small disadvantage is likely to be overstated if we factor in possible bias and model estimation errors. We therefore conclude the first-stage skill model (based on pseudo-counts arising from the DM models) is a viable alternative to the more general two-stage skill model that allows us to target any precise location on the dart-board.

We also show how the DM-based skill models can be used to analyze some specific situations that have arisen in real-world dart matches where a player's decisions were deemed surprising by some pundits. We consider two examples where our skill models allow us to argue that the player's decisions were in fact not surprising and suggest that the players themselves often have a better understanding of their own skills than the pundits / commentators.

We are certainly not the first to study the game of darts. \cite{Stern-CHANCE} and \cite{Percy99} consider the problem of where on the dartboard to aim in order to maximize the score of an individual dart. More recently, the aforementioned work of TPT proposed several models based on the normal and skew normal bivariate distributions to model the throwing skills of players. They develop an EM algorithm together with importance sampling to fit these distributions to dart-throwing data.
Utilizing the approach of TPT, \cite{MillerArch2021} consider where on the board the darts showed be aimed in order to estimate the player's skill model using as few darts as possible.
Another recent development is the work by \citet{HotHands_RSSA} who find some evidence for a weak-form of the so-called ``hot-hand'' phenomenon in darts. In particular, they find strong evidence for it during the three dart throws {\em within} a turn\footnote{We clarify what we mean by a ``turn'' in Section \ref{sec:Rules}.}, but they find little evidence for it persisting {\em across} turns.  
\cite{Significance-RSS-2015} consider whether the standard layout of a darts board disadvantages left-handers and propose an alternative layout to counter any such disadvantage. \cite{Otting_2020_Pressure} use darts as a vehicle for understanding and predicting how individuals will perform in high pressure situations. They find no evidence in favor of either ``choking'' or excelling under pressure.

Another stream of research is concerned with finding optimal {\em strategies} for playing darts. \cite{Kohler} uses a dynamic programming (DP) formulation to minimize the expected number of darts throws to {\em check out}, i.e. to reach a score of zero with the last throw being a double. More recently \citet{Baird2020} considers the slightly more general DP formulation where the goal is to minimize the expected number of {\em turns} until checking out. Both of these papers ignore their opponent's score when constructing a strategy and therefore produce sub-optimal policies. \cite{HaughWang-Darts-2021} formulate the game of darts as a ZSG and show how to solve the ZSG efficiently using DP methods. Employing the same data-set considered in this paper, they use their ZSG solutions to argue that the importance of playing strategically, i.e. taking an opponent's score into account, could improve a player's win-probability by approximately 2\% - 3\% in real-world matches. In contrast to this paper, however, their focus was on formulating and solving the ZSGs. In particular, they did not consider at all the various problems that can arise when using their data-set to estimate the players' skill models which is the focus of this work. \cite{CFW-2023} use\footnote{They also won the best research paper award at the 2024 MIT Sloan Sports Analytics Conference.} a DP formulation to develop a fair handicapping system that allows players of different skill levels to compete against one another.

The remainder of the paper is organized as follows. In Section \ref{sec:Rules} we describe the rules of darts and we describe the data-set in Section \ref{sec:Data}. We propose our DM-based models in Section \ref{sec:Empirical-Bayes} and then present the bivariate-normal skill model in Section \ref{sec:SkillModel}. We use proper scoring rules to compare the DM-based models versus benchmark models in Section \ref{sec:ScoringRules} and use dartboard visualizations in Section \ref{sec:Viz} to emphasize how well the DM-based models resolve the coverage issues in the data-set. In Section \ref{sec:Empirical Bayes-ZSG} we consider the DM-based skill models in the context of ZSGs and we conclude in Section \ref{sec:Conclusions}. A {\em Supplementary Material} file contains additional material including the EM algorithm for fitting our extension of the bivariate-normal skill model as well as complete versions, i.e. for all 16 players, of all figures and tables that appear in the main text.
Finally, we note that some sections, e.g. Sections \ref{sec:Rules} and \ref{sec:Data}, closely follow analogous sections in \citet{HaughWang-Darts-2021}.

\section{The Rules of Darts}
\label{sec:Rules}

The score obtained by a dart is determined by where it lands on the board. The small concentric circles in the middle of the board define the ``double bulls-eye'' (DB) and the ``single bulls-eye'' (SB) regions. If a dart lands in the DB region (the small red circle) then it scores 50 points while a dart landing in the SB region (the green annulus surrounding the DB region) scores 25. See Figure \ref{fig:DartB1}. Beyond the SB region, the dartboard is divided into 20 segments of equal area corresponding to the numbers 1 to 20. \\

\noindent
\begin{minipage}[b]{0.55\linewidth}
If a dart lands in the ``20'' segment, for example, then it will land in the treble twenty (T20) region, the double twenty region (D20), or the single twenty region (S20) for scores of 60, 40 or 20, respectively. The double region is the region between the two outermost circles on the dartboard whilst the treble region is the region between the two circles beyond the SB region. The single region is then the union of the two disjoint regions between the SB and treble region and between the treble and double regions. If a dart lands beyond the double region then it scores zero.
\end{minipage} \hspace{.5cm}
\begin{minipage}[t]{0.4\linewidth} \vspace{-6.4cm}
\centering
\includegraphics[width=.9\linewidth]{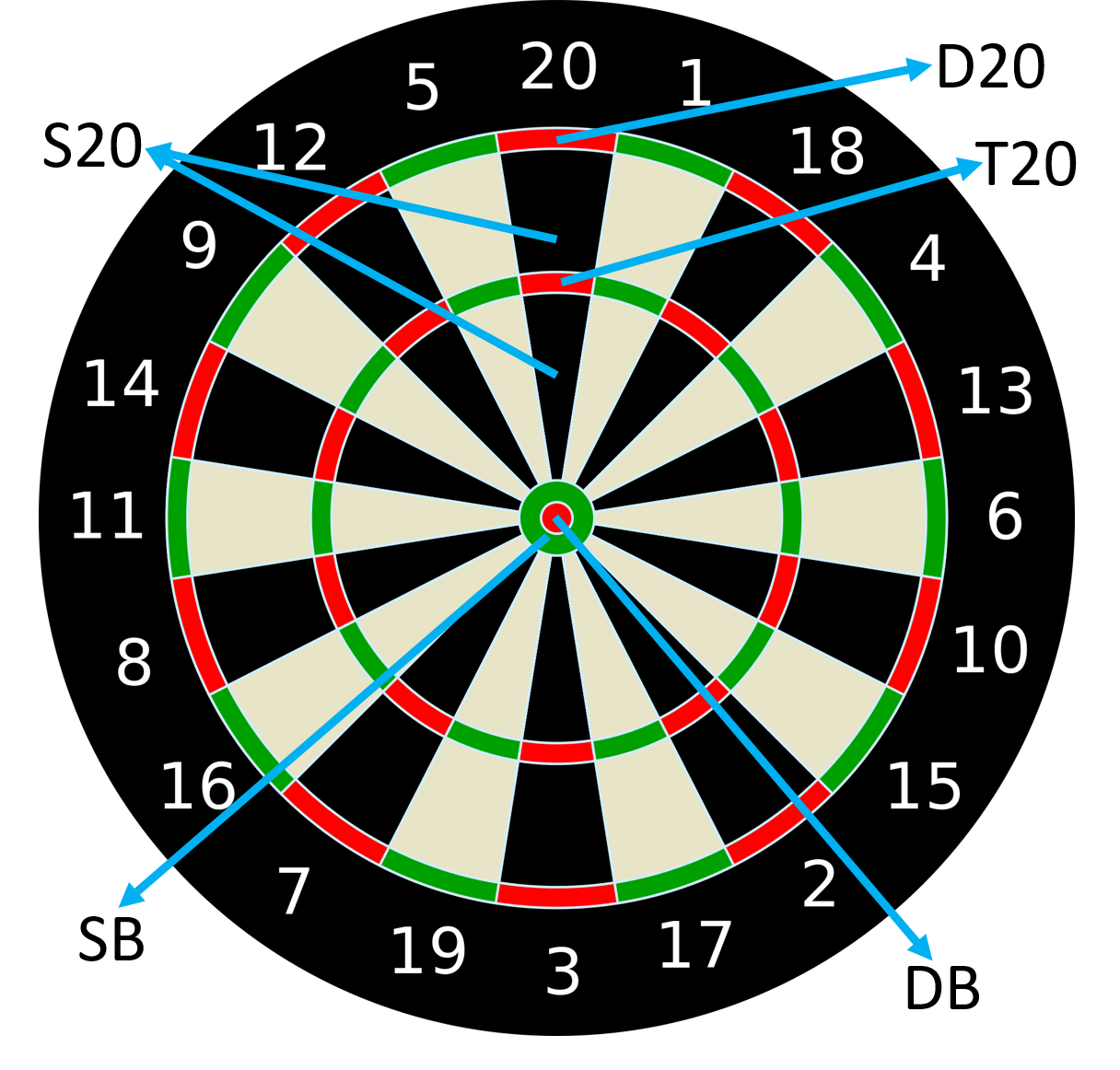}
\captionof{figure}{A standard dartboard}
\label{fig:DartB1}
\end{minipage}
\vspace{.05cm}
\noindent

The rules of darts that we describe here\footnote{The rules for 501 and other less common forms of the game are described at \url{https://www.mastersofgames.com/rules/darts-rules.htm}. Furthermore the geometry of the dartboard is described in Appendix \ref{sec:geom} of the {\em Supplementary Materials} file.} are for the most commonly played form of the game, namely ``501''. In 501, each player starts on a score of 501 and takes turns in throwing three darts. These three darts constitute a ``turn''. After a player's turn, his scores on the three darts are added together and subtracted from his score at the beginning of the turn. The first player to reach a score of {\em exactly} zero wins the game as long as his final dart is a double, i.e. a dart that lands on D1, D2, ..., D20, or DB in which case we say the player has {\em checked out}. If a player's turn would result in an updated score of 1, 0 (but not achieved via a double on his final dart) or a negative score, then the turn is invalidated, the player remains on the score he had prior to the turn, and his opponent then takes his turn. It's possible for a player to win without having to throw all three darts in his turn. For example, suppose a player has a score of 16 just prior to his turn. If he then scores D8 with the first dart of his turn he wins. Alternatively, if he scores S8 with his first dart, then he could still win by throwing a D4 with his second dart or (say) S4 and D2 with his second and third darts, respectively.

The game we have just described is known as a ``leg'' and in practice darts matches are typically played over many legs with the winner being the first to win some fixed number of legs. Alternatively, some tournaments have a legs and ``sets'' structure whereby the winner of the match is the first to win a fixed number of sets and the winner of a set is the first to win a fixed number of legs.
Finally, because the player that throws first has a considerable advantage in a given leg, players alternate in starting legs (and sets).

\section{The Data}
\label{sec:Data}

Our data-set relates to matches that were played by the top 16 professional players in the world during the 2019 season.
For each of the 16 players we have data of the form (TR, $z$, $n$) where $n$ is the number of darts that were aimed at the {\em target region} (TR) and achieved a score of $z$. There are a total of 62 possible target regions, namely the single regions S1, ..., S20, the double regions D1, ..., D20, the treble regions T1, ..., T20, and the single and double bulls-eye, i.e. SB and DB.
Because many regions of the dartboard are very rarely targeted, the target regions that appear in our data-set are the treble regions T20, T19, T18, and T17 together with all the double regions D1, ..., D20, and the double bulls-eye DB.

The possible realized value of the score $z$ depends on the target region TR since a professional darts player will only very rarely miss his target region TR by more than a small distance.
For each treble region, there are therefore 6 possible $z$ scores. In the case of TR = T20, for example, the possible values are $z \in \{\text{T20, S20, T5, S5, T1, S1} \}$ because the 5 and 1 segments of the dartboard are adjacent to the 20 segment; see Figure \ref{fig:DartB1}. Because we only have data for 4 treble regions, this means a total of $6 \times 4 = 24$ data-vectors (TR, $z$, $n$) for each player where TR is a treble region.
For each double target region TR = D$x$ for $1 \leq x \leq 20$, we have 7 possible $z$ values represented in the data-set. For example, if TR = D16, then the corresponding possible values of $z$ are $z\in \{\text{D16, S16, D8, S8, D7, S7, M}\}$ because 8 and 7 are adjacent to 16 on the dartboard and where M denotes a ``miss'', i.e. a dart that fails to score because it landed outside the double region; again, see Figure \ref{fig:DartB1}.
When TR = DB, we have 22 possible $z$'s corresponding to \{DB, SB, S1, S2, ..., S20\}.
Because targeting DB and some of the double regions is relatively rare, we note that for some of the (TR, $z$, $n$) combinations we often have $n=0$.

We provide the treble data in Appendix \ref{app:Data} and the entire data-set is available at \url{https://github.com/wangchunsem/OptimalDarts}. We now provide some summary statistics of the data. Aggregated across the 16 players, there were  117,600, 27,709, 7,717, and 2,461 attempts at T20, T19, T18, and T17, respectively, and 41.2\%, 41.7\%, 36.9\%, and 33.5\% of these attempts, respectively, were successful. However, the success rates at the individual level vary considerably more than the aggregate numbers. For example, several players have a success rate of less than 30\% when targeting T18 and over 40\% when targeting T20. There were a total of 16,777 attempts at the 21 double regions with D20 (4,399), D16 (2,338), and D10 (2,064) together accounting for over 50\% of this total. In contrast, there were only 1,866 attempts in total at the 10 odd doubles. These numbers reflect the preference for even doubles\footnote{If a player needs, say D20, to check out on the second dart of his turn, then if he hits an M and or S10 (the most likely fails) he can still check out with his third and final throw in the turn.} and, in particular, doubles that have powers of 2 in their factorizations. The average success rate at doubles across the 16 players was 40.2\%. Note this is smaller than the success rates for T20 and T19, despite the fact that the double beds are approximately 50\% larger than the treble areas or ``beds''. There are also substantial variations in success rates both across regions for a given player and across players. Some of this variation can certainly be explained by small number effects but even for the most popular double, i.e. D20, we see success rates vary from as low as 27\% to as high as 47\%.

\subsubsection*{Limitations of the Data-Set}
Unfortunately, the data-set has two important natural limitations. The first is that while we always know the realized value of the score $z$, we never know the precise point $(x,y)$ where each dart lands. This was also true of TPT's data-set and led to their development of EM algorithms for estimating skill models. The second natural limitation is that when a player targets a particular region, e.g. T20, we do not know the exact point in the region at which he was aiming. This was not true of TPT since they used self-generated data where they aimed at the center of the DB. This second limitation leads to some difficulty in interpreting the skill models from the second stage and in particular, raises the question of bias.

There are two other limitations of our data-set that we can assume away. The first one concerns so-called ``bounce-outs'' which are dart throws that fail to land in the dartboard. A ``bounce-out'' occurs because of a poor throw or because the tip of the dart strikes the wire that defines the boundaries between the different regions. With a sufficiently sharp dart tip, however, this rarely happens and in fact only approximately 3 in every 1,000 darts thrown by top players were bounce-outs in the 2019 season. We will assume there is no possibility of a bounce-out occurring when a dart is thrown.

The second limitation is that the data-set is not sufficiently granular to estimate a different skill model for each throw in a turn. It is known (see, for example, \citealp{HotHands_RSSA}) that the success rate of the first dart in a turn is generally smaller than the success rate of the second and third throws of the turn. This can be explained by the need of the dart-thrower to ``re-calibrate'' at the beginning of each turn.
If we had sufficiently granular data, then for each player we could fit one skill model for the first throw in a turn and a separate skill model for the combined second and third throws. As we do not have such data, however, we will assume that each player has the same skill model for each throw in his turn.

\section{An Empirical Bayes Approach via the DM Distribution (Stage 1)}
\label{sec:Empirical-Bayes}

The main idea behind our empirical Bayesian approach is to use a hierarchical Bayesian model that fits a skill model for a specific player-region pair by either (a) leveraging information across all players for the given region or (b) leveraging information across all regions for the given player. Model parameters are chosen\footnote{We use the {\em DirichletMultinomial} package \citet{DM-R_Package} in \texttt{R} to estimate the model parameters. We also note that we could have used a fully Bayesian hierarchical model rather than the empirical Bayesian approach we propose. We went with the latter, however, because it's simpler (it does not require posterior samples from an MCMC) and because we expect both approaches would yield very similar results.} via maximum likelihood estimation (MLE) rather than via the specification of a hyper-prior, and this results in updated ``raw'' data-sets which can, if necessary, then be used for fitting  (conditional) bivariate-normal skill models  for each player as discussed in Section \ref{sec:SkillModel}. Our specific model, the DM distribution, is a well-known generalization of the Beta-Binomial distribution and has been applied extensively in baseball analytics for example; see \citet{Jiang-Zhang2010} and \citet{Robinson_Emp-Bayes}. More generally, the DM distribution is often very useful for modelling proportional data and \citet{Minka2000} provides an excellent review of the model as well as various approaches to estimating it via MLE. Several chapters of \citet{EfronHastie} provide an overview of empirical Bayes as well as recent developments and applications of the methodology.
We first describe the Dirichlet distribution before proceeding in Sections \ref{sec:DM-Model} and \ref{sec:DM-Model-Regions} to describe how we use it to fit skill models for the darts players.

The Dirichlet distribution is a distribution on the $(K-1)$-dimensional simplex ${\cal S}_K :=\{{\bf p} \in \mathbb{R}^K_+ \, : \, \sum_{k=1}^K p_k = 1 \}$ and is therefore a distribution over probability vectors in $\mathbb{R}^K_+$. The density $f$ at a point $\bp = (p_1, \ldots , p_K) \in {\cal S}_K$ satisfies
\[
f(\bp) \sim {\cal D}(\balpha) := \frac{\Gamma \left( \sum_k \alpha_k \right)}{\prod_k \Gamma \left( \alpha_k \right)} \prod_k p_k^{\alpha_k - 1},
\]
where $\balpha := (\alpha_1, \ldots , \alpha_K) \in \mathbb{R}^K_+$ and $\Gamma(\cdot)$ denotes the Gamma function. It is well known that $\Ex[p_k] = \alpha_k/\sum_k \alpha_k$ and $\Var[p_k]=\Ex[p_k](1-\Ex[p_k])/\left(1+ \sum_k \alpha_k\right)$. The Dirichlet distribution is conjugate to the Multinomial distribution which makes the Dirichlet distribution very convenient for modelling proportions.

\subsection{The DM Model: Borrowing Strength from Players}
\label{sec:DM-Model}

Recall that there are 16 professional darts players in our data-set. In our first DM model, we assume that the skill model of the $j^{th}$ darts player targeting a particular region, e.g. D20, is represented by a draw $\bp_j \sim {\cal D}(\balpha)$. In this case $K=7$ because there are $7$ possible outcomes of a dart aimed at D20. Conditional on $\bp_j$, the dart scores of this player are $\mbox{Multinomial}(\bp_j)$. That is
\begin{eqnarray*}
\bp_j & \sim & {\cal D}(\balpha), \label{eq:Post-DM} \\
\bz_j & \sim & \mbox{Mult}(n_j,\bp_j), \label{eq:Post-DM0}
\end{eqnarray*}
where $\bz_j := (z_{j1}, \ldots , z_{jK})$ is the $K \times 1$ vector of dart scores and $n_j = \sum_k z_{jk}$ is the total number of darts thrown by the $j^{th}$ player at the target region. In our example with D20 as the target, the possible outcomes are D20, S20, D5, S5, D1, S1, and M. In this case, $z_{jk}$ is the number of times the $k^{th}$ possible score, e.g. D20, S20, etc., was achieved. It is easily seen that the posterior distribution is Dirichlet, i.e.
\begin{equation}  \label{eq:Post-DM1}
f(\bp_j \mid \bz_j) \sim {\cal D}(\balpha + \bz_j).
\end{equation}
Unfortunately, we do not know $\balpha$ but we can easily estimate\footnote{The alternative is to adopt a fully Bayesian approach which would require us to specify a hyper-prior distribution for $\balpha$.} it via an empirical Bayes approach.
Towards this end, it is easily shown that, unconditional on $\bp$, the dart scores have a DM distribution with PMF
\[
f(\bz_j \, ; \balpha) = \frac{\Gamma \left(\sum_k \alpha_k \right)}{\Gamma \left(\sum_k z_{jk}+\alpha_k \right)} \, \prod_k \frac{\Gamma \left(z_{jk} + \alpha_k \right)}{\Gamma \left(\alpha_k \right)}.
\]
Let $D := \{\bz_1, \ldots , \bz_{16} \}$ denote the data from the 16 professional darts players when targeting the region in question. The likelihood is then given by
\begin{equation}  \label{eq:Post-DM2}
L(\balpha \, ; D) = \prod_{j=1}^{16} f(\bz_j \, ; \balpha).
\end{equation}
We can maximize (\ref{eq:Post-DM2}) to obtain the MLE $\widehat{\balpha}:= (\widehat{\alpha}_1, \ldots , \widehat{\alpha}_K)$,
and then substitute $\widehat{\balpha}$ into (\ref{eq:Post-DM1}) to take this as our posterior for player $j$. That is, we take
\begin{equation}  \label{eq:Post-DM3}
f(\bp_j \mid \bz_j) \sim {\cal D}(\widehat{\balpha} + \bz_j).
\end{equation}
Under this posterior model, the fraction of new throws from the $j^{th}$ player would have the $k^{th}$ outcome as $(\widehat{\alpha}_k + z_{jk})/\sum_{l=1}^K(\widehat{\alpha}_l + z_{jl})$. If necessary, we can therefore use these fractions as inputs to the second stage of the model-fitting as described in Section \ref{sec:SkillModel}. The  $(\widehat{\alpha}_k + z_{jk})$ quantities are often called {\em pseudo-counts} and the {\em pseudo-fractions}  $(\widehat{\alpha}_k + z_{jk})/\sum_{l=1}^K(\widehat{\alpha}_l + z_{jl})$ are shrinkage estimators that shrink the raw estimators  $z_{jk}/\sum_{l=1}^K z_{jl}$ towards their population means. It is easy to see that the amount of shrinkage that takes place for the $j^{th}$ player decreases in the amount of data $\sum_{l=1}^K z_{jl}$ that we have for that player. This of course is a desirable property of these estimators.

\subsection{The DM Model: Borrowing Strength from Target Regions}
\label{sec:DM-Model-Regions}

Instead of leveraging information across players for a given target region, an interesting alternative is to leverage information across regions for a given player. The DM model does not immediately adapt to this setting because the target regions do not all share the same value of $K$. We can circumvent this problem by using one DM model for the doubles (with $K=7$ and excluding DB) and another for the trebles (with $K=6$). Even then an issue arises, however. Consider the targets T20 and T19, for example. Should a dart targeted at T20 that hits T1 be mapped to darts targeted at T19 that hit T7 or mapped to darts targeted at T19 that hit T3? While we have no way of answering this we propose a simple heuristic that resolves this orientation issue.

We will describe our approach for the doubles regions and note that the trebles are fitted similarly.  Consider, for example, the D20 target which has 7 possible outcomes represented in the data-set: D20, S20, D1, S1, D5, S5, and M (miss).
We combine D20's double neighbor regions (D1 and D5) and single neighbour regions (S1 and S5), respectively, and hence collapse these 7 outcomes into the following 5 outcomes: D20, S20, \{D1,D5\}, \{S1,S5\}, and M. We repeat this for all players and all double targets (D1, ..., D20). For each player, we then apply the DM model as in Section \ref{sec:DM-Model} but now with $K=5$ and $J=20$ regions. The DM model results in new pseudo-counts for each double in the lower $K=5$ dimensional setting. The final step is to split the pseudo-counts (in the case of target D20) for \{D1,D5\} into pseudo-counts for D1 and D5, and split the pseudo-counts for \{S1,S5\} into pseudo-counts for S1 and S5. The simplest and probably best way to do this is to divide the new additional counts equally. So for example, if S1 and S5 had 3 and 2 hits in the raw data, respectively, when the target was D20, and we have 3 new pseudo-counts for \{S1,S5\} then we would divide these 3 equally between S1 and S5 resulting in pseudo-counts of 3+1.5=4.5 and 2+1.5=3.5 for S1 and S5, respectively.

\section{The Bivariate-Normal Skill Model (Stage 2)}
\label{sec:SkillModel}

In this section we describe the bivariate-normal skill model of TPT and our simple extension of it to accommodate inference of the target $\bvartheta$ (as discussed below). This model can be fitted using either the raw count data or the pseudo-counts from Section \ref{sec:Empirical-Bayes}. The bivariate-normal skill models serve three purposes:
\begin{enumerate}
\item We require them to fit the benchmark models that we use in Section \ref{sec:ScoringRules} for evaluating the performance (via proper scoring rules) of the DM-based models.
\item They allow for clear visualisations that enable us to compare the performance of the DM-based models with models based on the raw data counts. See, for example, many of the figures in Section \ref{sec:Viz}.
\item We require them in some of the ZSG settings of Section \ref{sec:Empirical Bayes-ZSG} when we want to be able to target any location (e.g. every square millimeter) of a particular target region.
\end{enumerate}

Let $\bvartheta \in \mathbb{R}^2$ and $(x,y) \in \mathbb{R}^2$ denote the location coordinates of the intended target and outcome, respectively, of a dart throw. We use $z:=g(x,y)$ to denote the function $g$ that maps $(x,y) \in \mathbb{R}^2$ to the dart score $z$, e.g. D16, SB, T20, S7, etc. We will also let $h(z) \in \mathbb{N}$ denote the actual numerical score achieved by the dart, so for example, $h(\text{D16}) = 32$, $h(\text{T20})=60$, $h(\text{SB}) = 25$, $h(\text{S7}) = 7$, etc. Our goal then is to construct a model for $p(x,y;\, \bvartheta)$, the distribution of $(x,y)$ given $\bvartheta$, that can be easily estimated with the available data.
TPT proposed several such models including the model $[x \ y]^\top  \sim  \mbox{N}_2(\bvartheta,\sigma^2 \bI)$, i.e. a bivariate-normal distribution with mean $\bvartheta$ and covariance matrix $\sigma^2 \bI$ where $\bI$ is the $2 \times 2$ identity matrix. They also proposed $ [x \ y]^\top  \sim \mbox{N}_2(\bvartheta, \bSigma)$
where $\bSigma$ is an arbitrary covariance matrix. They developed an EM algorithm based on importance-sampling for estimating $\bSigma$. They assumed the intended target $\bvartheta$ was known for each data-point and that only the result $z=g(x,y)$ was observed\footnote{To be precise, TPT only assumed the score $h(z)$ was observed. Their EM algorithm still applies in this case although there may be more than one region $z$ corresponding to $s(z)$.} rather than the realized location $(x,y)$. That only the $z$'s were observed is a very reasonable assumption given the difficulty of measuring the $(x,y)$-coordinates of dart throws and, as mentioned earlier, this is also true for the data-set we consider in this paper.

In their models TPT implicitly assumed the mean outcome $\bmu$, say, and target  $\bvartheta$ were one and the same. While this seems like a very reasonable assumption for professional players, we suspect it is unlikely to hold for amateurs. Even for professionals, it's possible there may be some {\em bias}, i.e. discrepancy between the target $\bvartheta$ and the mean $\bmu$, especially for less frequently targeted parts of the dartboard. We defer a discussion of this bias issue to Appendix \ref{sec:Bias} and will therefore assume that $\bvartheta = \bmu$ throughout the paper unless otherwise stated. This leads to a model of the form
\begin{eqnarray}  \label{eq:DM2}
[x \ y]^\top & \sim & \mbox{N}_2(\bmu, \bSigma).
\end{eqnarray}

A further complication is that in our data-set we do not know $\bmu$ even though we always know the target region. So for example, we will know if a dart was aimed at T20 but we will not know precisely where in the T20 bed the dart was aimed. We could simply assume that $\bmu$ is the center of the target region with the center being defined as the midpoint of the polar-coordinates defining the region.
This seems like the most natural assumption to make but it is not difficult to imagine a player deviating from this on occasion. For example, suppose a player has two darts remaining in his turn and needs a D5 to check out, i.e. win the game. Rather than aiming at the center of D5, he may prefer to aim a little closer to the outer circular boundary of the D5 region on the basis that if he is going to miss D5 he would prefer to miss the scoring region entirely rather than hit S5 which would leave him unable to check out on his final dart of the turn. This argument does not apply to even doubles such as D20, D18, D16, etc. because if the corresponding single is hit then the player can still exit on the next dart in his turn.

Regardless of how we interpret $\bmu$, a simple model such as (\ref{eq:DM2}) will not be sufficiently rich for modeling the skills of professional darts players. This is because these players tend to focus on (and practice throwing at) specific parts of the darts board, e.g. T20, T19, DB, etc. This means that a player's skill level, as determined by his $\bSigma$, is likely to be a function of $\bmu$. Indeed as we discussed in Section \ref{sec:Data}, this is what we observe in the data. To give another example, the Dutch player Michael van Gerwen was successful 45.3\% of the time when targeting T20 but only 30.2\% of the time when targeting T17. This difference is statistically significant\footnote{Moreover the approximate 99.9\% confidence intervals (CIs) ($\hat{p} \pm 2.578 \sqrt{\hat{p} (1-\hat{p} )/n}$) for Van Gerwen's success percentages when targeting T20 and T17 do not overlap.} and, more importantly, {\em practically significant}. To see this we ran the following simple experiment. (We omit some of the less relevant details.) We assumed two players have no bias so their skill levels are determined entirely by $\bSigma$. We assumed player A's skill level was constant throughout the board and such that, when aiming at any treble region, had a success rate of $(45.3+ 30.2)/2 = 37.75\%$. Player B's skill level was identical to A's throughout the board except on T20 and T17 where his success rates were $45.3\%$ and $30.2\%$, respectively. Both players were therefore equally skillful when {\em averaged} across the entire board and identical on all target areas except T17 and T20.  Over the course of a best-of-35 leg dart match, we found that player B would win approximately\footnote{The precise number depends on which player starts the first leg.} 70\% of the time. This is because T20 is far more important than T17 and so even if both players are equally skillful on average, the skill level on T20 is much more crucial than that on T17. Hence, assuming the same skill model throughout the board can result in a drastic underestimation of a player's skill level and should be avoided. Finally, van Gerwen is in no way exceptional. As mentioned in Section \ref{sec:Data}, on average the players are significantly more skillful at targeting T20 and T19 than T18 or T17. In Section \ref{sec:ScoringRules} we provide formal justification via the use of proper scoring rules for our rejection of the simple skill model in (\ref{eq:DM2}).

Before proceeding, we acknowledge that a player with the skill model (\ref{eq:DM2}) applying throughout the dartboard would have different success rates when targeting different treble (or double) regions. This simply reflects the fact that the treble regions have different orientations. For example, suppose a player's $\bSigma$ is a diagonal element with a relatively larger variance along the $x$-axis than on the $y$-axis. Then this player should have a larger success rate when targeting T20 (oriented horizontally along the $x$-axis) than when targeting T6 (oriented vertically along the $y$-axis). Unfortunately we found that this possibility does not account for the extreme differences in success rates. For example, as may be seen from Figure \ref{fig:DartB1}, T18 and T19 have very similar orientations and so the same skill model  (\ref{eq:DM2}) could not account for the different success probabilities when targeting T18 and T19 that we see in the data.

\subsection{A Conditional Normal Skill Model}
\label{sec:CondGaussian}

In light of the data-set that we do have and the heterogeneity of success rates across target regions for each player, it seems appropriate to assume a separate skill model for different target regions. We therefore generalize (\ref{eq:DM2}) to
\begin{eqnarray}
 \label{eq:DM4}
[x \ y]^\top & \sim & \mbox{N}_2(\bmu_i, \bSigma_i), \ \ \mbox{ for } \bmu_i \in R_i,
\end{eqnarray}
where $R_i$ denotes the $i^{th}$ possible target region in our data-set. That is, we will infer a separate skill model (\ref{eq:DM2}) for each player and each target region (T20, T19, DB, D20, etc) in our data-set.
There are several reasons for using a separate normal skill model for each target region. First, it is not at all clear how different target regions should be combined and, as we noted earlier when discussing van Gerwen's success rates at T20 and T17, a naive approach could lead to severe inaccuracies in estimating the win-probabilities of players in real-world matches. Second, as we shall see in Section \ref{sec:Viz}, using a separate skill model for each target area allows us to highlight the natural limitations of the data-set.

Using a model such as (\ref{eq:DM4}) (or indeed (\ref{eq:DM2})) allows us to consider any potential target (e.g. down to the square-millimeter) on the dartboard when solving ZSGs. We could not do this if we just used the players' raw-data counts (even if we had data for each player on all (TR, $z$) combinations) or their pseudo-data counts arising from the DM-based models.
We note that (\ref{eq:DM4}) is as easy to estimate as (\ref{eq:DM2}) since we can partition our data-set by target regions. For each player we can therefore fit the skill model (\ref{eq:DM4})  using a simple extension of the EM algorithm developed by TPT. This extension allows for the case where the mean $\bmu$ is also unknown and therefore needs to be inferred. We describe our extended EM algorithm in Appendix \ref{app:EM}.

\section{Evaluating the DM Models Via Proper Scoring Rules}
\label{sec:ScoringRules}

In order to evaluate the quality of our skill models we will utilize proper scoring rules. Proper scoring rules are mechanisms designed to assess the quality of a probabilistic forecast (or distribution) $F$ by assigning a numerical score that is a function of both $F$ and the realized outcome. Scoring rules are deemed ``proper'' when the expected score is optimized if and only if the forecast $F$ coincides with the true data-generating distribution. The objective of proper scoring rules is therefore the truthful reporting of probabilities.
Their use is intended to ensure coherence and reliability in probabilistic forecasting and to facilitate the evaluation and comparison of predictive models within statistical inference and decision theory frameworks. They have been used extensively in a wide variety of domains including weather forecasting and risk management among others. We refer the interested reader to \cite{Gneiting_Raftery} who provide a thorough review of proper scoring rules.

In the context of darts, our skill models can be viewed as probabilistic forecasts. For example, when a player targets T20 then each of our skill models reports $\Pb(z \mid \mbox{T20 target})$ for each $z \in \{\text{T20, S20, T5, S5, T1, S1} \}$ with $z$ denoting the outcome of the dart throw. The $(K-1)$-dimensional simplex ${\cal S}_K$ then represents the space of possible probabilistic forecasts for dart outcomes with $K=6$ for the treble targets, $K=22$ for the DB target and $K=7$ for all other doubles targets. We can therefore compare our DM skill models against each other and against benchmark skill models through the use of proper scoring rules. We will consider two well-known proper scoring rules which are suitable for categorical data. Letting $\bp = (p_1, \ldots , p_K) \in {\cal S}_K$ denote a skill model / forecast for a particular target and $z \in \{1,\ldots,K\}$ be a dart outcome, we have:
\begin{enumerate}
\item The {\bf quadratic} or {\bf Brier score} which is defined as
\begin{equation} \label{eq:Brier}
S^B(\bp,z):= -\sum_{k=1}^K(\delta_{zk}-p_k)^2, 
\end{equation}
where $\delta_{zk}=1$ if $z=k$ and 0 otherwise.
\item The {\bf Spherical score} which is defined as
\begin{equation} \label{eq:Spherical}
S^S(\bp,z):= \frac{p_z^{\alpha -1}}{\left(\sum_{k=1}^K p_k^{\alpha}\right)^{(\alpha -1)/\alpha} },
\end{equation}
for some fixed $\alpha > 1$. (We take $\alpha=2$ everywhere in Section \ref{sec:ScoresNum}.)
\end{enumerate}
The scores in (\ref{eq:Brier}) or (\ref{eq:Spherical}) apply to a single dart so we simply average them when we have multiple darts thrown at the same target. For example, if a particular player has thrown $M$ darts at T20 then his Brier score for T20 will be
$$\widehat{S}^B :=\frac{1}{M}\sum_{i=1}^{M} S^B(\bp_{\mbox{\tiny T20}},z_i),$$
where $z_i$ is the outcome of the $i^{th}$ dart and $\bp_{\mbox{\tiny T20}}$ is his probabilistic forecast when targeting T20.

\subsubsection*{The Skill Models Under Consideration}

We will consider the following skill models.
\begin{enumerate}
\item The {\bf Basic} model is the $\mbox{N}_2(\bmu, \bSigma)$ in \eqref{eq:DM2} that simultaneously applies to all target regions. This is the simplest model of all and only has five free parameters to estimate the skill model for the entire dartboard.

\item The {\bf IJOC} model is the model used in \cite{HaughWang-Darts-2021}.
They consider the conditional normal skill models $\mbox{N}_2(\bmu_i, \bSigma_i)$ of \eqref{eq:DM4} with six regions:
$R_1 = \{\mbox{T}20\}$, $R_2 = \{\mbox{T}19\}$, $R_3=\{\mbox{T}18\}$, $R_4 = \{\mbox{T}17\}$, $R_5 = \{\mbox{DB}\}$ and $R_6 = \{\mbox{D}1, \ldots , \mbox{D}20 \}$. In particular, they fitted a single bivariate-normal model simultaneously to all twenty double regions as they were unable (due to lack of data) to estimate separate skill models for each double region. Our DM-based models overcome this problem.

\item The {\bf borrow-strength-from-players (BSP)} model as proposed in Section \ref{sec:DM-Model}.

\item The {\bf borrow-strength-from-regions (BSR)} model as proposed in Section \ref{sec:DM-Model-Regions}.
\end{enumerate}
The first two models are our benchmark models in that neither of them uses the empirical Bayes approach of the BSP and BSR models. They are also simpler models in that the Basic model uses the same skill model for the entire dartboard whereas the IJOC model uses the same skill model for all twenty double regions. In contrast, the BSP and BSR models use pseudo-counts to construct a separate skill model for each possible target. In order to fit the Basic and IJOC models we need to apply the EM algorithm (just once for the Basic model and six times for the IJOC model) to estimate the various $(\bmu, \bSigma)$ parameters. As mentioned earlier, our EM algorithm is a simple extension of TPT's that allows the mean outcome $\bmu$ to also be inferred. Details are in Appendix \ref{app:EM}. Once the various $(\bmu, \bSigma)$ parameters have been estimated, it is a simple matter of numerical integration to obtain the probabilistic forecasts for use with the scoring rules.
In addition to the above four skill models, we also consider two additional skill models:
\begin{enumerate}
\addtocounter{enumi}{+4}
\item The {\bf borrow-strength-from-players-normal (BSPN)}  model uses the BSP pseudo-counts to fit a separate $\mbox{N}_2(\bmu, \bSigma)$ model to each of the double regions, treble regions and DB region.
\item The {\bf borrow-strength-from-players-normal-center(BSPNC)} is the same as the BSPN model but now the mean outcome $\bmu$ is assumed to be known and equal to the center of the target region.
\end{enumerate}
Models 5 and 6 are therefore fitted using the EM algorithm (with $\bmu$ fixed in Model 6) but applied to the BSP pseudo-counts rather than the raw-data counts\footnote{Indeed as noted above it would not be possible to fit a separate skill model using the raw data-counts to each double region due to the scarcity of data for most doubles.}. Our main reason for considering the BSPN and BSPNC models is that we may want to be able to target very specific locations (down to the square millimeter say) in ZSG settings, while the BSP and BSR skill models do not allow for this.
Finally, we note that we could also have considered versions of models 5 and 6 that use the BSR pseudo-counts. We did not consider these models because as we shall see, the BSP and BSR models have very similar performances (in terms of proper scoring rules) with the BSP model appearing to be slightly better than the BSR model.

\subsection{Numerical Results}
\label{sec:ScoresNum}
Before describing our numerical results we first describe how we split the data into training- and test-sets.

\subsubsection*{Splitting Data into Training- and Test-Sets}

We apply an 80\%-20\% training-test set split on the entire data-set. Specifically, we aim to take a random sample of 80\% of the throws targeted at each player-target region combination as the training-set and take the remaining 20\% of throws as the test-set. Because certain targets (D17, D15, D13 and D11) have very few data-points and for some players no data-points at all, we dropped these targets from the analysis. For other rarely player-target combinations where 20\% of the number of data-points was less than one, we ensured at least one data-point was in the test-set and then took the remaining points to be in the training-set.
We then estimated the various skill models (including the estimation of $\balpha$ for the DM-based models) using the training-set\footnote{An alternative, slightly more efficient approach to constructing the train-test set splits would work as follows. We could have used the training-data for player $j$ and {\em all} of the data for the remaining players when fitting player $j$'s BSP, BSPN and BSPNC models. We could have done something similar for the BSR model: when evaluating player $j$'s score on a particular target region, e.g. D20, we could have used all of player $j$'s data as the training data except for the throws in the test-set that were aimed at D20. While this would be a more efficient use of the data, it would require more work (e.g. in the number of times $\balpha$ has to be estimated) so we went with the simpler approach outlined above.} and estimated the Brier and Spherical scores via the test-set.
Finally, since the training-test set splits are random the scores on the test-sets are also random. For this reason we take $N_s=20$ random training-test set splits and compute average scores across the $N_s$ test-sets. Rather than reporting the Brier and Spherical scores for each player-target combination we aggregate them by trebles (T20 through T17), the DB and the 20 other double regions.

\subsubsection*{Results}
\label{sec:ProperScoreResults}

We display the average Brier score for all 16 players when targeting the double regions in Table \ref{table:Brier scores Double}. We average across darts rather than regions (D20, D19, etc.) so that each dart in the test-set gets equal weight regardless of where it was targeted. Brier score results for the trebles and DB as well as all Spherical score results are reported in Appendix \ref{sec:AppScoringRules}. The Spherical and Brier score results provide very similar rankings of the six skill models which is why we only display the Brier scores here. For each player, we display the highest-ranked score in bold font and do likewise for the average across all players. Our main interest is in the doubles regions since these regions are very important in the game of darts (recall players must check out on a double) but we often only have limited data available for them. Our DM-based estimators were therefore designed with the doubles in mind. (Indeed because the four treble regions have plentiful data the DM-based estimators only provide minimal shrinkage for these regions.)

\begin{table}[h]
\begin{center}
\captionsetup{justification=centering}
{
\renewcommand{\tabcolsep}{1.2mm}
\caption{Brier scores for skill models aggregated across double regions}
\vspace{-1mm}
\label{table:Brier scores Double}
\small
\begin{tabular}{lc cccccc}
\toprule
Player & ~& Basic & IJOC & BSP & BSR & BSPN & BSPNC \\
\midrule
Anderson& ~ &\textbf{-0.6588}&-0.6623&-0.6593&-0.6657&-0.6588&-0.6641 \\
Aspinall& ~ &-0.6638&-0.6713&-0.6583&-0.6597&\textbf{-0.6579}&-0.6664 \\
Chisnall& ~ &-0.6642&-0.6623&-0.6591&-0.6601&\textbf{-0.6582}&-0.6594 \\
Clayton& ~ &-0.6893&\textbf{-0.6826}&-0.6865&-0.6850&-0.6867&-0.6827 \\
Cross& ~ &-0.6680&-0.6634&\textbf{-0.6587}&-0.6593&-0.6595&-0.6650 \\
Cullen& ~ &-0.6625&-0.6617&\textbf{-0.6556}&-0.6582&-0.6570&-0.6655 \\
van Gerwen& ~ &-0.6611&-0.6580&-0.6538&-0.6538&\textbf{-0.6533}&-0.6581 \\
Gurney& ~ &-0.6684&-0.6727&-0.6648&\textbf{-0.6626}&-0.6643&-0.6681 \\
Lewis& ~ &-0.6726&-0.6706&-0.6674&-0.6703&\textbf{-0.6672}&-0.6698 \\
Price& ~ &-0.6621&-0.6611&-0.6565&-0.6575&\textbf{-0.6561}&-0.6607 \\
Smith& ~ &-0.6683&-0.6701&-0.6625&-0.6631&\textbf{-0.6620}&-0.6625 \\
Suljovic& ~ &-0.6685&-0.6649&-0.6556&-0.6560&\textbf{-0.6545}&-0.6637 \\
Wade& ~ &-0.6601&-0.6550&-0.6532&-0.6530&\textbf{-0.6527}&-0.6570 \\
White& ~ &-0.6715&-0.6689&-0.6674&\textbf{-0.6660}&-0.6670&-0.6693 \\
Whitlock& ~ &-0.6701&-0.6637&-0.6608&-0.6574&-0.6593&\textbf{-0.6550} \\
Wright& ~ &-0.6651&-0.6639&-0.6617&-0.6614&\textbf{-0.6611}&-0.6641 \\
\midrule
Average& ~ &-0.6672&-0.6658&-0.6613&-0.6618&\textbf{-0.6610}&-0.6645 \\
\bottomrule
\end{tabular}}~\\
\end{center}
\end{table}

Our first observation regarding Table \ref{table:Brier scores Double} is that the BSP, BSR and BSPN models all perform similarly and on average outperform the two benchmark models Basic and IJOC. The BSPNC model also does better than the benchmark models but not as well as the three other DM-based models. This is not surprising since the BSPNC model has two fewer parameters (than the BSPN model) per target region to fit the BSP pseudo-counts. We note that the BSP and BSPN models do marginally better than the BSR models and we also observe this for the Spherical scores reported in Appendix \ref{sec:AppScoringRules}. With regards to the trebles  we see that the Basic model performs the worst of all six models, and we see that the BSP, BSR and BSPN models are the best performing models with identical (to 4 decimal places) average scores for both Brier and Spherical scores. Given the abundance of data available for the trebles, it is not surprising to see the IJOC model perform well with an average score that is only marginally below the BSP, BSR and BSPN scores. Finally, with regards to the DB we note that there are no results for the BSR model as there are no regions from which the DB region can borrow. For both the Brier and Spherical scores, the BSPN model is ranked highest on the DB followed by the BSPNC and then the Basic model.

We emphasize that the main value of the scoring rules is to provide a {\em ranking} of the skill models. In particular, it is difficult to attach much meaning to the differences in reported scores. For example, a proper scoring rule remains a proper scoring rule under any positive affine transformation. Nonetheless, it is possible to provide some interpretation to the numbers in Table \ref{table:Brier scores Double}. For example, when targeting D20 James Wade hits D20, S20, M, D5, S5, D1 and S1 with frequencies of 38.9\%, 36.8\%,  23.5\%, 0\%, \ 0.3\%, 0.5\%, and 0\%, respectively. If we take these frequencies to be the ground truth probability distribution and assume our forecast is perfect, i.e. equal to the ground truth distribution, then the expected Brier score\footnote{The reported expected Brier scores in this paragraph are for the case where we have an infinite number of darts. The actual expected Brier score for a finite number of darts will differ due to the non-linearity of the Brier score. } will be -0.6580 which would be the best possible Brier score for {\em any} skill model for James Wade targeting D20.  In contrast, if we used a naive forecast that reported a probability of $1/7$ for each of the 7 possible outcomes, then the resulting expected Brier score would be approximately -0.8571. These values help to provide some context\footnote{James Wade's results when targeting D20 are typical of the average player in the data-set. That most of the reported numbers for him in Table \ref{table:Brier scores Double} are slightly better than -0.6580 is due to a number of possible reasons including the fact that the reported numbers are aggregated across all 20 double regions, the non-linearity of the Brier score mentioned in the previous footnote, and random noise due to the variability in the training-test set splits.} for the reported Brier scores in Table \ref{table:Brier scores Double}. In Appendix \ref{sec:sig} we argue that the mean Brier and Spherical scores for the DM-based skill models are statistically different from the benchmark models through the use of paired t-tests. Nonetheless, statistical significance does not equate to practical significance. We address practical significance in Section \ref{sec:ZSG_Misspecified} where we use our ZSG setting to show that if a player assumes an incorrect skill model (e.g. the Basic or IJOC model) when his true skill model is the BSP model, then his win-probability, i.e. probability of winning a best of $N=35$ leg darts match can decrease by as much as $10\%$ for some players. This follows because a very small difference in forecast quality for a single dart will be magnified and can become practically significant over the many darts thrown in an entire darts match.

\section{Visualizing the Fitted DM Models}
\label{sec:Viz}

In Section \ref{sec:ScoringRules} we saw that the DM-based models outperform the benchmark, i.e. non-DM, models in terms of proper scoring rules. In this section we will demonstrate this visually by comparing the BSPN model with an analogous model that is fitted using the raw data. In particular, recall that the BSPN model uses a separate skill model (\ref{eq:DM4}) for each double and treble region and fits each of these models using the pseudo-fractions $(\widehat{\alpha}_k + x_{jk})/\sum_{l=1}^K(\widehat{\alpha}_l + x_{jl})$. We will compare the BSPN model with the same model but with all the $\widehat{\alpha}_k$'s set to zero so that only the raw fractions are used. We will refer to this latter model as the raw-normal ({\bf RN}) model.

As discussed in Section \ref{sec:Data}, a professional darts player is quite likely to hit any of 22 regions when aiming at the DB. In contrast, when aiming at a treble or double region, a professional player is only likely to hit at most 6 or 7 regions, respectively. For a given target and player, we will use the term {\em coverage} to refer to the number of regions with non-zero observations in the raw data. So the maximum coverage for DB is 22 while the maximum coverage for trebles and doubles are 6 and 7, respectively. By their nature, the BSP- and BSR-based models generally have full coverage but this is certainly not true for the RN models and, as we shall see below, this has a severe impact on the quality of the RN model fits. Indeed we did not use the RN model as a benchmark model in Section \ref{sec:ScoringRules} as it would have performed extremely poorly because (as we shall see below) it would have attached almost zero probability to outcomes that were often represented in the hold-out test-sets. Nonetheless, it provides a useful point of comparison to see how well the DM-based models (such as the BSPN models) resolve the coverage issues in the raw data.

\subsubsection*{The Double-Bull (DB)}
\begin{figure}[h]
\begin{center}
\includegraphics[width=1\linewidth]{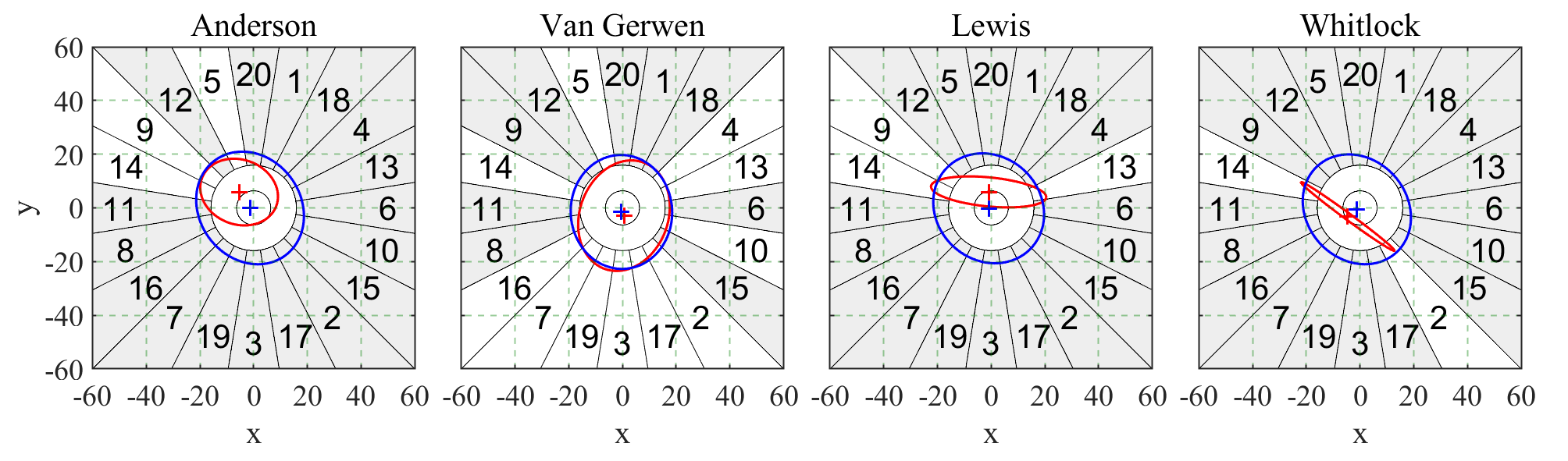}
\end{center}
\vspace{-0.5cm}
\caption{95\% CEs for the BSPN (blue) and RN (red) models when targeting DB. Regions with zero observations in the raw-data are shaded gray.}
\label{figure:FittedEllipseRealDB}
\end{figure}

We begin with the skill model for DB. Figure \ref{figure:FittedEllipseRealDB} displays 95\% confidence ellipses\footnote{Rather than present numerical values for $(\bmu_i, \bSigma_i)$, it is more informative to display results visually via CEs on the appropriate section of the dartboard. Due to space considerations, we only display results for 4 players here but Appendix \ref{sec:AppAllPlayersFigures} displays results for all 16 players.} (CEs) for the BSPN and RN models for 4 of the 16 players. The CEs for the BSPN fits look very reasonable whereas the RN CEs for
Anderson, Lewis, and Whitlock are clearly problematic and these problems arise due to insufficient coverage. The raw-data coverage for each player is displayed in Figure \ref{figure:FittedEllipseRealDB} by shading gray any regions with zero observations. (The pattern of zero observations then explains the shape of the RN CEs.) Besides DB and SB, the coverage of Anderson (S14, S9, S5), Lewis (S14, S13, S9), and Whitlock (S14, S2) in particular is insufficient to accurately estimate a model with 5 parameters. (The problem with Whitlock hints at the major problems to come when we consider fits to the double target regions using the raw-data counts.)
In contrast, the coverage of van Gerwen is 13 which results in a reasonable fit even for the RN model and this is also true of most of the other 12 players in the data-set. These problems do not arise with the BSPN model since the pseudo-counts result in full (pseudo-)coverage for all players.

\subsubsection*{The Other Doubles}
\begin{figure}[h]
\begin{center}
\includegraphics[width=1\linewidth]{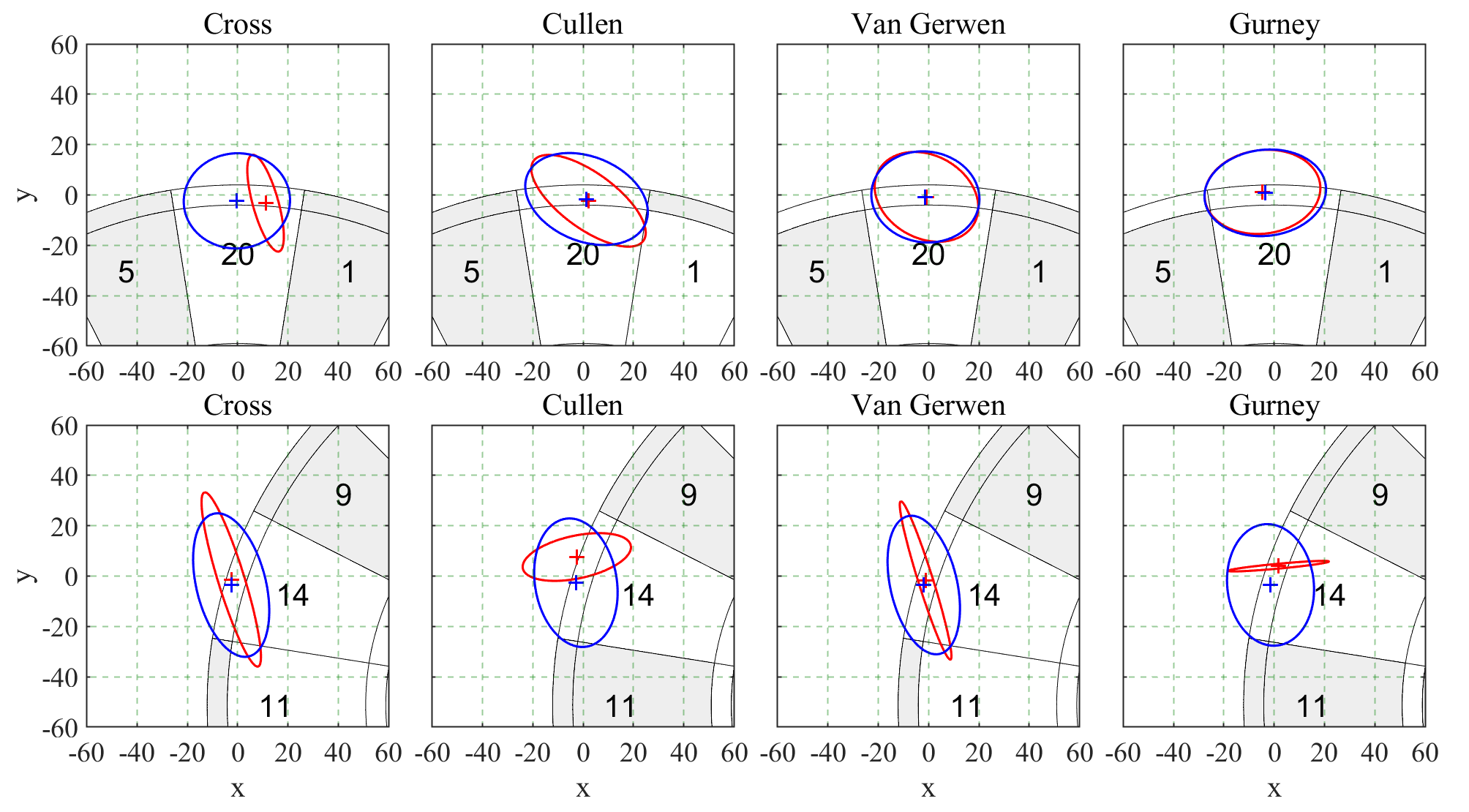}
\end{center}
\vspace{-0.5cm}
\caption{95\% CEs for the BSPN  (blue) and (red) RN models when targeting D20 and D14. Regions with zero observations in the raw-data are shaded gray.}
\label{figure:Doubles}
\end{figure}

Severe problems can occur when fitting the RN model to double regions, including even D20, which accounts for over 25\% of all doubles in the data-set. CEs for D20 and D14 are displayed in Figure \ref{figure:Doubles}.
In the case of D20, for example, while all 4 players have observations in D20, S20, and M, these are the only regions covered by Cross, while Cullen (S1), van Gerwen (D5), and Gurney (D5) only have a coverage of 4 when targeting D20. With such limited coverage the likelihood function (which the EM algorithm seeks to optimize) will be quite flat in large neighborhoods and terminating at a reasonable (local) maximum is quite unlikely although it can happen. For example, the RN CEs for van Gerwen and Gurney on D20 look reasonable despite only having a coverage of 4.  The quality of fits deteriorates as we consider doubles with even less data. There were only 312 darts targeted at D14, for example, which averages out to approximately 19.5 darts per player. It is no surprise then that the D14 coverages are low for all players including Cross (4), Cullen (3), van Gerwen (4), and Gurney (3) whose fits are displayed in the lower row of Figure \ref{figure:Doubles}. Matters are even worse for the odd doubles with D19 (the best of them) having only 154 darts targeting it for an average of 9.6 darts per player and all but 1 player having a coverage of just 2 or 3.

The shrinkage supplied by the empirical Bayes approach means that the coverage issues are resolved in the BSPN model and the corresponding CEs of Figure \ref{figure:Doubles} look much more reasonable. It is worth noting that the BSP model does not work for some of the very rarely targeted doubles. In particular, D11, D13, D15, and D17 are so rarely targeted\footnote{There were only 232 darts in total thrown by the 16 players at these 4 target areas in the data-set. Darts targeting D11, D13, D15, and D17 therefore represent less than .15\% of all dart throws in the data-set.} that even when we aggregate the data for these targets across all players, the pseudo-coverages are still too low to allow a reasonable estimation of (\ref{eq:DM4}). In that case of course, we could use the BSR pseudo-counts to fit these skill models.

\subsubsection*{The Trebles}

\begin{figure}[h]
\begin{center}
\includegraphics[width=1\linewidth]{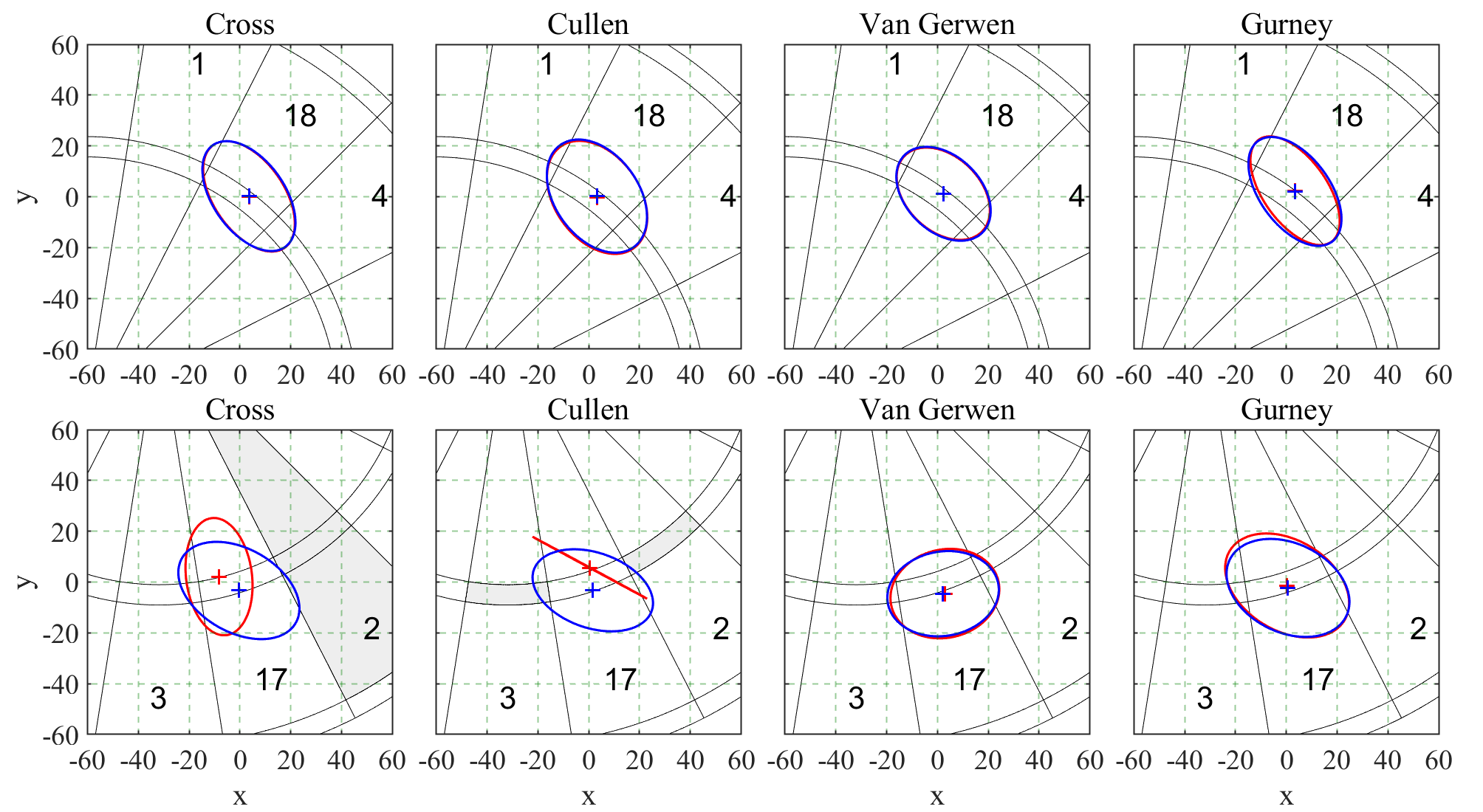}
\end{center}
\vspace{-0.5cm}
\caption{95\% CEs for the BSPN (blue) and RN (red) models when targeting T18 and T17. Regions with zero observations in the raw-data are shaded gray.}
\label{figure:FittedEllipseT18T17_RN_BSPN}
\end{figure}

Because there is ample data for players targeting T20, the pseudo-fractions for T20 are almost identical to the raw fractions and so minimal shrinkage takes place with the T20 data. The amount of shrinkage increases as we move progressively through T20, T19, T18, and T17 but is still quite minimal for T19 and T18. Some players (Cross and Cullen) have poor coverage when targeting T17, however, and so the RN model does a poor job with these players. This may be seen in Figure \ref{figure:FittedEllipseT18T17_RN_BSPN} where we display the BSPN and RN fits for T18 and T17. We see the BSPN and RN fits are almost identical for T18 (because minimal shrinkage takes place for T18) but this is not always the case for T17.

\subsection{An Aside on Correlation Identification}
\label{sec:CorrelIdent}
The superior fits provided by the BSPN models bring to light an interesting issue. We see from Figures \ref{figure:Doubles} and \ref{figure:FittedEllipseT18T17_RN_BSPN} that the main axis of the CEs (for the BSPN fits) tends to be aligned with the principal direction of the target region in question. For example, in the case of D14 we see the CEs tend to be oriented in a north-south direction which is also the general orientation of the D14 bed. Similarly, we see the CEs for D20 are oriented east-west which is also the orientation for the D20 bed. Indeed, as may be seen from Figures \ref{figure:T18_Fits}-\ref{figure:D14_Fits} in Appendix \ref{sec:AppAllPlayersFigures}, this behavior persists more generally.
This is a natural consequence of the bivariate-normal skill model as it stretches the variance along the main axis of the target area in order to fit the fractions of darts that landed in the neighboring doubles / trebles regions. A bivariate distribution with fatter tails, e.g. the bivariate $t$ distribution, might better fit this problem but owing to space considerations as well as other issues related to correlation identification (see below) and interpreting $\bmu$ (see Appendix \ref{sec:mu}) we leave this for future research.

\begin{figure}[H]
\begin{center}
\includegraphics[width=.8\linewidth]{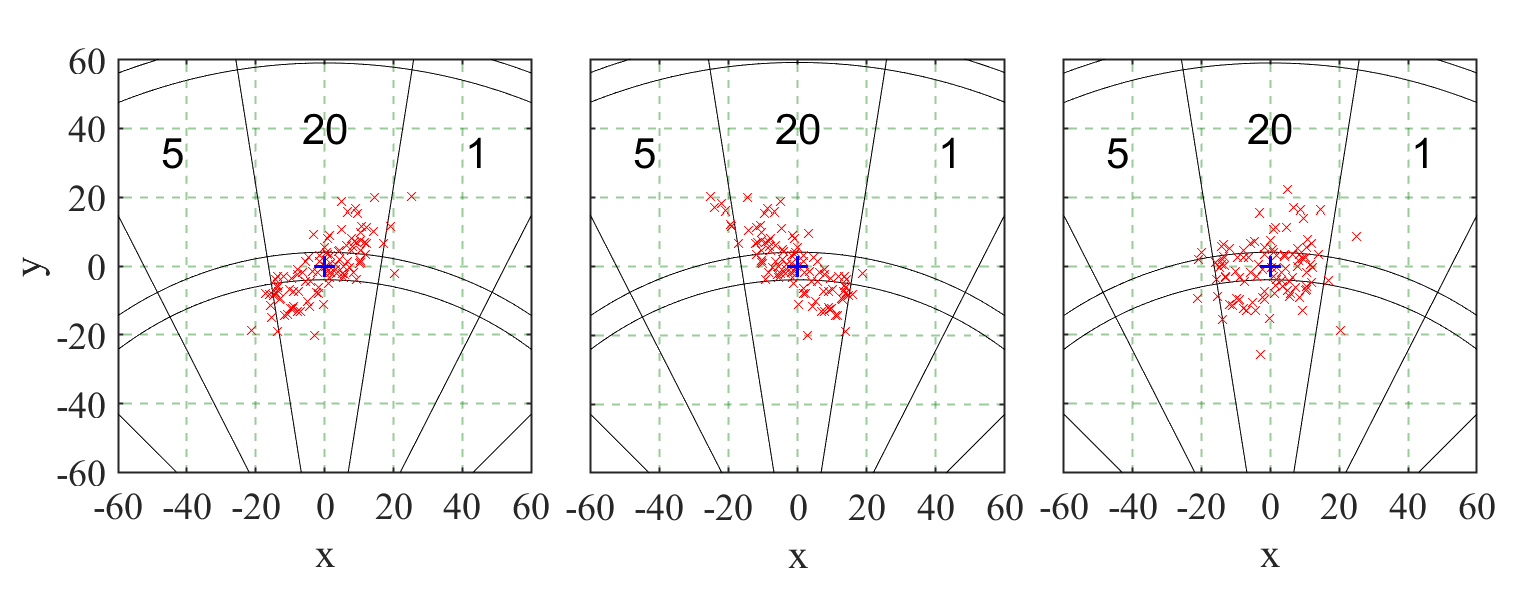}
\end{center}
\vspace{-0.5cm}
\small
{\em Note}: The blue ``+'' point is the intended target, and the red ``$\times$'' points are the landing locations.
The three panels have the same numbers of hits in each region:  37 in T20, 54 in S20, 6 in S5, 1 in T1, and 2 in S1.
\caption{Three synthetic T20 data-sets with identical outcomes but different correlations}
\label{figure:DataExample_T20}
\end{figure}

Considering this tendency to stretch the variance along the main axis of the target area, we believe the resulting (non-zero) correlations are spurious due to a problem intrinsic to the data-set.
In particular, we do not know the exact $(x,y)$-coordinates of where a dart lands. Instead we only know the region, e.g. S20, T20, etc., and this means identifying the correlation parameter is very difficult. To see this, consider Figure \ref{figure:DataExample_T20} which displays three synthetic data-sets each consisting of 100 darts aimed at T20. It is clear the skill-models generating these data-sets have very different correlations with the first, second, and third data-sets displaying positive, negative, and (approximately) zero correlations, respectively. However, all three data-sets have identical outcomes, i.e. 37, 54, 6, 1, and 2 darts hitting T20, S20, S5, T1, and S1, respectively. As such, the three data-sets would have identical likelihood functions and there is no way for us to identify the correct sign (never mind value) of the correlation. While we have no darts hitting T5 in these artificial data-sets, it should be clear that not having full coverage\footnote{It is true, however, that in the limit of infinite data eventually some darts would land outside the 20, 5, and 1 segments and allow for the identification of the correlation. But with professional darts players, we would be waiting a very long time for that to happen.} is not the cause of this problem. Moreover it is easy to see that this problem also applies to double regions.

Of course, this correlation identification issue does not arise with the Basic and (in the case of the double regions) IJOC benchmark models. In the Basic and IJOC models, we fit a single bivariate-normal skill model to all the (doubles and trebles) data and all the doubles data, respectively. Correlation identification is therefore feasible for these models but unfortunately, as we saw in Section \ref{sec:ScoringRules}, these models do not fit the data well and are dominated by the DM-based models.

In summary, it is perhaps best to view the bivariate-normal skill model as a mechanism that allows us to (i) fit the raw- or pseudo-counts data and (ii) then target any precise target (e.g. every square-millimeter) on the dartboard. In Table \ref{table:FittedProb_RealData_T20_nonfixedmu_fourplayers}, we display the observed versus fitted scoring percentages for four players when the bivariate-normal fits to the raw data when T20 is the target. This ability to fit the observed data is not surprising as the bivariate normal has five free parameters available to fit just six observed outcomes. While we could enforce zero correlations within the EM algorithm, we found this resulted (in unreported results) in a deterioration of the skill model as measured by proper scoring rules. As such, rather than attach any physical meaning to the fitted correlation parameters, we simply view them as additional degrees of freedom which allows us to improve the forecasting capabilities of the skill models.

\begin{table}[H]
\begin{center}
\captionsetup{justification=centering}
{
\renewcommand{\tabcolsep}{1.0mm}
\caption{Observed vs fitted (RN model) scoring percentages for T20. }
\vspace{-1mm}
\label{table:FittedProb_RealData_T20_nonfixedmu_fourplayers}
\small
\begin{tabular}{lc rrr rrr rrr rrr rrr rrr }  
\toprule
\multirow{2}{*}{Player} &~& \multicolumn{2}{c}{T20} &~& \multicolumn{2}{c}{S20} &~& \multicolumn{2}{c}{T5} &~& \multicolumn{2}{c}{S5} &~& \multicolumn{2}{c}{T1} &~& \multicolumn{2}{c}{S1} \\ 
\cline{3-4} \cline{6-7} \cline{9-10} \cline{12-13} \cline{15-16} \cline{18-19}  \noalign{\smallskip} 
~ &~& Obs. & Fit. &~& Obs. & Fit. &~& Obs. & Fit. &~& Obs. & Fit. &~& Obs. & Fit. &~& Obs. & Fit. \\ 
\midrule
Cross&~&42.3&42.3&~&51.6&51.6&~&1.0&0.7&~&1.4&1.6&~&1.1&1.2&~&2.6&2.5    \\ 
Cullen&~&38.0&38.0&~&52.6&52.8&~&1.9&2.0&~&2.6&2.3&~&1.3&1.1&~&3.5&3.8   \\ 
van Gerwen&~&45.3&45.3&~&48.2&48.3&~&2.0&2.0&~&2.1&1.9&~&0.9&0.7&~&1.5&1.7   \\ 
Gurney&~&40.4&40.6&~&52.8&52.8&~&2.2&2.2&~&2.7&2.4&~&0.5&0.3&~&1.4&1.7   \\ 
\midrule
Fitted Error&~&~&0.1&~&~&0.1&~&~&0.1&~&~&0.3&~&~&0.2&~&~&0.2  \\ 
\bottomrule
\end{tabular}}~\\
\end{center}
\end{table}

\section{Using the Empirical Bayes Model in Zero-Sum Game Settings}
\label{sec:Empirical Bayes-ZSG}

While the skill models of the professional players are of interest in their own right, they also play a vital role in the solution of the ZSG that models a game of darts between two players. \citet{HaughWang-Darts-2021} showed how a leg of darts could be formulated as a ZSG and used DP methods to find the unique Nash equilibrium of the game. In Section \ref{sec:ZSG_Misspecified} we consider the loss in win-probability when a player assumes the wrong-skill model, while in Section \ref{sec:SingleActionPerTarget} we consider the loss in win-probability when a player can only target a single location in each possible target area.
Finally in Section \ref{sec:RealWorld} we show via real-world examples how a player's skill model in conjunction with a player's optimal strategy can explain observed behavior that might otherwise appear surprising. Before proceeding, we provide a very brief review of ZSG concepts as they relate to the game of darts. The theory behind dynamic ZSGs was originally developed by \cite{Shapley1953} and was later extended by \cite{Patek97} for so-called stochastic shortest-path ZSGs. The game of darts was modelled as a stochastic shortest-path ZSG by \citet{HaughWang-Darts-2021} which should be consulted for further details and references to all results stated below.

The {\em state} of a game or leg is denoted by $\bs = (\sA,\sB,t,i,u)$ where $\sA, \sB \in \{0,2,3,\ldots,501\}$ are the scores of players A and B, respectively, at the beginning of the turn, $t \in \{\mbox{A},\mbox{B}\}$ denotes whose turn it is,
$i \in \{1,2,3\}$ denotes how many throws are left in player $t$'s turn and $u \in \{0,1,\ldots,(3-i)\times60\}$ is player $t$'s cumulative score thus far within the current turn. We use ${\cal S}_{\mbox{\tiny ZSG}}$ to denote the state-space for the ZSG and it contains all possible values of the state $\bs$ that can arise during the leg. The action-space ${\cal A}$ is the set of possible targets on the dartboard so, for example, if we assume each square millimeter on the board is a feasible target then it is easily seen that there are approximately 90,000 feasible targets or actions.

A strategy $\piA$ ($\piB$) for player A (B) is a rule telling him where in ${\cal A}$ he should target his next dart as a function of the current state $\bs$. Let $\JpiApib (\bs)$ be the probability that player A wins the leg when A and B play strategies $\piA$ and $\piB$, respectively, given the current state of the leg is $\bs$. The min-max and max-min values of the game are then defined to be
\begin{eqnarray*}
\underline{J}(\bs) &:=& \min_{\piB } \max_{\piA } \JpiApib (\bs), \\
\overline{J}(\bs) &:=& \max_{\piA} \min_{\piB }  \JpiApib (\bs).
\end{eqnarray*}
It can be shown that $\underline{J}(\bs) = \overline{J}(\bs) =: J^*(\bs)$ and that there exist strategies $\piA^*$ and $\piB^*$  satisfying
\begin{equation}\label{eq:EquilZSG1}
\JpisApibs(\bs) = \max_{\piA} \JpiApibs (\bs) =  \min_{\piB} \JpisApib (\bs) = \underline{J}(\bs) = \overline{J}(\bs) = J^*(\bs).
\end{equation}
It follows from (\ref{eq:EquilZSG1}) that $\piA^*$ and $\piB^*$ are optimal or Nash equilibrium strategies for A and B, respectively,
and we say $J^*(\bs)$ is the equilibrium value of the game starting from state $\bs$. $J^*(\bs)$ is the probability that player A wins the leg given both A and B play their optimal / Nash equilibrium strategies.
Let $\VA(\bs)$ denote player A's best-response (BR) value function to player B's strategy $\piB$ for arbitrary states $\bs \in {\cal S}_{\mbox{\tiny ZSG}}$. Then, player A's BR problem can be formulated as
\begin{equation} \label{eq:BR-PRobForm1}
\VA(\bs_1) = \max_{\piA } \JpiApib (\bs_1),
\end{equation}
where $\bs_1 := (501,501,\text{A},3,0)$ is the initial\footnote{For ease of exposition, we assume A is first to throw in the leg.} state of the game.
Problem (\ref{eq:BR-PRobForm1}) is a so-called stochastic shortest-path problem and standard DP techniques can be used to show that it has a unique solution.
The optimal solution / Nash equilibrium for the game can be found by repeatedly solving A's and B's BR problems until convergence occurs. Convergence is guaranteed and typically occurs within just 2 to 3 BR iterations if one player's initial strategy is the optimal strategy\footnote{This initial strategy takes no account of the opponent's score and can be computed via DP methods.} that minimizes the number of turns required to check out.

Given the optimal strategies $\piA^*$ and $\piB^*$, and A's equilibrium win-probability $J^*(\bs_1)$, it is straightforward to compute $\PA(N)$, the equilibrium probability that A wins an $N$-leg match against B. Specifically, we assume that in the match A starts the first leg, and then two players alternate in starting legs, e.g. B starts the second leg, A starts the third leg, etc. let $\VVA$ and $\VVB$ denote the numbers of legs won by A that are started by A and B, respectively.
As a typical darts match has an odd number of legs, we can assume $N=2K+1$ so that A needs to win at least $K+1$ legs (i.e. $\VVA+\VVB \ge K+1$) to win that match. A simple convolution calculation then yields
\begin{equation}
\PA(N) = \sum_{j=1}^{K+1} \Pb(\VVB \geq K+1-j  ) \Pb(\VVA = j),
\end{equation}
which is easily calculated since $\VVA \sim \mbox{Bin}(K+1,\pA^*)$ and $\VVB \sim \mbox{Bin}(K,\pB^*)$ where $\pA^* = J^*(\bs_1)$ and  $\pB^*$ is the equilibrium probability of A winning a leg given that B starts it.

\subsection{The Impact of Using Misspecified Models in ZSGs}
\label{sec:ZSG_Misspecified}
In this section we investigate the impact of using a misspecified skill model on the win-probability in a darts match. This will help us evaluate the practical significance of using the BSP model over the benchmark Basic and IJOC models when the BSP model is the ground truth model. (We recall from Section \ref{sec:ScoringRules} that the DM-based models, which of course includes the BSP model, outperform both benchmark Basic and IJOC models in terms of proper scoring rules.)

Suppose that skill model $\cal M$ is the ground truth model, and let $\pi_{\mbox{\tiny A}, {\cal M}}^{*}$ denote player A's equilibrium strategy when he plays against a replica\footnote{We only consider a player playing against a replica of himself in order to isolate the impact of a player using a misspecified skill model. The word ``replica'' simply means that the same data is used to fit the skill models of the two opponents.} of himself also playing $\pi_{\mbox{\tiny A}, {\cal M}}^{*}$.
Let $\cal M'$ be an alternative specification of A's skill model, and let $\pi_{\mbox{\tiny A}, {\cal M'}}^{*}$ be A's equilibrium strategy when he assumes his skill model is $\cal M'$ and he plays against a replica of himself also playing $\pi_{\mbox{\tiny A}, {\cal M'}}^{*}$.
We now want to evaluate the loss in win-probability when player A plays $\pi_{\mbox{\tiny A}, {\cal M}}^{*}$ against a replica of himself playing $\pi_{\mbox{\tiny A}, {\cal M'}}^{*}$ given that the ground truth skill model for players is $\cal M$. Towards this end, let $\PA(\piA,N)$ be the probability that A wins an $N$-leg match when A plays strategy $\piA$ throughout the match, A starts the first leg, and his opponent (a replica of A) always plays $\pi_{\mbox{\tiny A}, {\cal M}}^{*}$ throughout the match.  We define A's loss in win-probability from using ${\cal M'}$ to be
\begin{equation} \label{eq:Adv1}
\mbox{Loss}({\cal M}, {\cal M'}, N) := \PA(\pi_{\mbox{\tiny A}, {\cal M}}^{*},N) - \PA(\pi_{\mbox{\tiny A}, {\cal M'}}^{*},N).
\end{equation}
To be clear, because the ground truth skill model is $\cal M$, the {\em outcomes} of the dart throws of both A and his replica opponent are governed by $\cal M$.

In our numerical experiments, we assume the true underlying model ${\cal M}$ is BSP, and report the loss \eqref{eq:Adv1}  for $\cal M' \in \{\mbox{Basic}, \mbox{IJOC}, \mbox{BSPN}, \mbox{BSPNC} \}$. We assume\footnote{We also assume that players can only target DB, T17-T20, D1-D20, and S1-S20. We exclude SB, T1-T16, D11, D13, D15, and D17 as there was no (or almost no) data available to estimate the BSP skill model on these target regions. While we could have used the BSR model to estimate skill models for these regions, this would make no discernable difference since these regions are so rarely targeted in any case. We used data aggregated across all single regions for each player when determining a player's success probability when targeting a specific single region. If a player fails to hit a target single region (a relatively rare occurrence) then we assume he is equally likely to hit either of the two neighboring single regions.} a single target per target region is available to A and his replica so as not to unfairly disadvantage the BSP strategy when playing against the Basic, IJOC, BSPN, and BSPNC strategies. So, for example, if a player is aiming for T20, then we assume that the Basic, IJOC, BSPN, and BSPNC strategies can only target the center of the T20 bed.

Our results are displayed in Table \ref{table:winprob_loss model compare}.
For example, the first row of the table reports that Anderson's win-probability in a 35-leg match decreases by 3.5\% and 1.8\% when he uses equilibrium strategies based on the Basic and IJOC models, respectively.
Overall, we find that, in a 35-leg match, the average loss in win-probability across all sixteen players is 5.4\% and 3.7\% for the Basic and IJOC models, respectively, but for individual players, e.g. Clayton, these numbers can be much higher. In comparison, the average losses for using the BSPN and BSPNC models are both just 0.3\%, which simply reflects how well the bivariate normal can fit the pseudo-count data of the BSP model.

\begin{table}[h]
\begin{center}
\captionsetup{justification=centering}
{
\renewcommand{\tabcolsep}{0.7mm}
\caption{Losses of win-probabilities in $N$-leg matches for using misspecified models.\\ The BSP model is the ground truth model.}
\vspace{-1mm}
\label{table:winprob_loss model compare}
\small
\begin{tabular}{l ccc ccc ccc ccc ccc ccc}
\toprule
\multirow{2}{*}{Player} & ~ & \multicolumn{2}{c}{Basic} & ~ & \multicolumn{2}{c}{IJOC} & ~ & \multicolumn{2}{c}{BSPN} & ~ & \multicolumn{2}{c}{BSPNC} \\
\cline{3-4} \cline{6-7} \cline{9-10} \cline{12-13}  \noalign{\smallskip}
~ & ~& $N$=1 & $N$=35 & ~& $N$=1 & $N$=35 & ~& $N$=1 & $N$=35 & ~& $N$=1 & $N$=35 \\
\midrule
Anderson&~&0.8&3.5&~&0.4&1.8&~&0.0&0.1&~&0.0&0.1\\
Aspinall&~&0.3&1.8&~&1.1&4.9&~&0.0&0.2&~&0.1&0.3\\
Chisnall&~&2.1&9.3&~&0.4&1.7&~&0.0&0.2&~&0.0&0.2\\
Clayton&~&2.8&12.1~ &~&2.1&9.7&~&0.0&0.1&~&0.0&0.2\\
Cross&~&1.0&4.8&~&1.2&5.1&~&0.0&0.2&~&0.0&0.1\\
Cullen&~&1.5&6.8&~&0.8&3.4&~&0.0&0.1&~&0.0&0.2\\
van Gerwen&~&1.7&8.1&~&0.5&2.1&~&0.1&0.3&~&0.1&0.4\\
Gurney&~&0.4&1.9&~&0.9&3.5&~&0.0&0.2&~&0.0&0.2\\
Lewis&~&1.0&4.7&~&0.2&0.9&~&0.0&0.2&~&0.0&0.2\\
Price&~&0.7&3.3&~&0.5&2.5&~&0.0&0.2&~&0.0&0.2\\
Smith&~&0.6&2.9&~&0.8&3.4&~&0.1&0.2&~&0.1&0.4\\
Suljovic&~&1.7&7.4&~&0.7&3.4&~&0.1&0.4&~&0.1&0.4\\
Wade&~&0.3&1.4&~&0.6&2.8&~&0.1&0.5&~&0.1&0.5\\
White&~&2.0&9.0&~&0.6&2.7&~&0.0&0.1&~&0.1&0.2\\
Whitlock&~&1.4&6.4&~&1.7&7.4&~&0.3&1.6&~&0.3&1.6\\
Wright&~&0.5&2.4&~&0.8&3.9&~&0.0&0.1&~&0.0&0.2\\
\midrule
Average&~&1.2&5.4&~&0.8&3.7&~&0.1&0.3&~&0.1&0.3\\
\bottomrule
\end{tabular}}~\\
\end{center}
\justify
\small
{\em Notes.} Numbers are in percentages.
\end{table}

\begin{rem}
Instead of assuming player A plays $\pi_{\mbox{\tiny A}, {\cal M}}^{*}$ against his replica using $\pi_{\mbox{\tiny A}, {\cal M'}}^{*}$ we could instead have assumed A plays his best-response strategy versus $\pi_{\mbox{\tiny A}, {\cal M'}}^{*}$. This would have resulted in even larger losses in win-probability for A playing the sub-optimal $\pi_{\mbox{\tiny A}, {\cal M'}}^{*}$ against his replica opponent.
\end{rem}

\subsection{A Single Target Versus Multiple Targets Per Target Region}
\label{sec:SingleActionPerTarget}

\noindent
We now compare the performances of players having access to two different action sets. The first action set ${\cal A}_{\text{m}}$ is one where the player (the ``multi-action player'') can target every square millimeter on the dartboard and the second action set ${\cal A}_{\text{s}}$ is one where the player (the ``single-action player'') can only target the center of each possible target region. ${\cal A}_{\text{m}}$ and ${\cal A}_{\text{s}}$ therefore have approximately 90,000 and 61 actions, respectively. This comparison is of interest because the multi-action player must estimate the skill model (\ref{eq:DM4}) for each target region whereas the single-action player's skill model relies only on the pseudo-counts / pseudo-fractions for each target region and therefore he has no use for the skill model (\ref{eq:DM4}). The single-action player will inevitably be inferior to the corresponding multi-action player in a ZSG since he has a much smaller action set. If we find the degree of inferiority to be mild, however, then this would suggest that we lose little in the ZSG setting by restricting players to the action set ${\cal A}_{\text{s}}$ thereby finessing the issues of disentangling skill from bias (see Section \ref{sec:SkillModel} and Appendix \ref{sec:Bias}).

To make the comparison between the two players fair, we need to ensure that regardless of the target region, the success probabilities are equalized for the single-action and multi-action players when the latter targets the center of the target region. We do this as follows. Using the BSP pseudo-counts, we fit the skill model (\ref{eq:DM4}) to all\footnote{The D11, D13, D15, and D17 regions are very rarely targeted and so we do not have enough data to estimate the bivariate-normal skill models for them even when we use the BSP pseudo-fractions. While we could use the BSR pseudo-fractions, because they are so rarely targeted we simply assume the fitted skill model for DB applies to them, as well as all the single regions SB and S1-S20.}
target regions. That is, we fit the BSPNC model and take this to be the skill model of both the single-action and multi-action players. While the multi-action player can target every square-millimeter of the dartboard, the single-action player can only target the center of each target region. We also assume both players have full information and therefore know their own skill models and their opponent's skill models (both are BSPNC) as well as each other's action sets.  Finally we assume the multi-action player has zero bias. This obviously favors the multi-action player and, in light of our discussion in Appendix \ref{sec:Bias}, is not fully supported by the data.

Analogous to the analysis in \eqref{eq:Adv1}, we only consider players playing against replicas of themselves.
We let $\PA({\cal A}_{\text{s}}/{\cal A}_{\text{m}},N)$ be the probability that player A wins an $N$-leg match when A uses the single-action ${\cal A}_{\text{s}}$ and his replica opponent uses the multi-action ${\cal A}_{\text{m}}$ throughout the match. We let
$\PA({\cal A}_{\text{m}}/{\cal A}_{\text{m}},N)$ be the corresponding probability where both player A and his replica opponent use ${\cal A}_{\text{m}}$. We define the gain in win-probability to be
\begin{equation} \label{eq:Adv2}
\mbox{Gain}({\cal A}_{\text{m}}, {\cal A}_{\text{s}}, N) := \PA({\cal A}_{\text{m}}/{\cal A}_{\text{m}},N) - \PA({\cal A}_{\text{s}}/{\cal A}_{\text{m}},N).
\end{equation}
We report $\mbox{Gain}({\cal A}_{\text{m}}, {\cal A}_{\text{s}}, N)$ for each player in Table \ref{table:winprob_loss action set compare} for $N=1$ and $N=35$.
\begin{table}[ht]
\begin{center}
\captionsetup{justification=centering}
{
\renewcommand{\tabcolsep}{0.7mm}
\caption{Gains of win-probabilities in $N$-leg matches for using a multi-action set\\ rather than a single-action set}
\vspace{-1mm}
\label{table:winprob_loss action set compare}
\small
\begin{tabular}{l ccc}
\toprule
Player & ~ & $N=1$ & $N=35$\\
\midrule
Anderson&~&0.3&1.3\\
Aspinall&~&0.2&0.7\\
Chisnall&~&0.3&1.2\\
Clayton&~&0.2&0.9\\
Cross&~&0.3&1.1\\
Cullen&~&0.3&1.3\\
van Gerwen&~&0.2&0.9\\
Gurney&~&0.2&0.7\\
\bottomrule
\end{tabular} \hspace{1.cm}
\begin{tabular}{l ccc}
\toprule
Player & ~ & $N=1$ & $N=35$\\
\midrule
Lewis&~&0.3&1.0\\
Price&~&0.2&0.7\\
Smith&~&0.2&0.8\\
Suljovic&~&0.2&0.8\\
Wade&~&0.3&1.0\\
White&~&0.2&1.0\\
Whitlock&~&0.3&1.0\\
Wright&~&0.2&0.9\\
\bottomrule
\end{tabular}}~\\
\end{center}
\justify
\small
{\em Notes.} Numbers are in percentages. The average gains across 16 players are 0.2\% and 1.0\% for $N=1$ and $N=35$, respectively.
In \eqref{eq:Adv2}, ${\cal A}_{\text{m}}$ and ${\cal A}_{\text{s}}$ denote the multi-action and single-action players, respectively, with the player before the ``/'' throwing first in the first leg of the match.
\end{table}

We find that the gain numbers are very small, ranging from 0.2\% - 0.3\% in the single-leg case to 0.7\% - 1.3\% in the N=35 leg case. For any given player, we see $\mbox{Gain}({\cal A}_{\text{m}}, {\cal A}_{\text{s}}, N)$ increases in $N$ but this must be the case since any edge in win-probability over a single leg will be magnified as $N$ increases so that $\lim_{N \to \infty} \mbox{Gain}({\cal A}_{\text{m}}, {\cal A}_{\text{s}}, N)=50\%$. Note that the gain is relatively small even in the best of 35-leg case (a typical number of legs in finals of major competitions), which suggests that the simpler action set ${\cal A}_{\text{s}}$ is more than adequate in ZSG settings. Indeed, these gain numbers are almost surely overstated since we implicitly assumed (i) the bias was zero for each player and (ii) the skill model (\ref{eq:DM4}) was an accurate representation of the multi-action player's skill model. At best, these assumptions will only be approximately true.

\subsection{Analysis of Some Real-World Game Situations}
\label{sec:RealWorld}

In this section we consider two specific real-world match-play situations each of which featured a player's decision that was viewed by pundits and fans as being at least somewhat surprising. We show how our BSPNC multi-action models\footnote{In order to finesse the possibility of bias we simply assumed that when a player targeted a particular region in the data that he aimed at the center of the region. Hence we use the BSPNC model rather than the BSPN model.} can be used to analyze these decisions and in fact argue that the decisions were not surprising at all. To be clear, these situations are  only intended to be indicative of the types of analysis we could conduct given sufficient data, and for several reasons should not be viewed as definitive. For example, our models were fit using data from the 2019 season whereas our first example is from a match during the 2018 season. It is quite possible that the form (and therefore skill levels) of the players in 2018 were quite different from their average form in the 2019 season. This is also true for the second example even though it is from the 2019 season. It is possible that at the time of these matches the players' forms were different from their average 2019 form which is essentially what the skill models aim to capture. Moreover, and as discussed in Section \ref{sec:Data}, we have ignored the possibility of bounce-outs and assumed the same skill model applies to each of the throws within a turn. All of this reinforces that our fitted skill models are only approximations to reality and more granular data would be required to improve them.

\subsubsection{Price vs Wright in 2018 Shanghai Masters Quarter-Final}

The first example we consider is a quarter-final match between Gerwyn Price (player A) and Peter Wright (player B) in the 2018 Shanghai Masters. This was a best of 15 legs match and the situation occurred in the $14^{th}$ leg with Wright leading by 7 legs to 6. In the $14^{th}$ leg Wright was on a score of 18 and it was his turn to shoot with Price on 20. An obvious exit strategy for Wright was to aim at D9. Instead, however, he (successfully) aimed his first throw of the turn at S2 leaving him two throws to exit on D8. However, he missed both of these throws and Price went\footnote{Wright did win the final deciding leg, however, and therefore won the match. The incident in question and indeed the entire match may be seen via \cite{WrightPrice} on \texttt{YouTube}.} on to win the leg in his next turn.

\begin{figure}[h]
  \begin{center}
    \subfigure[Leg 14 of Price (player A) vs. Wright (player B) in QF of 2018 Shanghai Masters.]
    {\label{figure:PriceWrightExample1_DM}
    \includegraphics[width=0.48\linewidth]{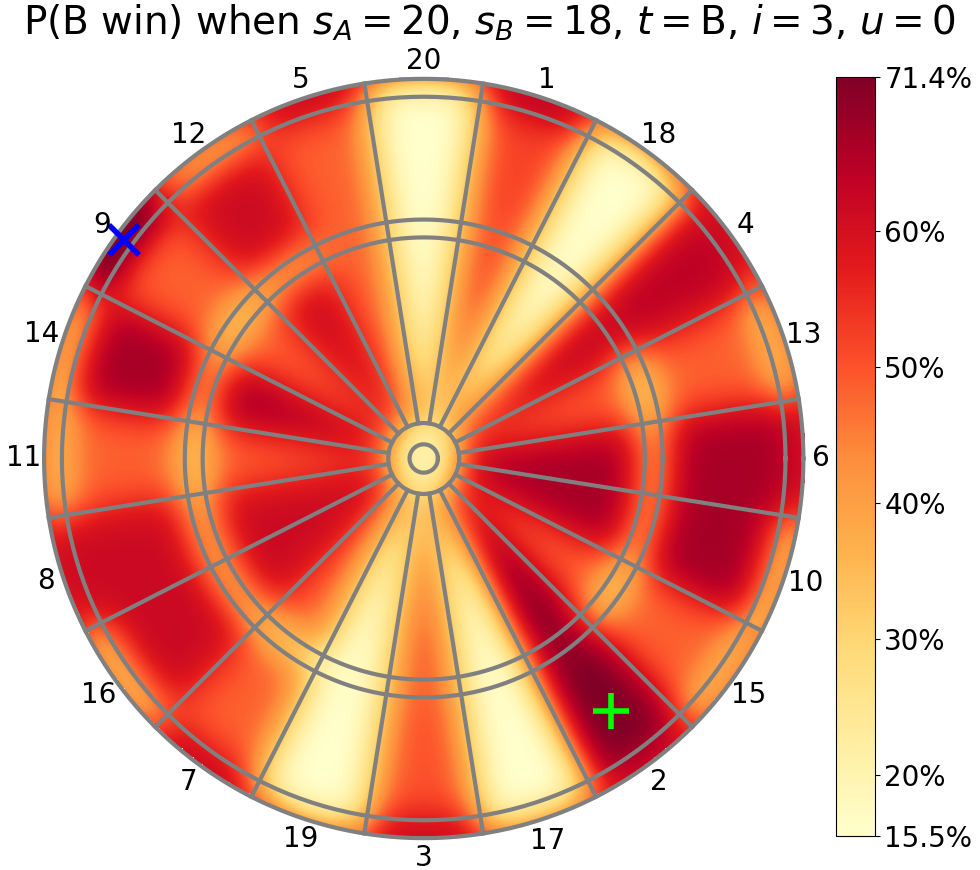}}
    \hspace{1mm}
    \subfigure[A hypothetical situation of Price (player A) vs. Smith (player B).]
    {\label{figure:PriceWrightExample2_DM}
    \includegraphics[width=0.48\linewidth]{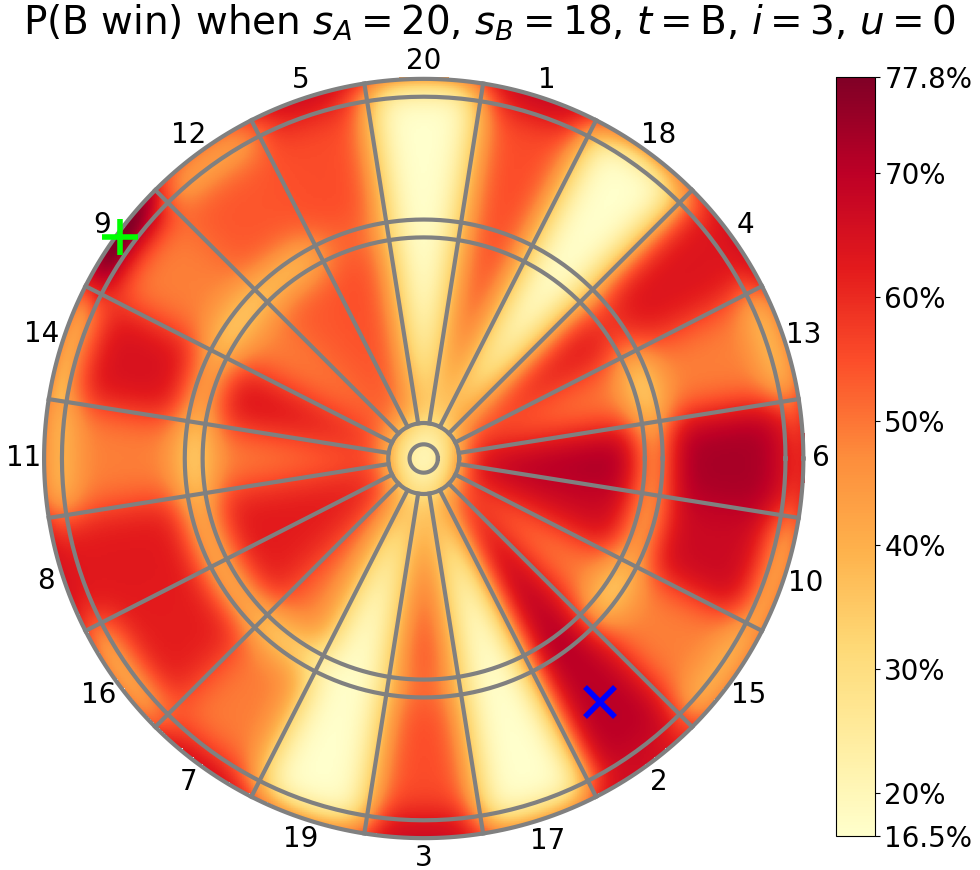}}
  \end{center}
  \vspace{-3mm}
  \small
  \emph{Note:}
  The heat-maps display win-probabilities as a function of where player B targets the first throw in his turn assuming both players play optimally thereafter. The left-hand heat-map displays the situation at the beginning of Wright's (player B's) turn.
  In the first throw, Wright wins the leg with probability 71.4\% if he targets S2 (the green ``+'') followed (if successful) by D8, or 70.5\% if he aims at D9 (the blue ``$\times$'') instead.
  The right-hand heat-map displays the same situation but where player B is now Michael Smith.
  In this case the optimal decision for Smith is to target D9 (the green ``+''), with a corresponding win-probability of 77.8\%,
  comparing to 70.3\% if he aims at S2 (the blue ``$\times$'').
  \caption{Comparing S2-D8 and D9 exit strategies. }
  \label{figure:PriceWrightExample_DM}
\end{figure}

After losing this leg there was some questioning of Wright's decision to go for the S2 (and then D8) rather than having three throws at D9. The analysis using our fitted skill model for Price may be seen in Figure \ref{figure:PriceWrightExample1_DM}.
We see that, according to the fitted DM skill model, the optimal decision for Wright was indeed to aim for S2 with his first throw (and to then check out via D8) rather than D9. While the difference in winning probabilities (71.4\% for S2-D8 vs 70.5\% for D9) is small, if anything the evidence points to his having made the correct decision.
It is interesting to consider what other players should have done had they been in the same situation as Wright.
For example, as displayed in Figure \ref{figure:PriceWrightExample2_DM}, we find that Michael Smith should have targeted a D9 exit rather than the S2-D8 strategy of Wright. In Smith's case the difference in winning probabilities (77.8\% for D9 vs 70.3\% for S2-D8) is substantial. The difference in optimal strategies for the two players is explained by their fitted BSPNC skill models when targeting D8 and D9. In particular, Smith succeeds in targeting D8 and D9 with probabilities 41.5\% and 44.7\%, respectively, while Wright succeeds in targeting D8 and D9 with probabilities 43.0\% and 36.4\%, respectively.

\vspace{-2mm}

\subsubsection{Anderson vs Price in 2019 Grand Slam of Darts Final}

Our second example comes from a match between Gary Anderson (player A) and Gerwyn Price (player B) in the quarter-final of the 2019 Grand Slam of Darts. This was a best of 31 legs match. In one of the legs Anderson was on a score of 80 and it was his turn while Price was on a score of 20. Anderson tried to exit via D20-D20 but missed outside (for a score of zero) with his first throw, failed to hit D20 with his second throw and therefore failed to exit on his turn. The most common approach to exiting from a score of 80 is to target T20-D10, however, and so we consider this situation using Anderson's fitted skill model.

\begin{figure}[h]
  \begin{center}
    \subfigure[Optimal to aim at S20 with first throw.]
    {\label{figure:AndersonPriceExample1_DM}\includegraphics[width=0.48\linewidth]{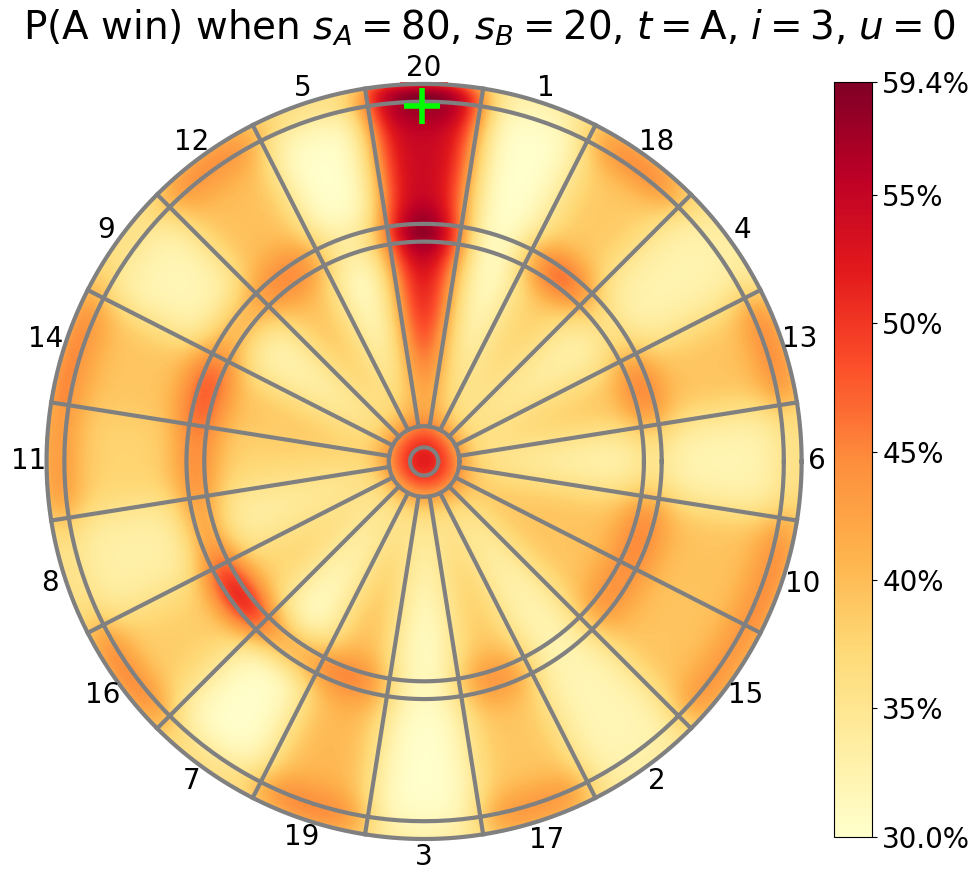}}
    \subfigure[Optimal to aim at S20 with second throw if first throw hits D20.]
    {\label{figure:AndersonPriceExample2_DM}\includegraphics[width=0.48\linewidth]{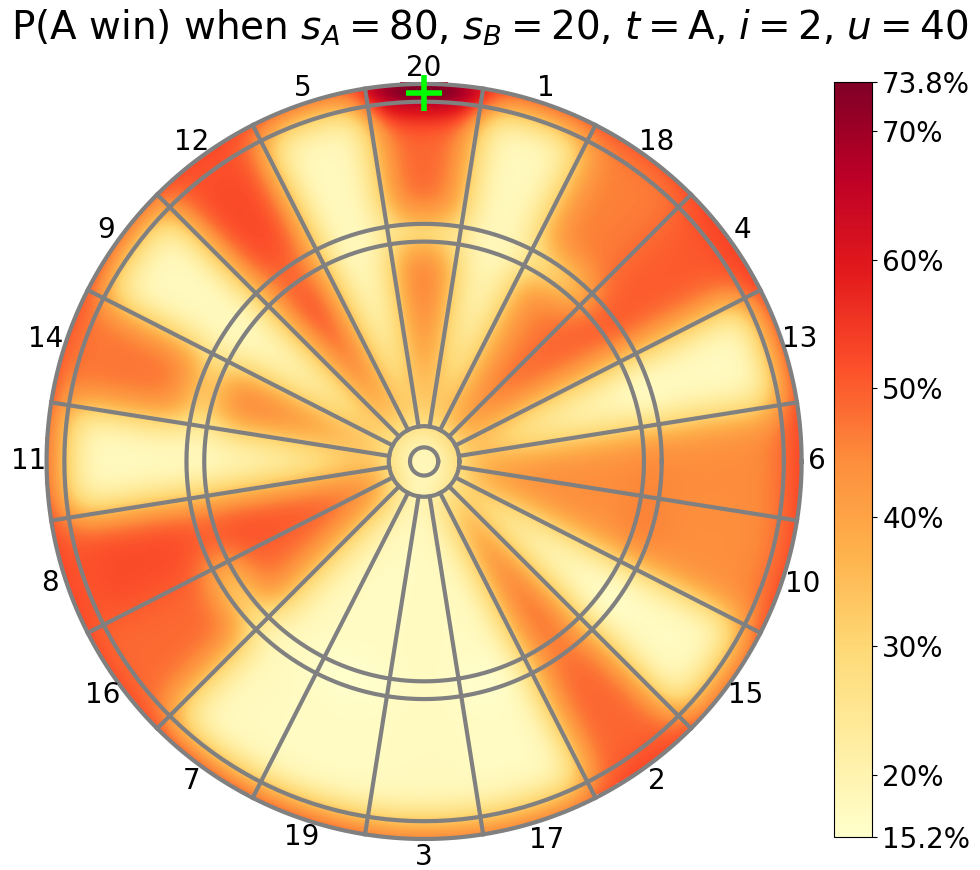}} \\
  \end{center}
  \small
  \emph{Note:} The main takeaway is that on the first throw of his turn, Anderson should hedge his bets by targeting the boundary between the S20 and D20 regions.
  \caption{Anderson (player A) vs. Price (player B) in QF of 2019 Grand Slam of Darts}
  \label{figure:AndersonPriceExample_DM}
\end{figure}

In Figure \ref{figure:AndersonPriceExample_DM} we display heat-maps for the various scenarios. Figure \ref{figure:AndersonPriceExample1_DM} considers the first throw of the turn and in fact it suggests that the optimal action for Anderson was to target S20. It is interesting to note, however, that the optimal target (as indicated by the green cross) is located on the boundary of the D20 and S20 regions. This suggests the real goal was a score of D20 but that the strategy wanted to ``hedge its bets'' so that if Anderson missed D20, it was important to miss and score S20 rather than miss outside and score nothing.
This highlights one of the advantages of building a skill model which can analyze any target within a region. In Figure \ref{figure:AndersonPriceExample2_DM} it is clearly optimal for Anderson to target D20 on his second throw if he had been successful in hitting D20 on his first throw. It is interesting to note that the green cross in Figure \ref{figure:AndersonPriceExample2_DM} is now located in the center of the D20 region rather than on the boundary as in Figure \ref{figure:AndersonPriceExample1_DM}.
We note that Anderson's success rates (based on his fitted BSPNC model) are 39.1\% and 45.5\% when targeting D10 and D20, respectively. In contrast, James Wade, for example, has success rates (again based on his fitted BSPNC model) of 48.3\% and 39.4\% when targeting D10 and D20, respectively. It is therefore not surprising that he would have preferred\footnote{And we confirmed this using our ZSG analysis.} the more typical T20-D10 checkout attempt than the D20-D20 preference of Anderson in the scenario above.

\section{Conclusions and Further Research}
\label{sec:Conclusions}

We performed an exploratory data analysis on a data-set for the top 16 professional players from the 2019 season. We identified several problems that arose due to natural limitations in the data and proposed an empirical Bayesian approach based on the DM distribution to overcome some of these problems. Specifically we introduce two DM-based skill models where the first model borrows strength from other darts players and the second model borrows strength from other regions of the dartboard. We evaluate these models through the use of proper scoring rules and find that they outperform other benchmark models in terms of both statistical and practical significance. We also used our DM-based models to estimate the importance of being able to target every square millimeter of the dartboard, and also to analyze specific situations that arose in real-world darts matches during the 2019 season.

There are several directions for future research. First, if we had access to the professional players we could resolve the issue regarding the decomposition of  $\bmu$ into the target $\bvartheta$ and the bias $\btheta$ as discussed briefly in Section \ref{sec:SkillModel} and in further detail in Appendix \ref{sec:Bias} of the {\em Supplementary Material} file. Second, the technology to capture the precise landing location of a dart now exists and we anticipate that at some point this technology will be used in professional competitions. If this occurs and the resulting data is made available, then fitting second-stage skill models will no longer be a missing-data problem requiring the use of the EM algorithm. In that event we could certainly resolve the bias and correlation identification issues and develop skill models beyond (\ref{eq:DM2}) that might more accurately reflect the players' skill levels.

\section*{Acknowledgements}
We are very grateful to Christopher Kempf (\texttt{Twitter}  @ochepedia), Statistical Analyst for the PDC  for providing us with the data underlying this work and for some very insightful conversations. All errors are our own.

\bibliographystyle{ormsv080}
\bibliography{References_Darts_Stats}

\newpage


\newpage
\thispagestyle{empty}
\begin{center}
\textbf{\large Supplementary Materials for: \\ \vspace{.25cm} An Empirical Bayes Approach for Estimating Skill Models for Professional Darts Players}

\vspace{.75cm}

\author{\sffamily Martin B. Haugh \\
\sffamily Department of Analytics, Marketing \& Operations \\
\sffamily Imperial College Business School, Imperial College \\
\texttt{m.haugh@imperial.ac.uk}
\vspace{.75cm}

\and \sffamily Chun Wang \\
\sffamily  Department of Management Science and Engineering\\
\sffamily School of Economics and Management, Tsinghua University \\
\texttt{wangchun@sem.tsinghua.edu.cn}}
\end{center}

\makeatletter

\setcounter{page}{1}
\renewcommand{\thepage}{S-\arabic{page}}
\setcounter{section}{0}
\setcounter{rem}{0}
\appendix

\section{Dartboard Geometry}
\label{sec:geom}

A standard competition dartboard has the following measurements (in millimeters):\vspace{.2cm}

{\small
\noindent
\begin{minipage}[b]{0.55\linewidth}
\vspace{.25cm}
{ \begin{tabular}{l c }
  \toprule
  Distance & Measurement \\
  \midrule
  Center to DB wire & 6.35 \\
  Center to SB wire & 15.9 \\
  Center to inner triple wire & 99 \\
  Center to outer triple wire & 107 \\
  Center to inner double wire & 162 \\
  Center to outer double wire & 170 \\
  \bottomrule
\end{tabular}\vspace{.25cm}
}
\end{minipage} 
\begin{minipage}[t]{0.45\linewidth}    \vspace{-4.3cm}
\begin{center}
\includegraphics[width=.65\linewidth]{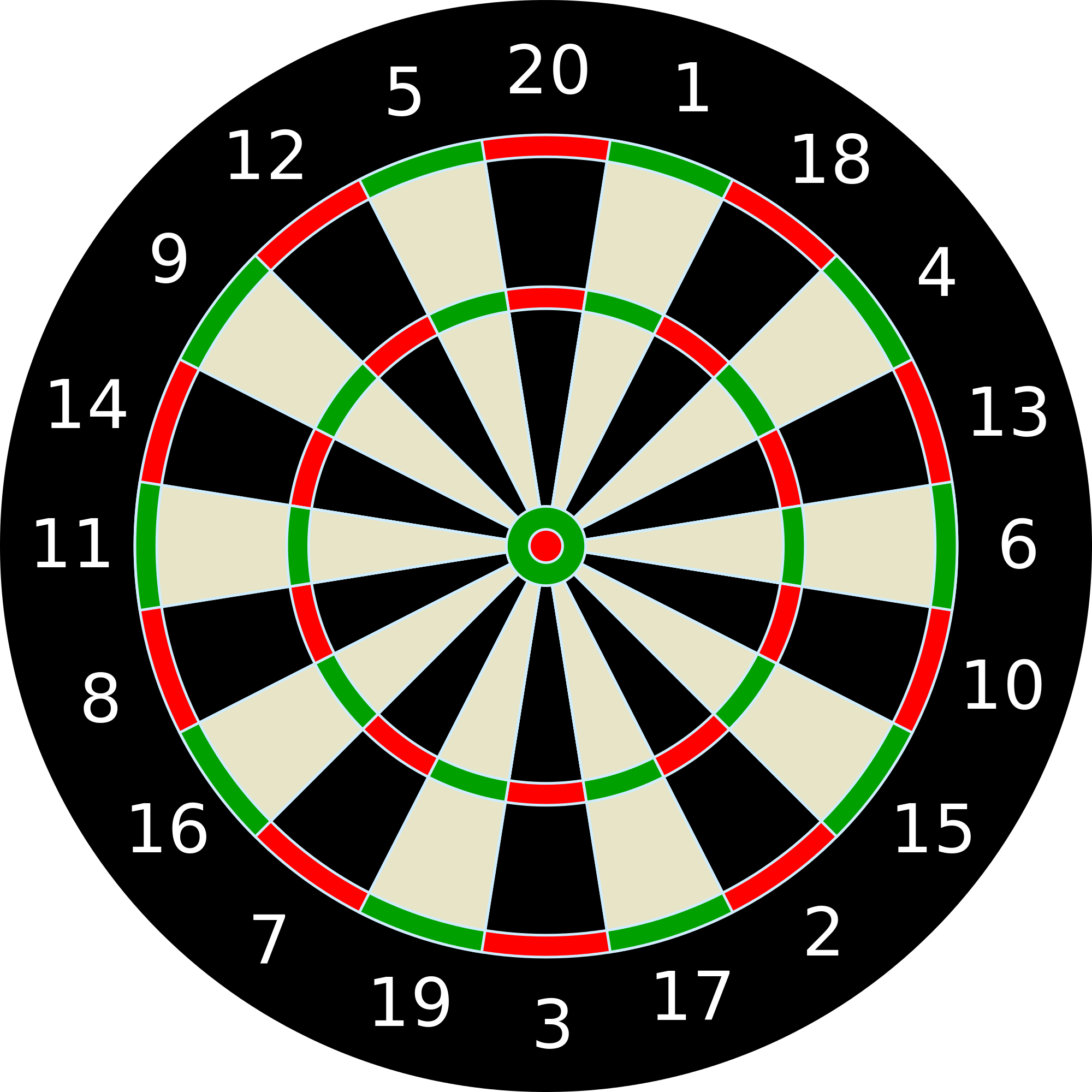}
\end{center}
\end{minipage}
\vspace{.2cm}
}

\noindent
Given these measurements and taking the center of DB to be $(0,0)$, it is straightforward to compute the map $z=g(x,y)$ that maps any $(x,y)$-coordinate on the dartboard to the corresponding dart score.

\section{Fitting the Bivariate-Normal Skill Model via the EM Algorithm}
\label{app:EM}

Our derivation of the EM algorithm here closely follows the derivation in \cite{HaughWang-Darts-2021}. The algorithm was developed by \cite{Tibshirani} and we provide an easy extension here to allow for the case where the mean $\bmu$ is also unknown and therefore needs to be inferred.

Assume then that we wish to fit each component of the Gaussian skill model (\ref{eq:DM4}) for a particular player. We will fix a particular target region and let the data for this component be $\{(z_k,n_k )\}_{k=1}^K$ where $K$ is the number of possible outcomes for the target region
and $n_k$ is the number of darts that were aimed at the target region and achieved an outcome of $z_k$.
As discussed in Section \ref{sec:Data}, we have that $K=$ 22, 7, and 6 for DB, double, and treble regions, respectively.
The log-likelihood then satisfies
\vspace{-0.1cm}
\begin{equation} \label{eq:CompleteLogLik1}
L(\bmu,\bSigma; \, \{(z_k,n_k )\}_{k=1}^K ) \propto \sum_{k=1}^K n_k \, \log \left(\Pb \left((x,y) \in  R(z_k) \mid (x,y) \sim \mbox{N}_2\left(\bmu, \bSigma\right)\right)\right),
\end{equation}
where $R(z_k)$ is the region on the dartboard defined by the outcome $z_k$, i.e. $R(z_k) = \{(x,y) \in \mathbb{R}^2 \, : \, g(x,y) = z_k \}$.
We omit the dependence of $(\bmu,\bSigma)$ on the target region, and also omit the dependence of $L$ on the data. We note that $L(\bmu,\bSigma)$ can be easily computed via numerical integration but rather than attempting to maximize it w.r.t. $(\bmu,\bSigma)$ directly, we can instead use the EM algorithm.

Let $n=n_1 + \cdots + n_K$ denote the total number of darts and let  $\bnu_i \in \mathbb{R}^2$ denote the unobserved realized position of the $i^{th}$ dart. The complete-data log-likelihood then\footnote{There is a slight abuse of notation here as $z_i$ now represents the outcome of the $i^{th}$ dart whereas in (\ref{eq:CompleteLogLik1}) $z_k$ represents the $k^{th}$ possible outcome of which there are only $K$. The particular interpretation we have in mind should be clear from the context and also from the subscript: we use $z_i$ to denote the outcome of the $i^{th}$ dart and $z_k$ to denote the $k^{th}$ possible outcome.} satisfies
\vspace{-0.3cm}
\begin{eqnarray}
L(\bmu,\bSigma; \, \{(\bnu_i,z_i,)\}_{i=1}^n) &=& -\frac{n}{2} \log | \bSigma | - \frac{1}{2} \sum_{i=1}^n \left(\bmu - \bnu_i\right)^\top \bSigma^{-1} \left(\bmu - \bnu_i \right) \nonumber \\
&=& -\frac{n}{2} \log | \bSigma | - \frac{1}{2} \mbox{tr} \left(\bSigma^{-1} \sum_{i=1}^n \left(\bmu - \bnu_i\right) \left(\bmu - \bnu_i \right)^\top \right),  \label{eq:LogLike1}
\end{eqnarray}
where $z_i = g(\bnu_i)$ for $i=1, \ldots , n$ and $\mbox{tr}(\cdot)$ denotes the trace operator. The E-step of the EM algorithm requires the calculation of
\begin{equation} \label{eq:E-step}
\Ex_{\bmuold,\bSigmaOld}\left[L(\bmu,\bSigma) \mid \bz \right] = -\frac{n}{2} \log | \bSigma | - \frac{1}{2} \mbox{tr} \left(\bSigma^{-1} \sum_{i=1}^n \Ex_{\bmuold,\bSigmaOld}\left[ \left(\bmu - \bnu_i\right) \left(\bmu - \bnu_i \right)^\top \mid z_i\right]\right)
\end{equation}
where $\Ex_{\bmuold,\bSigmaOld}[\cdot \mid \bz]$ denotes the expectation operator conditional on the observed dart scores $\bz = (z_1, \ldots , z_n)$, and using the current parameter estimates $(\bmuold,\bSigmaOld)$. The M-step requires the maximization of (\ref{eq:E-step}) over $(\bmu,\bSigma)$.
This is similar to the usual maximum likelihood estimation for a multivariate Gaussian and has solution
\begin{eqnarray}
\bmuNew &=& \frac{1}{n} \sum_{i=1}^n \Ex_{\bmuold,\bSigmaOld} \left[\bnu_i  \mid z_i  \right] \nonumber \\
&=& \frac{1}{n} \sum_{k=1}^{K} n_k \Ex_{\bmuold,\bSigmaOld}\left[\bnu_k  \mid z_k  \right], \label{eq:M-step0}\\
\bSigmaNew &=& \frac{1}{n} \sum_{i=1}^n \Ex_{\bmuold,\bSigmaOld}\left[ \left(\bmuNew - \bnu_i\right) \left(\bmuNew - \bnu_i \right)^\top \mid z_i\right] \nonumber \\
&=& \frac{1}{n} \sum_{k=1}^{K} n_k \Ex_{\bmuold,\bSigmaOld}\left[ \left(\bmuNew - \bnu_k\right) \left(\bmuNew - \bnu_k \right)^\top \mid z_k\right]. \label{eq:M-step2}
\end{eqnarray}
The expectations in (\ref{eq:M-step0}) and (\ref{eq:M-step2}) cannot be computed in closed form and so we compute them numerically via importance-sampling (IS). Specifically, we compute the expectations as
\begin{eqnarray}
\Ex_{\bmuold,\bSigmaOld}\left[\bnu_k  \mid z_k  \right] & \approx & \sum_{l=1}^m w_{k,l} \bnu_{k,l},   \label{eq:E-step20} \\
\Ex_{\bmuold,\bSigmaOld}\left[ \left(\bmuNew - \bnu_k\right) \left(\bmuNew - \bnu_k \right)^\top \mid z_k\right] & \approx & \sum_{l=1}^m w_{k,l} \left(\bmuNew - \bnu_{k,l}\right) \left(\bmuNew - \bnu_{k,l} \right)^\top,  \label{eq:E-step2}
\end{eqnarray}
where $m$ is the number\footnote{We used $m=50,000$ when fitting our skill models. We found that a lower number of samples such as $m=3,000$ sometimes created non-convergence issues for the EM algorithm.} of samples we used, the $\bnu_{k,l}$'s are samples of $\bnu_k$ and the $w_{k,l}$'s are corresponding IS weights. Specifically
\begin{equation} \label{eq:IS1}
w_{k,l} = \frac{p_{k,l}/q_{k,l}}{\sum_{i=1}^m p_{k,i}/q_{k,i}}
\end{equation}
where $p_{k,l}$ is the density of $\bnu_{k,l}$ under the true distribution of $\bnu_{k}$ conditional on $z_k$ and using the parameter estimate $(\bmuold,\bSigmaOld)$, and $q_{k,l}$ is the sampling density of $\bnu_{k,l}$. Because we only know $p_{k,l}$ up to a multiplicative constant, we normalize in (\ref{eq:IS1}) to ensure the weights sum to 1. TPT suggested taking the sampling distribution $q_{k,l}$ to be uniform on $R(z_k)$ and this of course satisfies the IS requirement that $p=0$ whenever $q=0$. Moreover with this choice (\ref{eq:IS1}) is easy to compute since by assumption $q_{k,l} \propto 1$ and $p_{k,l}$ is the bivariate normal PDF with mean $\bmuold$ and variance-covariance matrix $\bSigmaOld$ evaluated at $\bnu_{k,l}$.
The EM algorithm then proceeds by iterating (\ref{eq:M-step0}) to (\ref{eq:IS1}) until convergence.

\begin{rem}
It is clear from (\ref{eq:M-step0}) and (\ref{eq:M-step2}) that we only need the {\em fractions} $n_k/n$ of darts that fell in each of the possible outcome areas for the target under consideration. In our two-stage modelling approach, this means we should use the pseudo-fractions $(\widehat{\alpha}_k + x_{jk})/\sum_{l=1}^K(\widehat{\alpha}_l + x_{jl})$
 for fitting the second-stage skill model of player $j$. (This was also noted near the end of Section \ref{sec:DM-Model}.)
\end{rem}

\begin{rem}
If $\bmu$ is known then we only need to iterate (\ref{eq:M-step2}), (\ref{eq:E-step2}), and (\ref{eq:IS1}) with $\bmuold=\bmuNew$ set equal to the known value of $\bmu$ throughout. This is the case for the unbiased model.
\end{rem}

\section{Interpreting $\mu$: Is There a Bias in the Players' Skill Models?}
\label{sec:Bias}

As mentioned in Section \ref{sec:SkillModel}, TPT implicitly assumed the mean outcome $\bmu$ and target  $\bvartheta$ were one and the same. Even for professional players, it is quite possible there may be some {\em bias}, i.e. discrepancy between the target $\bvartheta$ and the mean $\bmu$, especially for less frequently targeted parts of the dartboard. A further complication is that in our data-set we do not know $\bvartheta$ even though we always know the target region. So for example, we will know if a dart was aimed at T20 but we will not know precisely where in the T20 bed the dart was aimed. In light of this limitation in our data-set, it seems natural to make the following assumption.

\begin{assum} \label{ass:Center}
Given a target region TR, the specific target $\bvartheta$ that a player aims for is the center of the target region with the center being defined as the midpoint of the polar-coordinates defining the region.
\end{assum}
While Assumption \ref{ass:Center} is likely to hold quite generally, it's not difficult to imagine a player deviating from this assumption on occasion. For example, suppose a player has two darts remaining in his turn and needs a D5 to check out, i.e. win the leg. Rather than aiming at the center of D5, he may prefer to aim a little closer to the outer circular boundary of the D5 region on the basis that if he is going to miss D5 he would prefer to miss the scoring region entirely rather than hit S5 which would leave him unable to check out on his final dart of the turn. This argument does not apply to even doubles such as D20, D18, D16, etc. because if the corresponding single is hit then the player can still exit on the next dart in his turn. It's also true that Assumption \ref{ass:Center} may not always hold with the T20 region. In particular, if darts from earlier throws in the turn have already landed in T20, then a player may adjust his intended target within the T20 region to account for the positioning of the earlier darts in the throw. Unfortunately, we cannot identify such situations within our data-set. Nonetheless, we believe this to be at most a second-order issue when it comes to the ultimate goal of modelling darts matches between players.

Even if we invoke Assumption \ref{ass:Center}, however, there remains the possibility of {\em bias}. It therefore makes sense to decompose the mean outcome $\bmu$ according to $\bmu = \bvartheta +\btheta$ where $\btheta$ denotes a bias term. Given sufficient data, we can infer $\bmu$ but if we do not know the target $\bvartheta$ then we cannot infer the bias $\btheta = \bmu-\bvartheta$. TPT implicitly assumed $\btheta = {\bf 0}$ and since they knew $\bvartheta$, they therefore also knew $\bmu=\bvartheta$.
Regardless then of whether we invoke Assumption \ref{ass:Center}, $\bmu$ will be unknown and therefore must be estimated in our skill models. From a model estimation point of view, our decomposition of $\bmu$ into $\bvartheta +\btheta$ has no bearing: we simply fit $[x \ y]^\top \sim \mbox{N}_2(\bmu, \bSigma).$
But the decomposition of $\bmu$ into $\bvartheta+\btheta$ has significant ramifications for using the skill model in a ZSG.
To see this, consider two cases: (a) $\btheta = {\bf 0}$ so $\bmu = \bvartheta$ and (b)  $\btheta$ is ``large'' in magnitude. In case (b) the player is presumably unaware of this large bias since if he was aware of it then he could make adjustments to counteract it. Therefore, if employing the skill model in a game of darts we would have to assume the darts player under interpretation (a) would be considerably more skillful than the same dart player under interpretation (b). We finesse this issue in Section \ref{sec:Empirical Bayes-ZSG} by implicitly invoking Assumption \ref{ass:Center} and assuming $\bmu = \bvartheta$ so there is no bias.
In Appendix \ref{sec:mu} below we will in fact provide some evidence to suggest the bias $\btheta$ might be a significant component of $\bmu$.

\subsection[Interpreting Mu]{Interpreting $\bmu$}
\label{sec:mu}

\begin{figure}[h]
\begin{center}
\includegraphics[width=1\linewidth]{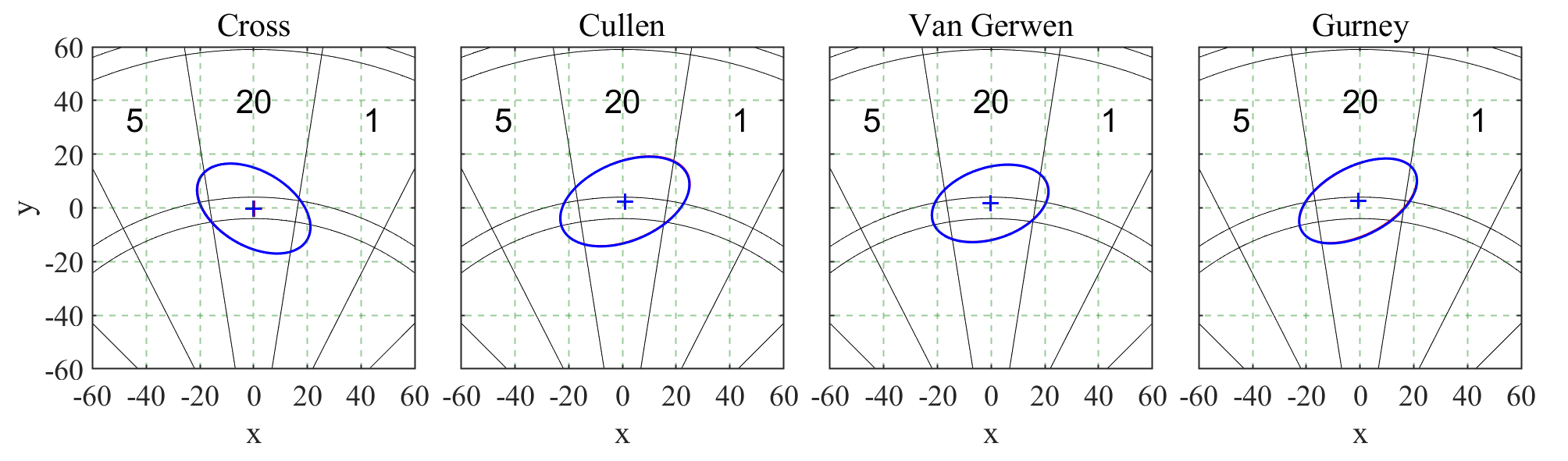}
\end{center}
\vspace{-0.5cm}
\caption{95\% CEs for the BSPN (blue) and RN (red) models when targeting T20. The two models are essentially identical with minimal shrinkage taking place due to the abundance of T20 data.}
\label{figure:FittedEllipseRealT20}
\end{figure}

We now briefly describe two points of evidence to suggest there is a significant non-zero bias $\bmu$.
First, if we assumed the inferred $\bmu$ was indeed the intended target then that would imply players are intentionally lowering their chances of hitting the intended target. Consider, for example, Gurney when he is targeting T20. As may be seen from the right-hand panel of Figure \ref{figure:FittedEllipseRealT20},  the inferred $\bmu$ for Gurney is very close to the boundary separating T20 from the upper S20 bed. As may be seen from Table \ref{table:FittedProb_RealData_T20_nonfixedmu_fourplayers} in Section \ref{sec:CorrelIdent}, his estimated probability of hitting T20 in his fitted BSPN model is 40.6\%. However, if the fitted $\bmu$ was indeed his intended target, then by moving his target to the center of T20 he could increase this success probability to 44.5\%. This would be a significant increase and it is difficult to imagine a player is intentionally behaving as sub-optimally as this. That said, we must also allow for the possibility that some of this is due to the limitations of the data-set and the use of the bivariate-normal skill model.

Second, we can compute the average distance between the inferred $\bmu$ and the center of the target region. We display these distances (averaged across the sixteen players) for the RN and BSPN models for each of the four treble regions in Table \ref{table:distance_change}. It is interesting to see that they increase as we move successively through T20, T19, T18, and T17. Because players throw considerably more often at the higher trebles, we would expect the magnitude of any bias to also increase as we move successively through these trebles and so our data is consistent with this observation. It is also gratifying to see that the phenomenon (of the distance increasing as we move from T20 to T17) persists (albeit it to a lesser extent) when we go from the RN model to the BSPN model. Overall then, these two observations suggest the bias term $\btheta$ might account for a considerable component of $\bmu$.

\begin{table}[h]
    \begin{center}
    \captionsetup{justification=centering}
    {
    \renewcommand{\tabcolsep}{1.0mm}
    \caption{Average absolute distance (in mm's) between the inferred $\bmu$ and the center of the target region}
    \vspace{-1mm}
    \label{table:distance_change}
    \small
    \begin{tabular}{lc rrr rrr rrr rrr rrr rrr r}
    \toprule
    Model Fitted &~& T20 & T19 & T18 & T17\\
    \midrule
    RN  &~& 1.50 & 1.62 & 3.56 & 5.23 \\
    BSPN        &~& 1.47 & 1.48 & 3.42 & 3.88 \\
    \bottomrule
    \end{tabular}}~\\
    \end{center}
\end{table}

\section{Additional Figures and Scoring Rules Results}
\label{sec:AllFits}

In this Appendix we present additional results on the various skill models that we considered in the main text. We begin in Appendix \ref{sec:AppScoringRules} where we display Brier scores and Spherical scores for the sixteen players and various skill models. (In Section \ref{sec:ScoringRules} we only presented the Brier scores for the double regions.) Then in Appendix \ref{sec:sig} we report some results on the statistical significance of the estimated Brier and Spherical scores when compared across the various skill models.
Finally  in Appendix \ref{sec:AppAllPlayersFigures} we display all the figures and tables from Section \ref{sec:Viz} in the main text but for all sixteen players rather than just four of them. We also include there the confidence ellipses (CEs) for D19 to demonstrate just how poor the bivariate-normal skill model can be when we use the raw-data-counts rather than the pseudo-counts arising from the DM-based models.

\subsection{Additional Results for Brier and Spherical Scores}
\label{sec:AppScoringRules}
As in Section \ref{sec:ScoringRules}, we highlight the best skill model for each player in bold and do likewise for the overall best performing model.

\begin{table}[H]
\begin{center}
\captionsetup{justification=centering}
{
\renewcommand{\tabcolsep}{1.2mm}
\caption{Brier scores for skill models aggregated across treble regions}
\vspace{-1mm}
\label{table:BrierScoresTriple}
\small
\begin{tabular}{lc cccccc}
\toprule
Player & ~& Basic & IJOC & BSP & BSR & BSPN & BSPNC \\
\midrule
Anderson& ~ &\textbf{-0.5895}&-0.5898&-0.5896&-0.5898&-0.5896&-0.5897 \\
Aspinall& ~ &-0.5633&-0.5632&\textbf{-0.5631}&-0.5632&-0.5631&-0.5631 \\
Chisnall& ~ &-0.5915&-0.5905&\textbf{-0.5903}&-0.5904&-0.5903&-0.5904 \\
Clayton& ~ &-0.5810&-0.5806&-0.5806&\textbf{-0.5805}&-0.5807&-0.5807 \\
Cross& ~ &-0.5592&-0.5582&\textbf{-0.5580}&-0.5581&-0.5580&-0.5581 \\
Cullen& ~ &-0.5708&-0.5713&-0.5709&-0.5710&\textbf{-0.5708}&-0.5709 \\
van Gerwen& ~ &-0.5562&-0.5559&-0.5559&-0.5559&-0.5559&\textbf{-0.5559} \\
Gurney& ~ &-0.5550&-0.5546&\textbf{-0.5544}&-0.5544&-0.5544&-0.5545 \\
Lewis& ~ &-0.5876&-0.5863&-0.5862&\textbf{-0.5862}&-0.5863&-0.5864 \\
Price& ~ &-0.5776&-0.5775&-0.5774&-0.5775&\textbf{-0.5774}&-0.5774 \\
Smith& ~ &-0.5604&-0.5594&-0.5593&-0.5593&\textbf{-0.5593}&-0.5593 \\
Suljovic& ~ &-0.5591&-0.5585&\textbf{-0.5583}&-0.5584&-0.5583&-0.5584 \\
Wade& ~ &\textbf{-0.5466}&-0.5469&-0.5468&-0.5468&-0.5469&-0.5469 \\
White& ~ &-0.5868&-0.5861&\textbf{-0.5859}&-0.5861&-0.5860&-0.5860 \\
Whitlock& ~ &-0.5619&-0.5619&\textbf{-0.5617}&-0.5619&-0.5617&-0.5617 \\
Wright& ~ &-0.5847&-0.5845&\textbf{-0.5845}&-0.5845&-0.5845&-0.5845 \\
\midrule
Average& ~ &-0.5707&-0.5703&\textbf{-0.5702}&-0.5702&-0.5702&-0.5703 \\
\bottomrule
\end{tabular}}~\\
\end{center}
\end{table}

\begin{table}[H]
\begin{center}
\captionsetup{justification=centering}
{
\renewcommand{\tabcolsep}{1.2mm}
\caption{Brier scores for skill models on DB region}
\vspace{-1mm}
\small
\begin{tabular}{lc ccccccc}
\toprule
Player & ~& Basic & IJOC & BSP & BSPN & BSPNC \\
\midrule
Anderson& ~ &\textbf{-0.5960}&-0.6132&-0.6029&-0.5974&-0.5978 \\
Aspinall& ~ &-0.5840&\textbf{-0.5836}&-0.5866&-0.5876&-0.5879 \\
Chisnall& ~ &\textbf{-0.6641}&-0.6671&-0.6719&-0.6658&-0.6656 \\
Clayton& ~ &\textbf{-0.6821}&-0.6905&-0.6925&-0.6845&-0.6854 \\
Cross& ~ &\textbf{-0.6531}&-0.6567&-0.6567&-0.6550&-0.6551 \\
Cullen& ~ &-0.4742&-0.5020&-0.4759&-0.4700&\textbf{-0.4698} \\
van Gerwen& ~ &\textbf{-0.6261}&-0.6270&-0.6273&-0.6269&-0.6272 \\
Gurney& ~ &-0.6544&\textbf{-0.6440}&-0.6542&-0.6451&-0.6450 \\
Lewis& ~ &-0.5201&-0.5268&-0.5246&\textbf{-0.5192}&-0.5194 \\
Price& ~ &\textbf{-0.6472}&-0.6504&-0.6480&-0.6512&-0.6511 \\
Smith& ~ &\textbf{-0.5553}&-0.5564&-0.5602&-0.5573&-0.5573 \\
Suljovic& ~ &-0.6029&-0.6018&-0.6054&-0.5993&\textbf{-0.5989} \\
Wade& ~ &-0.6369&-0.6253&\textbf{-0.6232}&-0.6385&-0.6383 \\
White& ~ &-0.6917&\textbf{-0.6834}&-0.6883&-0.6857&-0.6857 \\
Whitlock& ~ &-0.5317&-0.5456&-0.5329&\textbf{-0.5301}&-0.5301 \\
Wright& ~ &-0.5872&-0.5873&-0.5894&-0.5861&\textbf{-0.5859} \\
\midrule
Average& ~ &-0.6067&-0.6101&-0.6087&\textbf{-0.6062}&-0.6063 \\
\bottomrule
\end{tabular}}~\\
\end{center}
\justify
\small
{\em Notes.} BSR does not apply on DB region as there are no comparable regions from which we can borrow strength.
\end{table}

\begin{table}
\begin{center}
\captionsetup{justification=centering}
{
\renewcommand{\tabcolsep}{1.2mm}
\caption{Spherical scores for skill models aggregated across treble regions}
\vspace{-1mm}
\small
\begin{tabular}{lc cccccc}
\toprule
Player & ~& Basic & IJOC & BSP & BSR & BSPN & BSPNC \\
\midrule
Anderson& ~ &\textbf{0.6407}&0.6405&0.6406&0.6405&0.6406&0.6406 \\
Aspinall& ~ &0.6608&0.6609&\textbf{0.6610}&0.6610&0.6610&0.6610 \\
Chisnall& ~ &0.6391&0.6399&\textbf{0.6401}&0.6400&0.6400&0.6400 \\
Clayton& ~ &0.6473&0.6476&0.6476&\textbf{0.6476}&0.6475&0.6475 \\
Cross& ~ &0.6639&0.6646&\textbf{0.6648}&0.6648&0.6648&0.6647 \\
Cullen& ~ &\textbf{0.6552}&0.6548&0.6551&0.6550&0.6551&0.6550 \\
van Gerwen& ~ &0.6662&0.6664&0.6664&0.6664&\textbf{0.6664}&0.6664 \\
Gurney& ~ &0.6671&0.6674&\textbf{0.6675}&0.6675&0.6675&0.6674 \\
Lewis& ~ &0.6422&0.6432&0.6432&\textbf{0.6432}&0.6431&0.6430 \\
Price& ~ &0.6499&0.6499&0.6500&0.6500&\textbf{0.6500}&0.6500 \\
Smith& ~ &0.6630&0.6638&0.6638&0.6638&\textbf{0.6638}&0.6638 \\
Suljovic& ~ &0.6640&0.6644&\textbf{0.6646}&0.6646&0.6646&0.6645 \\
Wade& ~ &\textbf{0.6733}&0.6731&0.6732&0.6732&0.6732&0.6732 \\
White& ~ &0.6428&0.6433&\textbf{0.6435}&0.6434&0.6434&0.6434 \\
Whitlock& ~ &0.6619&0.6619&\textbf{0.6620}&0.6619&0.6620&0.6620 \\
Wright& ~ &0.6444&0.6446&\textbf{0.6446}&0.6446&0.6446&0.6446 \\
\midrule
Average& ~ &0.6551&0.6554&\textbf{0.6555}&0.6555&0.6555&0.6554 \\
\bottomrule
\end{tabular}}~\\
\end{center}
\end{table}

\begin{table}
\begin{center}
\captionsetup{justification=centering}
{
\renewcommand{\tabcolsep}{1.2mm}
\caption{Spherical scores for skill models aggregated across double regions}
\vspace{-1mm}
\small
\begin{tabular}{lc cccccc}
\toprule
Player & ~& Basic & IJOC & BSP & BSR & BSPN & BSPNC \\
\midrule
Anderson& ~ &\textbf{0.5841}&0.5811&0.5835&0.5783&0.5839&0.5795 \\
Aspinall& ~ &0.5798&0.5734&0.5846&0.5835&\textbf{0.5849}&0.5777 \\
Chisnall& ~ &0.5795&0.5812&0.5839&0.5830&\textbf{0.5847}&0.5836 \\
Clayton& ~ &0.5578&0.5635&0.5605&0.5617&0.5603&\textbf{0.5637} \\
Cross& ~ &0.5763&0.5801&\textbf{0.5844}&0.5838&0.5837&0.5789 \\
Cullen& ~ &0.5809&0.5816&\textbf{0.5869}&0.5847&0.5858&0.5785 \\
van Gerwen& ~ &0.5822&0.5849&0.5884&0.5884&\textbf{0.5888}&0.5848 \\
Gurney& ~ &0.5759&0.5722&0.5791&\textbf{0.5810}&0.5796&0.5762 \\
Lewis& ~ &0.5722&0.5740&0.5768&0.5744&\textbf{0.5770}&0.5747 \\
Price& ~ &0.5814&0.5821&0.5861&0.5853&\textbf{0.5865}&0.5826 \\
Smith& ~ &0.5760&0.5744&0.5811&0.5805&\textbf{0.5815}&0.5811 \\
Suljovic& ~ &0.5758&0.5789&0.5868&0.5863&\textbf{0.5877}&0.5800 \\
Wade& ~ &0.5830&0.5873&0.5888&0.5890&\textbf{0.5893}&0.5856 \\
White& ~ &0.5732&0.5754&0.5769&\textbf{0.5779}&0.5773&0.5752 \\
Whitlock& ~ &0.5744&0.5798&0.5822&0.5848&0.5835&\textbf{0.5872} \\
Wright& ~ &0.5787&0.5798&0.5818&0.5820&\textbf{0.5824}&0.5796 \\
\midrule
Average& ~ &0.5769&0.5781&0.5820&0.5815&\textbf{0.5823}&0.5793 \\
\bottomrule
\end{tabular}}~\\
\end{center}
\end{table}

\begin{table}
\begin{center}
\captionsetup{justification=centering}
{
\renewcommand{\tabcolsep}{1.2mm}
\caption{Spherical scores for skill models on DB region}
\vspace{-1mm}
\small
\begin{tabular}{lc ccccc}
\toprule
Player & ~& Basic & IJOC & BSP & BSPN & BSPNC \\
\midrule
Anderson& ~ &\textbf{0.6358}&0.6246&0.6315&0.6349&0.6346 \\
Aspinall& ~ &0.6450&\textbf{0.6454}&0.6437&0.6422&0.6420 \\
Chisnall& ~ &\textbf{0.5817}&0.5792&0.5757&0.5803&0.5804 \\
Clayton& ~ &\textbf{0.5666}&0.5630&0.5606&0.5661&0.5654 \\
Cross& ~ &\textbf{0.5915}&0.5868&0.5870&0.5886&0.5886 \\
Cullen& ~ &0.7322&0.7140&0.7319&0.7357&\textbf{0.7359} \\
van Gerwen& ~ &\textbf{0.6132}&0.6123&0.6114&0.6118&0.6116 \\
Gurney& ~ &0.5904&\textbf{0.5976}&0.5888&0.5963&0.5965 \\
Lewis& ~ &\textbf{0.6969}&0.6911&0.6932&0.6964&0.6963 \\
Price& ~ &\textbf{0.5955}&0.5940&0.5943&0.5926&0.5927 \\
Smith& ~ &\textbf{0.6672}&0.6664&0.6644&0.6659&0.6659 \\
Suljovic& ~ &0.6302&0.6325&0.6293&0.6333&\textbf{0.6336} \\
Wade& ~ &0.6038&\textbf{0.6151}&0.6144&0.6029&0.6031 \\
White& ~ &0.5596&\textbf{0.5639}&0.5611&0.5635&0.5635 \\
Whitlock& ~ &0.6871&0.6763&0.6869&\textbf{0.6876}&0.6876 \\
Wright& ~ &0.6426&0.6433&0.6418&0.6437&\textbf{0.6439} \\
\midrule
Average& ~ &0.6275&0.6253&0.6260&\textbf{0.6276}&0.6276 \\
\bottomrule
\end{tabular}}~\\
\end{center}
\justify
\small
{\em Notes.} BSR does not apply on DB region as there are no comparable regions from which we can borrow strength.
\end{table}

\subsection{Statistical Significance of Scoring Results}
\label{sec:sig}

We also conducted a series of paired t-tests to test whether the differences in test scores were significantly different from zero. We did this for each proper scoring rule (Brier or Spherical) and for the trebles, doubles and DB regions. For example, referring to the results in Table \ref{table:BrierScoresTriple} we could compare the Brier Score aggregated across the treble regions for the BSP skill model versus the IJOC skill model by running a paired t-test with 16 observations. Because each player borrows strength from other players in the BSP, BSPN and BSPN skill models, the samples for the t-tests involving these skill models are not strictly independent. Nonetheless, the dependence is likely to be very weak owing to the fact that the test sets for all the proper scoring rule estimates were independent across players. We report our results in Tables \ref{table:BrierScoresT-test} and \ref{table:SphericalScoresT-test} for the Brier and Spherical scores, respectively.
\begin{itemize}
\item[] Each hypothesis test compares
\item[] $H_0$: the means of the scores underlying the row skill model and the column skill model are equal

vs

\item[] $H_1$: the mean of the scores underlying the row skill model is higher (or lower depending on whether the estimated difference in scores is higher or lower) than that of the column model.
\end{itemize}
The difference of means (row skill model score - column skill model score) is reported with an asterisk denoting that $H_0$ is rejected at the 5\% level. Our main takeaways are that scores of the benchmark models (Basic and IJOC) are statistically lower than (and thus inferior to) the DM-based models on the doubles and trebles regions. There is little difference among the models when targeting DB.

\begin{table}[ht]
\begin{center}
\captionsetup{justification=centering}
{
\renewcommand{\tabcolsep}{1.2mm}
\caption{Paired t-tests on Brier scores of skill models}
\vspace{-1mm}
\label{table:BrierScoresT-test}
\small
\begin{tabular}{r rrrrrrrr}
\toprule
~ & \multicolumn{6}{c}{treble regions}\\
\cline{2-7}  \noalign{\smallskip}
~ & \multicolumn{1}{c}{Basic} & \multicolumn{1}{c}{IJOC} & \multicolumn{1}{c}{BSP} & \multicolumn{1}{c}{BSR} & \multicolumn{1}{c}{BSPN} & \multicolumn{1}{c}{BSPNC} \\
\midrule
Basic&~&-0.0004*&-0.0005*&-0.0005*&-0.0005*&-0.0005*\\
IJOC&~&~&-0.0001*&-0.0001*&-0.0001*&-0.0001*\\
BSP&~&~&~&0.0001*&0.0000\textcolor{white}{*}&0.0001*\\
BSR&~&~&~&~&-0.0000*&0.0000\textcolor{white}{*}\\
BSPN&~&~&~&~&~&0.0001*\\
\toprule
~ & \multicolumn{6}{c}{double regions}\\
\cline{2-7}  \noalign{\smallskip}
~ & \multicolumn{1}{c}{Basic} & \multicolumn{1}{c}{IJOC} & \multicolumn{1}{c}{BSP} & \multicolumn{1}{c}{BSR} & \multicolumn{1}{c}{BSPN} & \multicolumn{1}{c}{BSPNC} \\
\midrule
Basic&~&-0.0014\textcolor{white}{*}&-0.0058*&-0.0053*&-0.0062*&-0.0027*\\
IJOC&~&~&-0.0045*&-0.0040*&-0.0048*&-0.0013\textcolor{white}{*}\\
BSP&~&~&~&0.0005\textcolor{white}{*}&-0.0004*&0.0031*\\
BSR&~&~&~&~&-0.0009\textcolor{white}{*}&0.0026*\\
BSPN&~&~&~&~&~&0.0035*\\
\toprule
~ & \multicolumn{6}{c}{DB region}\\
\cline{2-7}  \noalign{\smallskip}
~ & \multicolumn{1}{c}{Basic} & \multicolumn{1}{c}{IJOC} & \multicolumn{1}{c}{BSP} & \multicolumn{1}{c}{BSR} & \multicolumn{1}{c}{BSPN} & \multicolumn{1}{c}{BSPNC} \\
\midrule
Basic&~&0.0034\textcolor{white}{*}&0.0021\textcolor{white}{*}&~&-0.0005\textcolor{white}{*}&-0.0004\textcolor{white}{*}\\
IJOC&~&~&-0.0013\textcolor{white}{*}&~&-0.0038\textcolor{white}{*}&-0.0038\textcolor{white}{*}\\
BSP&~&~&~&~&-0.0025\textcolor{white}{*}&-0.0025\textcolor{white}{*}\\
BSPN&~&~&~&~&~&0.0000\textcolor{white}{*}\\
\bottomrule
\end{tabular}}~\\
\end{center}
\small
{\em Notes.} We conduct paired samples t-tests:
$H_0$: the means of the scores underlying the row skill model and the column skill model are equal.
vs
$H_1$: the mean of the scores underlying the row skill model is higher (or lower depending on whether the estimated difference in scores is higher or lower) than that of the column model.
The difference of means (row model - column model) is reported, and a ``*'' is marked if $H_0$ is rejected with a p-value smaller than 0.05.
Treble regions include T17-T20. Double regions include D1-D20 except for D11, D13, D15, and D17.
\end{table}

\begin{table}[ht]
\begin{center}
\captionsetup{justification=centering}
{
\renewcommand{\tabcolsep}{1.2mm}
\caption{Paired t-tests on Spherical scores of skill models}
\vspace{-1mm}
\label{table:SphericalScoresT-test}
\small
\begin{tabular}{r rrrrrrrr}
\toprule
~ & \multicolumn{6}{c}{treble regions}\\
\cline{2-7}  \noalign{\smallskip}
~ & \multicolumn{1}{c}{Basic} & \multicolumn{1}{c}{IJOC} & \multicolumn{1}{c}{BSP} & \multicolumn{1}{c}{BSR} & \multicolumn{1}{c}{BSPN} & \multicolumn{1}{c}{BSPNC} \\
\midrule
Basic&~&-0.0003*&-0.0004*&-0.0003*&-0.0004*&-0.0003*\\
IJOC&~&~&-0.0001*&-0.0001*&-0.0001*&-0.0000*\\
BSP&~&~&~&0.0000*&0.0000\textcolor{white}{*}&0.0001*\\
BSR&~&~&~&~&-0.0000*&0.0000\textcolor{white}{*}\\
BSPN&~&~&~&~&~&0.0000*\\
\toprule
~ & \multicolumn{6}{c}{double regions}\\
\cline{2-7}  \noalign{\smallskip}
~ & \multicolumn{1}{c}{Basic} & \multicolumn{1}{c}{IJOC} & \multicolumn{1}{c}{BSP} & \multicolumn{1}{c}{BSR} & \multicolumn{1}{c}{BSPN} & \multicolumn{1}{c}{BSPNC} \\
\midrule
Basic&~&-0.0012\textcolor{white}{*}&-0.0050*&-0.0046*&-0.0054*&-0.0024*\\
IJOC&~&~&-0.0039*&-0.0034*&-0.0042*&-0.0012\textcolor{white}{*}\\
BSP&~&~&~&0.0005\textcolor{white}{*}&-0.0003*&0.0027*\\
BSR&~&~&~&~&-0.0008\textcolor{white}{*}&0.0022*\\
BSPN&~&~&~&~&~&0.0030*\\
\toprule
~ & \multicolumn{6}{c}{DB region}\\
\cline{2-7}  \noalign{\smallskip}
~ & \multicolumn{1}{c}{Basic} & \multicolumn{1}{c}{IJOC} & \multicolumn{1}{c}{BSP} & \multicolumn{1}{c}{BSR} & \multicolumn{1}{c}{BSPN} & \multicolumn{1}{c}{BSPNC} \\
\midrule
Basic&~&0.0021\textcolor{white}{*}&0.0015\textcolor{white}{*}&~&-0.0002\textcolor{white}{*}&-0.0001\textcolor{white}{*}\\
IJOC&~&~&-0.0007\textcolor{white}{*}&~&-0.0023\textcolor{white}{*}&-0.0022\textcolor{white}{*}\\
BSP&~&~&~&~&-0.0016\textcolor{white}{*}&-0.0016\textcolor{white}{*}\\
BSPN&~&~&~&~&~&0.0000\textcolor{white}{*}\\
\bottomrule
\end{tabular}}~\\
\end{center}
\small
{\em Notes.} We conduct paired samples t-tests:
$H_0$: the means of the scores underlying the row skill model and the column skill model are equal.
vs
$H_1$: the mean of the scores underlying the row skill model is higher (or lower depending on whether the estimated difference in scores is higher or lower) than that of the column model.
The difference of means (row model - column model) is reported, and a ``*'' is marked if $H_0$ is rejected with a p-value smaller than 0.05.
Treble regions include T17-T20. Double regions include D1-D20 except for D11, D13, D15, and D17.
\end{table}


\clearpage
\subsection{Figures and Tables from Section \ref{sec:Viz} for all Sixteen Players}
\label{sec:AppAllPlayersFigures}

\begin{figure}[h]
\begin{center}
\includegraphics[width=1\linewidth]{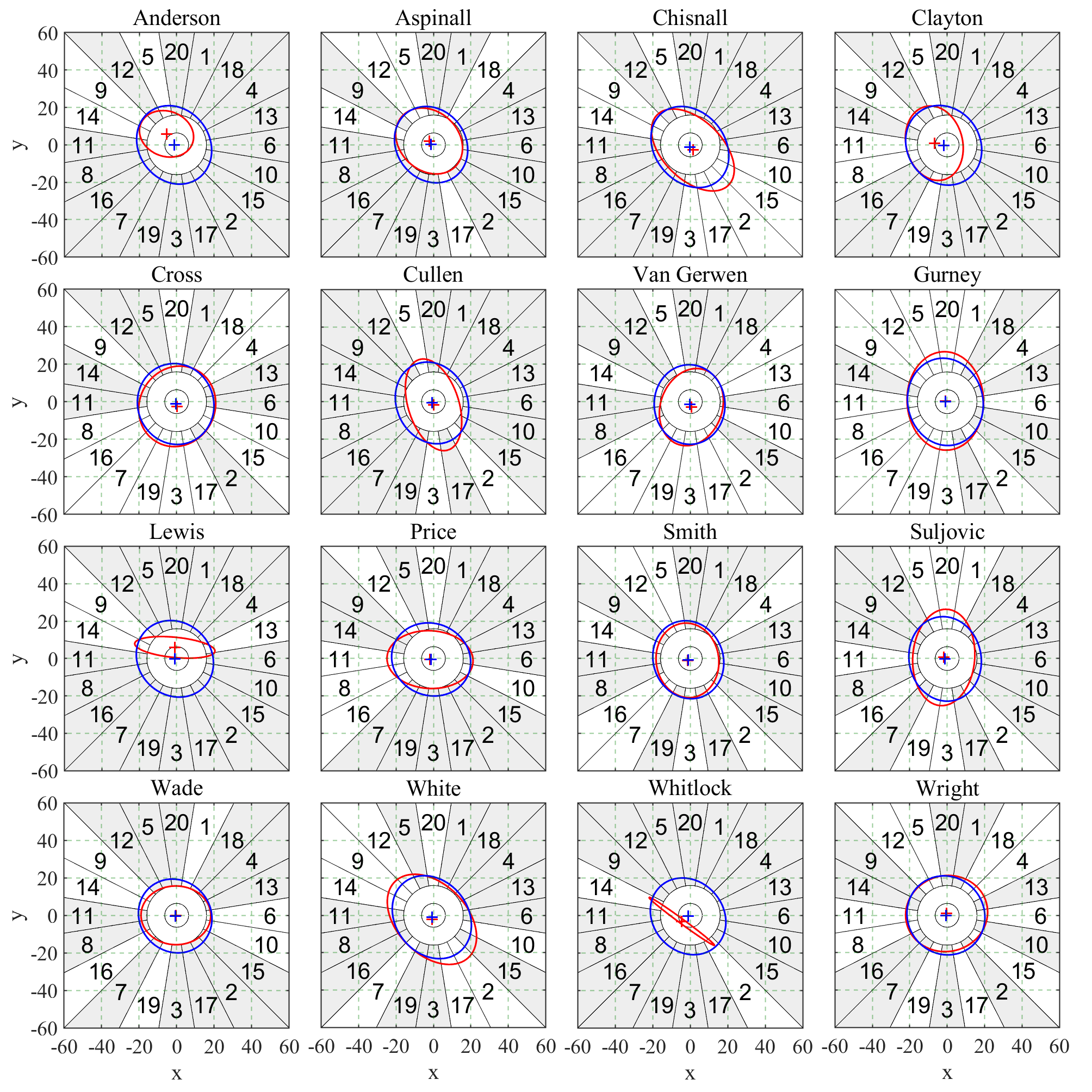}
\end{center}
\vspace{-0.5cm}
\caption{95\% CEs for the BSPN (blue) and RN (red) models when targeting DB. Regions with zero observations in the raw-data are shaded gray.}
\label{figure:DB_Fits}
\end{figure}

\begin{figure}[h]
\begin{center}
\includegraphics[width=1\linewidth]{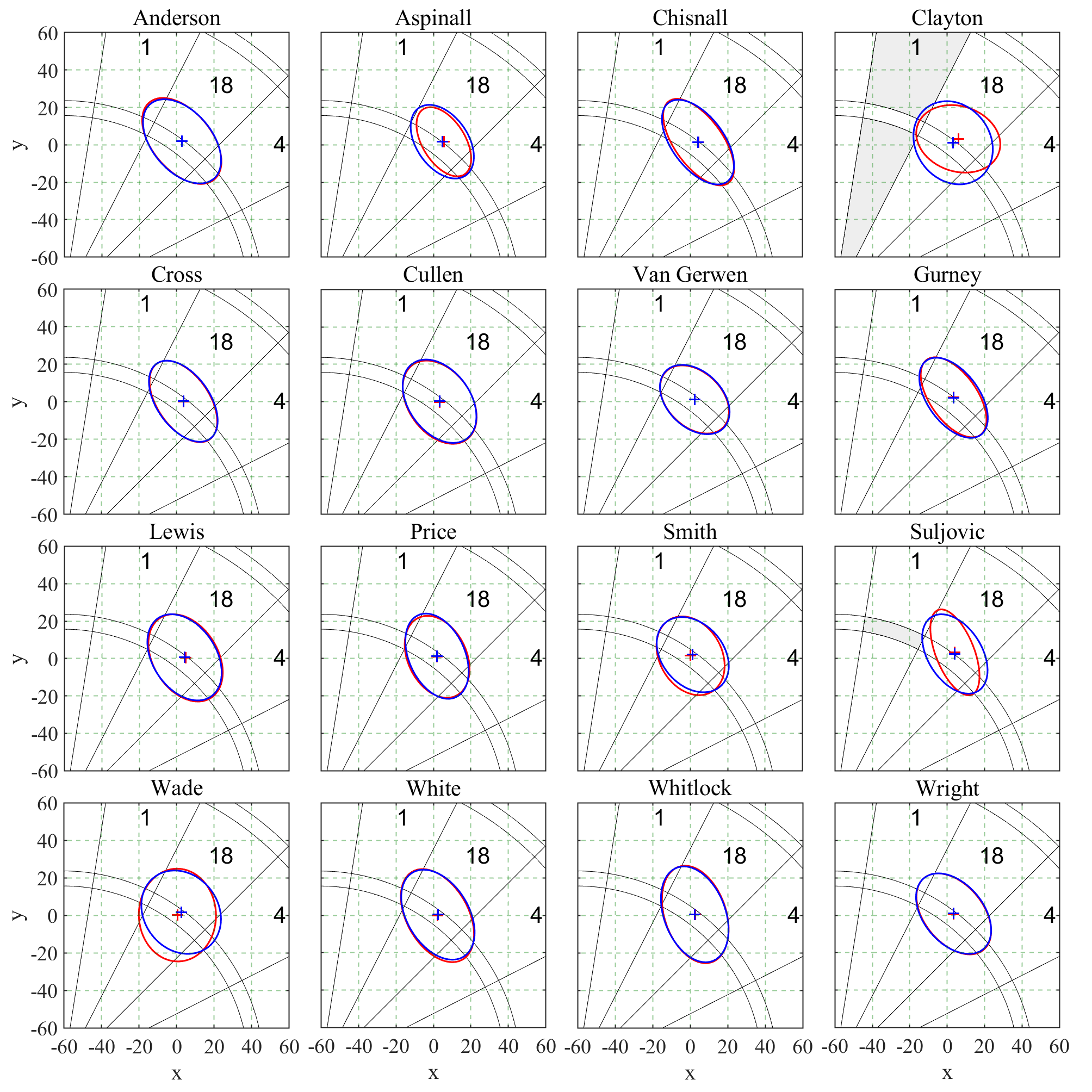}
\end{center}
\vspace{-0.5cm}
\caption{95\% CEs for the BSPN (blue) and RN (red) models when targeting T18. Regions with zero observations in the raw-data are shaded gray.}
\label{figure:T18_Fits}
\end{figure}

\begin{figure}[h]
\begin{center}
\includegraphics[width=1\linewidth]{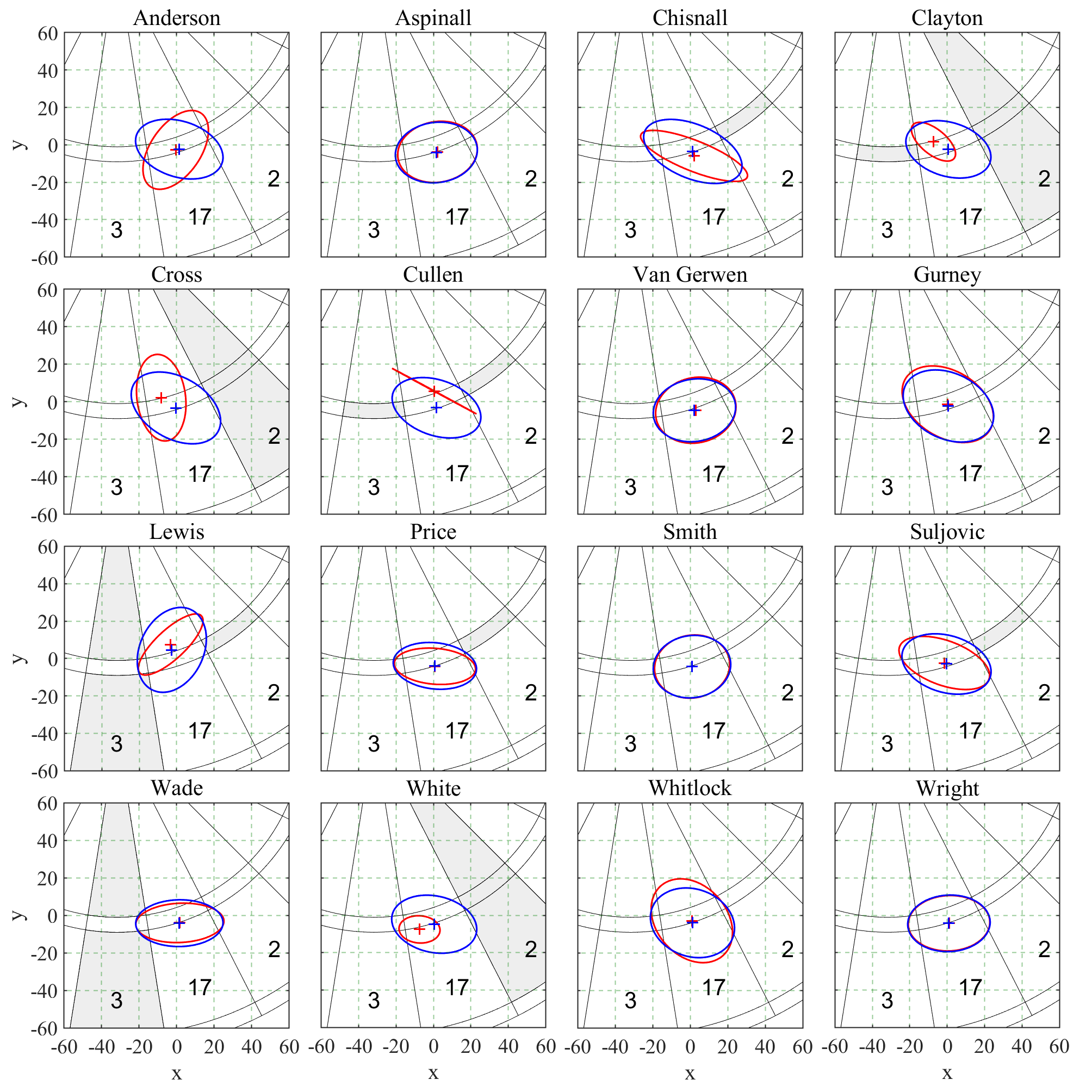}
\end{center}
\vspace{-0.5cm}
\caption{95\% CEs for the BSPN (blue) and RN (red) models when targeting T17. Regions with zero observations in the raw-data are shaded gray.}
\label{figure:T17_Fits}
\end{figure}

\begin{figure}[h]
\begin{center}
\includegraphics[width=1\linewidth]{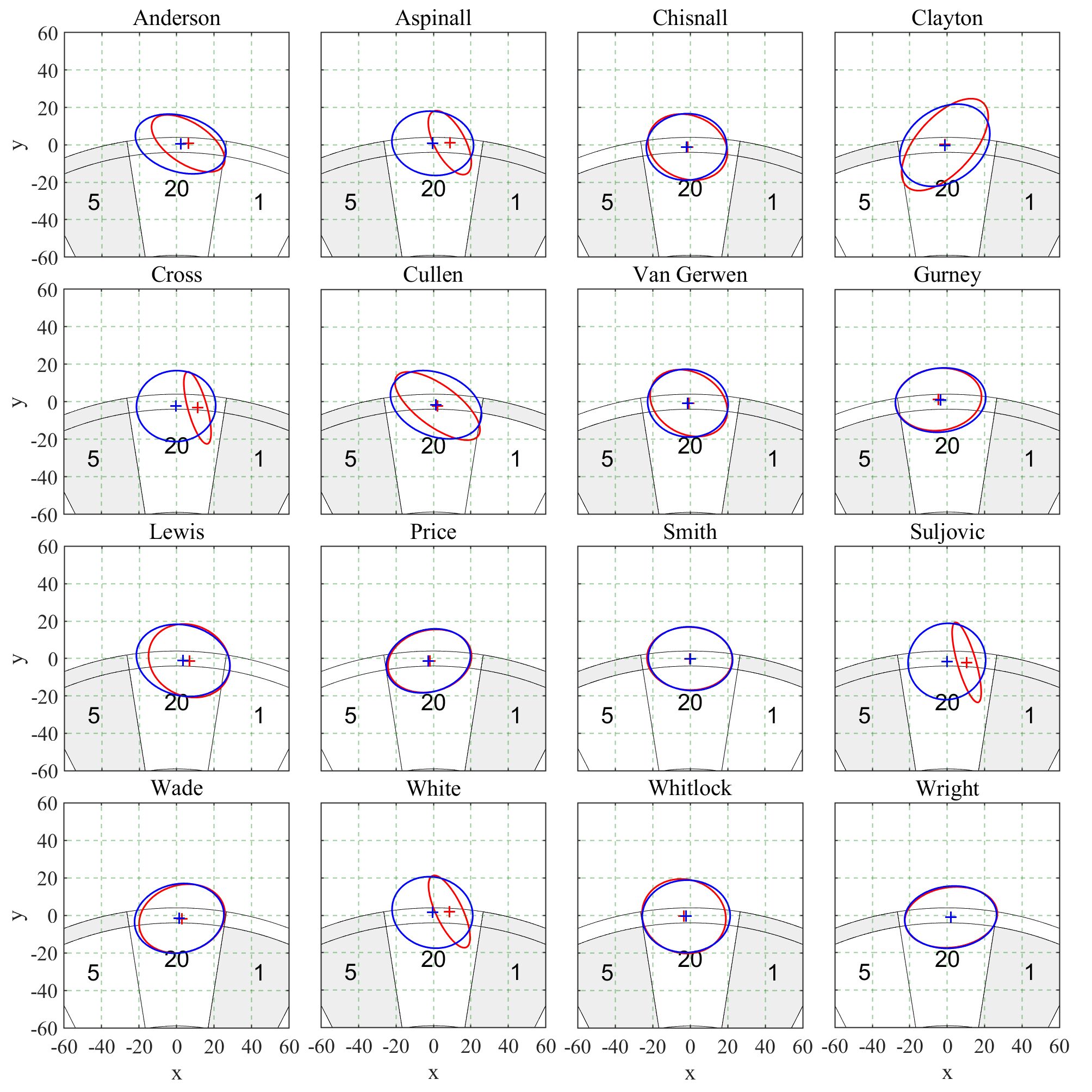}
\end{center}
\vspace{-0.5cm}
\caption{95\% CEs for the BSPN (blue) and RN (red) models when targeting D20. Regions with zero observations in the raw-data are shaded gray.}
\label{figure:D20_Fits}
\end{figure}

\begin{figure}[h]
\begin{center}
\includegraphics[width=1\linewidth]{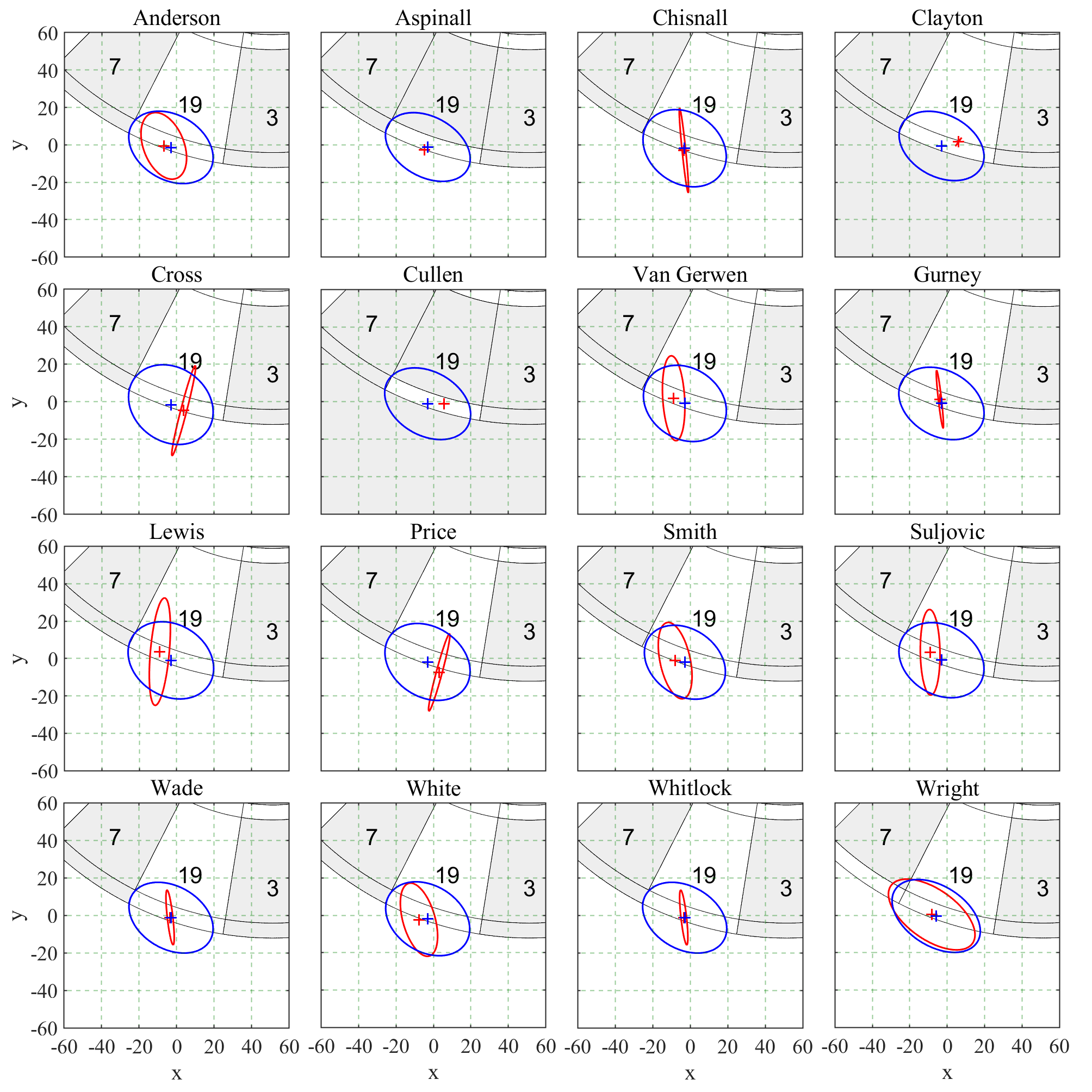}
\end{center}
\vspace{-0.5cm}
\caption{95\% CEs for the BSPN (blue) and RN (red) models when targeting D19. Regions with zero observations in the raw-data are shaded gray.}
\label{figure:D19_Fits}
\end{figure}

\begin{figure}[h]
\begin{center}
\includegraphics[width=1\linewidth]{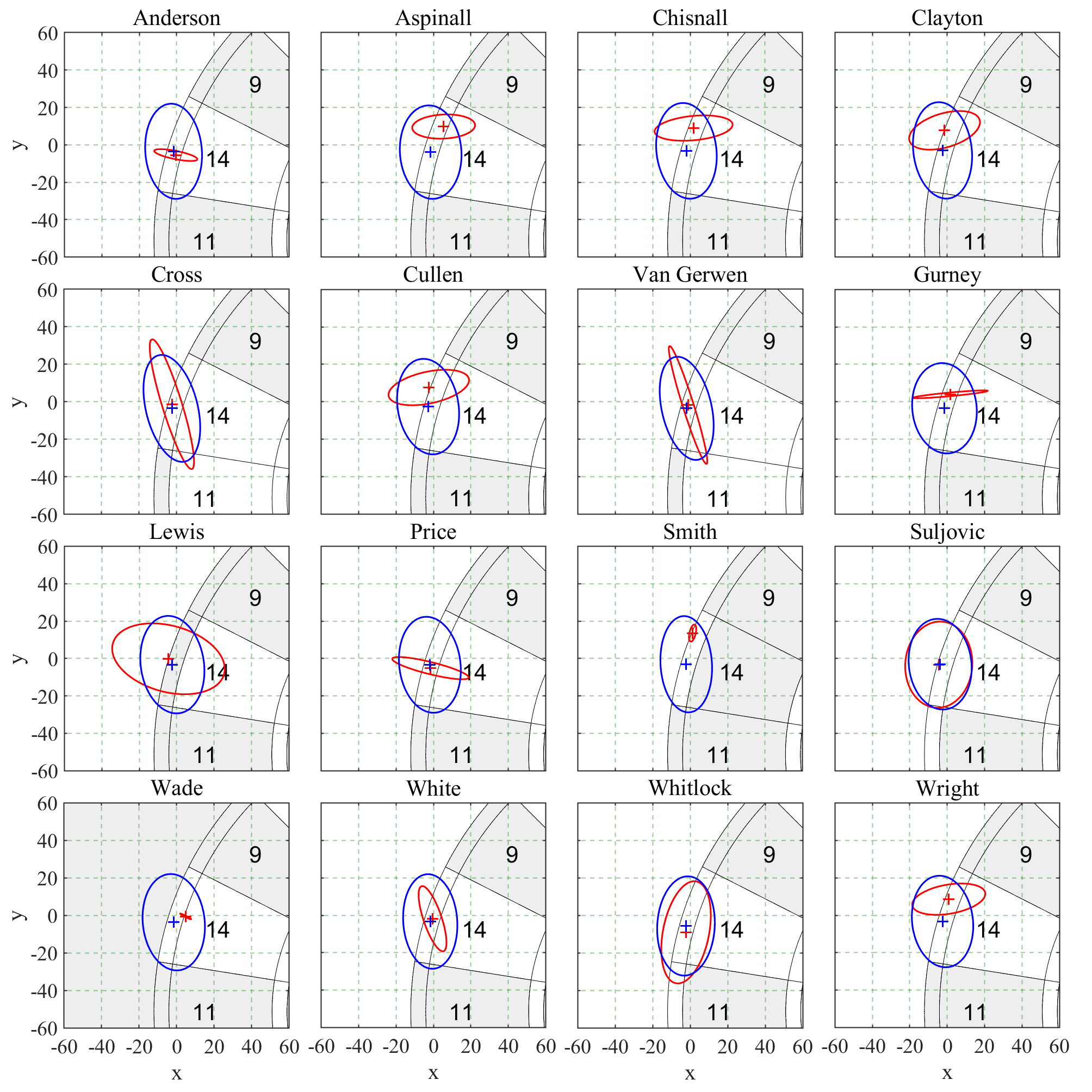}
\end{center}
\vspace{-0.5cm}
\caption{95\% CEs for the BSPN (blue) and RN (red) models when targeting D14. Regions with zero observations in the raw-data are shaded gray.}
\label{figure:D14_Fits}
\end{figure}

\section{The Treble Data and Fitted RN Models} 
\label{app:Data}

In Tables \ref{tab:T20} to \ref{tab:T17} we display\footnote{We only display the data for the four treble regions here. The entire data-set is available at \url{https://github.com/wangchunsem/OptimalDarts}, however.} the data from the 2019 season for the top 16 players in the world when they were targeting T20, T19, T18, and T17, respectively. We see the success rates vary from approx. 36\% to 45\% for T20, 36\% to 46\% for T19, 26\% to 42\% for T18, and 20\% to 48\%for T17. The average success rate across all players varied from  41.7\% for T19 to  33.5\% for T17. Then in Tables \ref{table:FittedProb_RealData_T20_nonfixedmu-s} to \ref{table:FittedProb_RealData_T17_nonfixedmu} we display the observed versus fitted scoring percentages (using the RN model) for each possible outcome when the targets are T20, T19, T18, and T17, respectively. Because of the minimal shrinkage that takes place when we target the trebles, Tables \ref{table:FittedProb_RealData_T20_nonfixedmu-s} to \ref{table:FittedProb_RealData_T17_nonfixedmu} would be very similar if we had displayed the BSPN fitted scoring percentages instead.

\begin{table}[h]
\begin{center}
\captionsetup{justification=centering}
{
\caption{Data for T20 as Target} \vspace{-1mm}
\label{tab:T20}
\small
\begin{tabular}{l c rrrrrr c cc}
\toprule
\multirow{2}{*}{Player} & ~ & \multicolumn{6}{c}{Hitting Score} & ~ & Success & Expected \\
\cline{3-8} \noalign{\smallskip}
~ & ~ & \multicolumn{1}{c}{T20} & \multicolumn{1}{c}{S20} & \multicolumn{1}{c}{T5} & \multicolumn{1}{c}{S5} & \multicolumn{1}{c}{T1} & \multicolumn{1}{c}{S1} & ~ & Rate & Score \\
\midrule
Anderson        & ~ &	1,849	&	2,075	&	85	&	126	&	91	&	137	& ~ &	42.4\%	&	35.5 \\
Aspinall		& ~ &	2,572	&	3,143	&	103	&	148	&	75	&	136	& ~ &	41.6\%	&	35.6 \\
Chisnall		& ~ &	2,985	&	3,191	&	154	&	207	&	104	&	209	& ~ &	43.6\%	&	36.0 \\
Clayton         & ~ &	1,295	&	1,822	&	85	&	149	&	41	&	83	& ~ &	37.3\%	&	33.5 \\
Cross		    & ~ &	4,117	&	5,017	&	99	&	134	&	108	&	248	& ~ &	42.3\%	&	36.0 \\
Cullen		    & ~ &	1,163	&	1,608	&	59	&	80	&	39	&	108	& ~ &	38.0\%	&	33.8 \\
van Gerwen      & ~ &	5,134	&	5,464	&	227	&	240	&	107	&	169	& ~ &	45.3\%	&	37.2 \\
Gurney		    & ~ &	4,014	&	5,243	&	217	&	264	&	53	&	143	& ~ &	40.4\%	&	35.3 \\
Lewis           & ~ &	1,573	&	2,027	&	66	&	103	&	88	&	173	& ~ &	39.0\%	&	34.0 \\
Price           & ~ &	4,077	&	4,472	&	253	&	276	&	175	&	185	& ~ &	43.2\%	&	36.0 \\
Smith           & ~ &	3,761	&	4,624	&	203	&	293	&	58	&	109	& ~ &	41.6\%	&	35.7 \\
Suljovic		& ~ &	3,141	&	4,095	&	144	&	291	&	51	&	94	& ~ &	40.2\%	&	35.1 \\
Wade		    & ~ &	3,840	&	5,584	&	116	&	271	&	121	&	170	& ~ &	38.0\%	&	34.2 \\
White		    & ~ &	2,784	&	3,259	&	111	&	206	&	103	&	212	& ~ &	41.7\%	&	35.3 \\
Whitlock		& ~ &	1,261	&	1,920	&	73	&	106	&	42	&	77	& ~ &	36.2\%	&	33.3 \\
Wright		    & ~ &	4,926	&	6,024	&	254	&	319	&	212	&	357	& ~ &	40.7\%	&	34.9 \\
\midrule
Total & ~ &	48,492	&   59,568	& 2,249	& 3,213	& 1,468	& 2,610	& ~ &   41.2\%	&   35.4 \\
\bottomrule
\end{tabular}}~\\
\end{center}
\small
{\em Notes.} The table displays the number of darts that achieved a particular score when the target was T20. For example, Cross succeeded in hitting T20 4,117 times and therefore scored 60 for each of these darts. He hit S20 5,017 times, T5 (for a score of 15) 99 times etc. The ``Success Rate'' column contains the percentage of darts that succeeded in hitting T20 while the ``Expected Score'' column yields the expected score using the data's empirical distribution.
\end{table}

\begin{table}
\begin{center}
\captionsetup{justification=centering}
{
\caption{Data for T19 as Target} \vspace{-1mm}
\label{tab:T19}
\small
\begin{tabular}{l c rrrrrr c cc}
\toprule
\multirow{2}{*}{Player} & ~ & \multicolumn{6}{c}{Hitting Score} & ~ & Success & Expected \\
\cline{3-8} \noalign{\smallskip}
~ & ~ & \multicolumn{1}{c}{T19} & \multicolumn{1}{c}{S19} & \multicolumn{1}{c}{T7} & \multicolumn{1}{c}{S7} & \multicolumn{1}{c}{T3} & \multicolumn{1}{c}{S3} & ~ & Rate & Score \\
\midrule
Anderson& ~ & 379 & 467 & 12 & 21 & 25 & 31 & ~ &40.5\%&33.4\\
Aspinall& ~ & 593 & 707 & 26 & 25 & 9 & 24 & ~ &42.8\%&34.8\\
Chisnall& ~ & 671 & 680 & 46 & 46 & 26 & 31 & ~ &44.7\%&35.2\\
Clayton& ~ & 147 & 200 & 10 & 8 & 5 & 11 & ~ &38.6\%&32.9\\
Cross& ~ & 1,531 & 1,882 & 33 & 49 & 73 & 99 & ~ &41.8\%&34.1\\
Cullen& ~ & 498 & 626 & 16 & 17 & 22 & 26 & ~ &41.3\%&34.0\\
van Gerwen& ~ & 2,259 & 2,401 & 58 & 83 & 38 & 54 & ~ &46.2\%&36.1\\
Gurney& ~ & 772 & 1,109 & 39 & 54 & 18 & 25 & ~ &38.3\%&33.0\\
Lewis& ~ & 350 & 491 & 18 & 27 & 17 & 22 & ~ &37.8\%&32.5\\
Price& ~ & 1,156 & 1,423 & 38 & 60 & 30 & 71 & ~ &41.6\%&34.1\\
Smith& ~ & 982 & 1,251 & 35 & 64 & 12 & 27 & ~ &41.4\%&34.2\\
Suljovic& ~ & 376 & 585 & 17 & 27 & 2 & 11 & ~ &36.9\%&32.6\\
Wade& ~ & 336 & 550 & 7 & 13 & 13 & 19 & ~ &35.8\%&32.0\\
White& ~ & 499 & 553 & 35 & 34 & 14 & 21 & ~ &43.2\%&34.7\\
Whitlock& ~ & 464 & 684 & 25 & 50 & 9 & 24 & ~ &36.9\%&32.2\\
Wright& ~ & 532 & 625 & 25 & 43 & 24 & 36 & ~ &41.4\%&33.7\\
\midrule
Total& ~ & 11,545 & 14,234 & 440 & 621 & 337 & 532 & ~ &41.7\%&34.2\\
\bottomrule
\end{tabular}}~\\
\end{center}
\small
{\em Notes.} The table displays the number of darts that achieved a particular score when the target was T19 together with the success rate and the expected score using the data's empirical distribution.
\end{table}

\begin{table}
\begin{center}
\captionsetup{justification=centering}
{
\caption{Data for T18 as Target} \vspace{-1mm}
\label{tab:T18}
\small
\begin{tabular}{l c rrrrrr c cc}
\toprule
\multirow{2}{*}{Player} & ~ & \multicolumn{6}{c}{Hitting Score} & ~ & Success & Expected \\
\cline{3-8} \noalign{\smallskip}
~ & ~ & \multicolumn{1}{c}{T18} & \multicolumn{1}{c}{S18} & \multicolumn{1}{c}{T4} & \multicolumn{1}{c}{S4} & \multicolumn{1}{c}{T1} & \multicolumn{1}{c}{S1} & ~ & Rate & Score \\
\midrule
Anderson        & ~ &	104	&	171	&	14	&	10	&	5	&	10	& ~ &	33.1\%	&	28.4 \\
Aspinall        & ~ &	135	&	229	&	15	&	6	&	1	&	1	& ~ &	34.9\%	&	30.0 \\
Chisnall		& ~ &	105	&	165	&	20	&	11	&	2	&	7	& ~ &	33.9\%	&	28.8 \\
Clayton		    & ~ &	28	&	72	&	2	&	4	&	2	&	0	& ~ &	25.9\%	&	26.4 \\
Cross		    & ~ &	508	&	640	&	59	&	48	&	7	&	19	& ~ &	39.7\%	&	31.1 \\
Cullen		    & ~ &	131	&	197	&	16	&	16	&	4	&	4	& ~ &	35.6\%	&	29.6 \\
van Gerwen	    & ~ &	536	&	658	&	30	&	19	&	13	&	14	& ~ &	42.2\%	&	32.5 \\
Gurney		    & ~ &	136	&	205	&	15	&	8	&	2	&	8	& ~ &	36.4\%	&	30.1 \\
Lewis		    & ~ &	106	&	195	&	18	&	17	&	3	&	3	& ~ &	31.0\%	&	27.9 \\
Price		    & ~ &	300	&	434	&	15	&	21	&	5	&	12	& ~ &	38.1\%	&	30.9 \\
Smith		    & ~ &	284	&	406	&	8	&	13	&	12	&	13	& ~ &	38.6\%	&	31.0 \\
Suljovic		& ~ &	80	&	152	&	7	&	5	&	0	&	3	& ~ &	32.4\%	&	29.0 \\
Wade		    & ~ &	83	&	174	&	4	&	8	&	5	&	4	& ~ &	29.9\%	&	27.7 \\
White		    & ~ &	90	&	136	&	10	&	13	&	3	&	6	& ~ &	34.9\%	&	29.1 \\
Whitlock		& ~ &	118	&	206	&	10	&	18	&	2	&	7	& ~ &	32.7\%	&	28.5 \\
Wright		    & ~ &	106	&	158	&	13	&	11	&	3	&	5	& ~ &	35.8\%	&	29.7 \\
\midrule
Total   			& ~ &	2,850&	4,198&	256	&   228	&   69	&   116	& ~ &   36.9\%	&   30.3 \\
\bottomrule
\end{tabular}}~\\
\end{center}
\small
{\em Notes.} The table displays the number of darts that achieved a particular score when the target was T18 together with the success rate and the expected score using the data's empirical distribution.
\end{table}

\begin{table}
\begin{center}
\captionsetup{justification=centering}
{
\caption{Data for T17 as Target} \vspace{-1mm}
\label{tab:T17}
\small
\begin{tabular}{l c rrrrrr c cc}
\toprule
\multirow{2}{*}{Player} & ~ & \multicolumn{6}{c}{Hitting Score} & ~ & Success & Expected \\
\cline{3-8} \noalign{\smallskip}
~ & ~ & \multicolumn{1}{c}{T17} & \multicolumn{1}{c}{S17} & \multicolumn{1}{c}{T3} & \multicolumn{1}{c}{S3} & \multicolumn{1}{c}{T2} & \multicolumn{1}{c}{S2} & ~ & Rate & Score \\
\midrule
Anderson& ~ &59&88&1&3&1&3& ~ &38.1\%&29.3\\
Aspinall& ~ &49&87&2&1&2&3& ~ &34.0\%&27.9\\
Chisnall& ~ &42&91&2&6&0&4& ~ &29.0\%&25.7\\
Clayton& ~ &22&23&0&1&0&0& ~ &47.8\%&33.0\\
Cross& ~ &33&77&2&5&0&0& ~ &28.2\%&25.9\\
Cullen& ~ &18&32&0&1&0&1& ~ &34.6\%&28.2\\
van Gerwen& ~ &68&144&2&2&3&6& ~ &30.2\%&26.5\\
Gurney& ~ &72&147&3&7&2&4& ~ &30.6\%&26.5\\
Lewis& ~ &13&48&2&0&0&3& ~ &19.7\%&22.8\\
Price& ~ &82&108&4&2&0&3& ~ &41.2\%&30.5\\
Smith& ~ &114&216&6&2&4&3& ~ &33.0\%&27.8\\
Suljovic& ~ &35&51&2&4&0&1& ~ &37.6\%&28.9\\
Wade& ~ &54&79&5&0&1&4& ~ &37.8\%&29.1\\
White& ~ &27&57&2&2&0&0& ~ &30.7\%&26.9\\
Whitlock& ~ &34&79&1&2&1&1& ~ &28.8\%&26.3\\
Wright& ~ &102&172&5&4&2&5& ~ &35.2\%&28.3\\
\midrule
Total& ~ &824&1,499&39&42&16&41& ~ &33.5\%&27.7\\
\bottomrule
\end{tabular}}~\\
\end{center}
\small
{\em Notes.} The table displays the number of darts that achieved a particular score when the target was T17 together with the success rate and the expected score using the data's empirical distribution.
\end{table}

\newpage

\begin{table}
\begin{center}
\captionsetup{justification=centering}
{
\renewcommand{\tabcolsep}{1.0mm}
\caption{Observed vs fitted (RN model) scoring percentages for all 16 players on T20.}
\vspace{-1mm}
\label{table:FittedProb_RealData_T20_nonfixedmu-s}
\small
\begin{tabular}{lc rrr rrr rrr rrr rrr rrr |r}
\toprule
\multirow{2}{*}{Player} &~& \multicolumn{2}{c}{T20} &~& \multicolumn{2}{c}{S20} &~& \multicolumn{2}{c}{T5} &~& \multicolumn{2}{c}{S5} &~& \multicolumn{2}{c}{T1} &~& \multicolumn{2}{c}{S1}  \\
\cline{3-4} \cline{6-7} \cline{9-10} \cline{12-13} \cline{15-16} \cline{18-19} \noalign{\smallskip}
~ &~& Obs. & Fit. &~& Obs. & Fit. &~& Obs. & Fit. &~& Obs. & Fit. &~& Obs. & Fit. &~& Obs. & Fit.  \\
\midrule
Anderson&~&42.4&42.3&~&47.6&47.6&~&1.9&1.6&~&2.9&3.2&~&2.1&2.2&~&3.1&3.1 \\
Aspinall&~&41.6&41.6&~&50.9&51.1&~&1.7&1.8&~&2.4&2.1&~&1.2&1.0&~&2.2&2.4 \\
Chisnall&~&43.6&43.4&~&46.6&46.9&~&2.2&2.4&~&3.0&2.7&~&1.5&1.3&~&3.1&3.2 \\
Clayton&~&37.3&37.3&~&52.4&52.6&~&2.4&2.6&~&4.3&3.9&~&1.2&0.9&~&2.4&2.7 \\
Cross&~&42.3&42.3&~&51.6&51.6&~&1.0&0.7&~&1.4&1.6&~&1.1&1.2&~&2.6&2.5 \\
Cullen&~&38.0&38.0&~&52.6&52.8&~&1.9&2.0&~&2.6&2.3&~&1.3&1.1&~&3.5&3.8 \\
van Gerwen&~&45.3&45.3&~&48.2&48.3&~&2.0&2.0&~&2.1&1.9&~&0.9&0.7&~&1.5&1.7 \\
Gurney&~&40.4&40.6&~&52.8&52.8&~&2.2&2.2&~&2.7&2.4&~&0.5&0.3&~&1.4&1.7 \\
Lewis&~&39.0&39.0&~&50.3&50.3&~&1.6&1.3&~&2.6&2.8&~&2.2&2.2&~&4.3&4.3 \\
Price&~&43.2&43.3&~&47.4&47.4&~&2.7&2.6&~&2.9&2.8&~&1.9&1.7&~&2.0&2.2 \\
Smith&~&41.6&41.7&~&51.1&51.2&~&2.2&2.3&~&3.2&2.9&~&0.6&0.4&~&1.2&1.5 \\
Suljovic&~&40.2&40.3&~&52.4&52.6&~&1.8&2.0&~&3.7&3.3&~&0.7&0.4&~&1.2&1.5 \\
Wade&~&38.0&38.1&~&55.3&55.4&~&1.1&1.3&~&2.7&2.3&~&1.2&0.9&~&1.7&2.0 \\
White&~&41.7&41.5&~&48.8&49.2&~&1.7&1.8&~&3.1&2.7&~&1.5&1.3&~&3.2&3.4 \\
Whitlock&~&36.2&36.3&~&55.2&55.3&~&2.1&2.2&~&3.0&2.7&~&1.2&1.0&~&2.2&2.5 \\
Wright&~&40.7&40.7&~&49.8&50.1&~&2.1&2.2&~&2.6&2.3&~&1.8&1.6&~&3.0&3.1 \\
\midrule
Fitted Error&~&~&0.1&~&~&0.2&~&~&0.1&~&~&0.3&~&~&0.2&~&~&0.2 \\
\bottomrule
\end{tabular}}~\\
\end{center}
\end{table}

\begin{table}
\begin{center}
\captionsetup{justification=centering}
{
\renewcommand{\tabcolsep}{1.0mm}
\caption{Observed vs fitted (RN model) scoring percentages for all 16 players on T19.}
\vspace{-1mm}
\label{table:FittedProb_RealData_T19_nonfixedmu}
\small
\begin{tabular}{lc ccc ccc ccc ccc ccc ccc |r}
\toprule
\multirow{2}{*}{Player} &~& \multicolumn{2}{c}{T19} &~& \multicolumn{2}{c}{S19} &~& \multicolumn{2}{c}{T7} &~& \multicolumn{2}{c}{S7} &~& \multicolumn{2}{c}{T3} &~& \multicolumn{2}{c}{S3}  \\
\cline{3-4} \cline{6-7} \cline{9-10} \cline{12-13} \cline{15-16} \cline{18-19} \noalign{\smallskip}
~ &~& Obs. & Fit. &~& Obs. & Fit. &~& Obs. & Fit. &~& Obs. & Fit. &~& Obs. & Fit. &~& Obs. & Fit.  \\
\midrule
Anderson&~&40.5&40.5&~&49.9&50.2&~&1.3&1.1&~&2.2&2.5&~&2.7&2.7&~&3.3&3.1 \\
Aspinall&~&42.8&42.8&~&51.1&51.2&~&1.9&1.9&~&1.8&1.9&~&0.7&0.6&~&1.7&1.6 \\
Chisnall&~&44.7&44.8&~&45.3&45.2&~&3.1&3.1&~&3.1&3.1&~&1.7&1.4&~&2.1&2.4 \\
Clayton&~&38.6&38.7&~&52.5&52.4&~&2.6&2.4&~&2.1&2.5&~&1.3&1.2&~&2.9&2.9 \\
Cross&~&41.8&41.8&~&51.3&51.5&~&0.9&0.7&~&1.3&1.6&~&2.0&2.0&~&2.7&2.4 \\
Cullen&~&41.3&41.4&~&52.0&52.0&~&1.3&1.2&~&1.4&1.7&~&1.8&1.7&~&2.2&2.0 \\
van Gerwen&~&46.2&46.2&~&49.1&49.1&~&1.2&1.3&~&1.7&1.7&~&0.8&0.7&~&1.1&1.0 \\
Gurney&~&38.3&38.5&~&55.0&54.8&~&1.9&1.9&~&2.7&2.8&~&0.9&0.6&~&1.2&1.5 \\
Lewis&~&37.8&38.0&~&53.1&52.9&~&1.9&1.9&~&2.9&3.0&~&1.8&1.4&~&2.4&2.7 \\
Price&~&41.6&41.5&~&51.2&51.3&~&1.4&1.5&~&2.2&2.2&~&1.1&1.0&~&2.6&2.5 \\
Smith&~&41.4&41.6&~&52.8&52.7&~&1.5&1.6&~&2.7&2.7&~&0.5&0.4&~&1.1&1.1 \\
Suljovic&~&36.9&37.2&~&57.5&57.3&~&1.7&1.7&~&2.7&2.7&~&0.2&0.1&~&1.1&1.0 \\
Wade&~&35.8&36.0&~&58.6&58.6&~&0.7&0.5&~&1.4&1.7&~&1.4&1.3&~&2.0&1.8 \\
White&~&43.2&43.4&~&47.8&47.7&~&3.0&3.0&~&2.9&3.0&~&1.2&0.9&~&1.8&2.1 \\
Whitlock&~&36.9&37.1&~&54.5&54.4&~&2.0&2.1&~&4.0&4.0&~&0.7&0.6&~&1.9&1.9 \\
Wright&~&41.4&41.5&~&48.6&48.5&~&1.9&2.1&~&3.3&3.4&~&1.9&1.7&~&2.8&2.8 \\
\midrule
Fitted Error&~&~&0.1&~&~&0.1&~&~&0.1&~&~&0.1&~&~&0.2&~&~&0.2 \\
\bottomrule
\end{tabular}}~\\
\end{center}
\end{table}

\begin{table}
\begin{center}
\captionsetup{justification=centering}
{
\renewcommand{\tabcolsep}{1.0mm}
\caption{Observed vs fitted (RN model) scoring percentages for all 16 players on T18.}
\vspace{-1mm}
\label{table:FittedProb_RealData_T18_nonfixedmu}
\small
\begin{tabular}{lc ccc ccc ccc ccc ccc ccc |r}
\toprule
\multirow{2}{*}{Player} &~& \multicolumn{2}{c}{T18} &~& \multicolumn{2}{c}{S18} &~& \multicolumn{2}{c}{T4} &~& \multicolumn{2}{c}{S4} &~& \multicolumn{2}{c}{T1} &~& \multicolumn{2}{c}{S1}  \\
\cline{3-4} \cline{6-7} \cline{9-10} \cline{12-13} \cline{15-16} \cline{18-19} \noalign{\smallskip}
~ &~& Obs. & Fit. &~& Obs. & Fit. &~& Obs. & Fit. &~& Obs. & Fit. &~& Obs. & Fit. &~& Obs. & Fit.  \\
\midrule
Anderson&~&33.1&34.0&~&54.5&53.7&~&4.5&3.5&~&3.2&4.2&~&1.6&1.2&~&3.2&3.4 \\
Aspinall&~&34.9&35.8&~&59.2&58.5&~&3.9&3.1&~&1.6&2.2&~&0.3&0.0&~&0.3&0.4 \\
Chisnall&~&33.9&34.7&~&53.2&52.6&~&6.5&5.5&~&3.5&4.5&~&0.6&0.3&~&2.3&2.4 \\
Clayton&~&25.9&27.1&~&66.7&65.6&~&1.9&1.3&~&3.7&4.4&~&1.9&0.6&~&0.0&1.0 \\
Cross&~&39.7&39.9&~&50.0&49.8&~&4.6&4.3&~&3.7&4.1&~&0.5&0.4&~&1.5&1.5 \\
Cullen&~&35.6&36.3&~&53.5&52.9&~&4.3&3.8&~&4.3&4.9&~&1.1&0.7&~&1.1&1.5 \\
van Gerwen&~&42.2&42.7&~&51.8&51.4&~&2.4&1.9&~&1.5&2.0&~&1.0&0.8&~&1.1&1.2 \\
Gurney&~&36.4&36.8&~&54.8&54.5&~&4.0&3.4&~&2.1&2.8&~&0.5&0.3&~&2.1&2.2 \\
Lewis&~&31.0&32.0&~&57.0&56.1&~&5.3&4.3&~&5.0&5.9&~&0.9&0.4&~&0.9&1.3 \\
Price&~&38.1&38.2&~&55.1&55.2&~&1.9&2.0&~&2.7&2.7&~&0.6&0.5&~&1.5&1.4 \\
Smith&~&38.6&38.7&~&55.2&55.0&~&1.1&1.2&~&1.8&1.7&~&1.6&1.2&~&1.8&2.2 \\
Suljovic&~&32.4&32.8&~&61.5&61.4&~&2.8&2.6&~&2.0&2.1&~&0.0&0.0&~&1.2&1.1 \\
Wade&~&29.9&30.3&~&62.6&62.2&~&1.4&1.4&~&2.9&3.1&~&1.8&1.0&~&1.4&2.2 \\
White&~&34.9&35.2&~&52.7&52.4&~&3.9&3.6&~&5.0&5.3&~&1.2&0.9&~&2.3&2.6 \\
Whitlock&~&32.7&32.8&~&57.1&56.9&~&2.8&2.8&~&5.0&5.0&~&0.6&0.4&~&1.9&1.9 \\
Wright&~&35.8&36.4&~&53.4&52.9&~&4.4&3.8&~&3.7&4.4&~&1.0&0.7&~&1.7&1.8 \\
\midrule
Fitted Error&~&~&0.5&~&~&0.4&~&~&0.5&~&~&0.5&~&~&0.4&~&~&0.3 \\
\bottomrule
\end{tabular}}~\\
\end{center}
\end{table}

\begin{table}
\begin{center}
\captionsetup{justification=centering}
{
\renewcommand{\tabcolsep}{1.0mm}
\caption{Observed vs fitted (RN model) scoring percentages for all 16 players on T17.}
\vspace{-1mm}
\label{table:FittedProb_RealData_T17_nonfixedmu}
\small
\begin{tabular}{lc ccc ccc ccc ccc ccc ccc |r}
\toprule
\multirow{2}{*}{Player} &~& \multicolumn{2}{c}{T17} &~& \multicolumn{2}{c}{S17} &~& \multicolumn{2}{c}{T3} &~& \multicolumn{2}{c}{S3} &~& \multicolumn{2}{c}{T2} &~& \multicolumn{2}{c}{S2}  \\
\cline{3-4} \cline{6-7} \cline{9-10} \cline{12-13} \cline{15-16} \cline{18-19} \noalign{\smallskip}
~ &~& Obs. & Fit. &~& Obs. & Fit. &~& Obs. & Fit. &~& Obs. & Fit. &~& Obs. & Fit. &~& Obs. & Fit.  \\
\midrule
Anderson&~&38.1&38.0&~&56.8&57.1&~&0.6&0.4&~&1.9&2.2&~&0.6&0.7&~&1.9&1.6 \\
Aspinall&~&34.0&34.6&~&60.4&60.0&~&1.4&0.9&~&0.7&1.3&~&1.4&1.1&~&2.1&2.1 \\
Chisnall&~&29.0&27.8&~&62.8&64.1&~&1.4&1.5&~&4.1&4.2&~&0.0&0.0&~&2.8&2.4 \\
Clayton&~&47.8&46.9&~&50.0&51.3&~&0.0&0.0&~&2.2&1.8&~&0.0&0.0&~&0.0&0.0 \\
Cross&~&28.2&28.2&~&65.8&66.0&~&1.7&1.8&~&4.3&4.1&~&0.0&0.0&~&0.0&0.0 \\
Cullen&~&34.6&28.1&~&61.5&30.4&~&0.0&0.0&~&1.9&0.5&~&0.0&0.0&~&1.9&1.3 \\
van Gerwen&~&30.2&30.4&~&64.0&63.9&~&0.9&0.7&~&0.9&1.2&~&1.3&1.2&~&2.7&2.6 \\
Gurney&~&30.6&30.7&~&62.6&62.5&~&1.3&1.2&~&3.0&3.2&~&0.9&0.6&~&1.7&1.9 \\
Lewis&~&19.7&18.0&~&72.7&74.5&~&3.0&2.4&~&0.0&1.2&~&0.0&0.0&~&4.5&4.0 \\
Price&~&41.2&40.5&~&54.3&55.0&~&2.0&2.1&~&1.0&1.0&~&0.0&0.1&~&1.5&1.3 \\
Smith&~&33.0&34.2&~&62.6&61.8&~&1.7&0.9&~&0.6&1.4&~&1.2&0.6&~&0.9&1.1 \\
Suljovic&~&37.6&38.4&~&54.8&54.2&~&2.2&1.9&~&4.3&4.5&~&0.0&0.1&~&1.1&0.9 \\
Wade&~&37.8&38.4&~&55.2&54.6&~&3.5&2.3&~&0.0&1.3&~&0.7&0.3&~&2.8&3.0 \\
White&~&30.7&31.2&~&64.8&64.5&~&2.3&2.2&~&2.3&2.1&~&0.0&0.0&~&0.0&0.0 \\
Whitlock&~&28.8&28.9&~&66.9&66.9&~&0.8&0.8&~&1.7&1.8&~&0.8&0.5&~&0.8&1.2 \\
Wright&~&35.2&35.4&~&59.3&59.2&~&1.7&1.5&~&1.4&1.7&~&0.7&0.6&~&1.7&1.6 \\
\midrule
Fitted Error&~&~&1.0&~&~&2.5&~&~&0.3&~&~&0.5&~&~&0.2&~&~&0.2 \\
\bottomrule
\end{tabular}}~\\
\end{center}
\end{table}

\end{document}